\documentclass[lettersize,journal]{IEEEtran}
\usepackage{amsmath,amsfonts}
\usepackage{algorithmic}
\usepackage{algorithm}
\usepackage{array}
\usepackage[caption=false,font=normalsize,labelfont=rm,textfont=rm]{subfig}
\usepackage{textcomp}
\usepackage{stfloats}
\usepackage{url}
\usepackage{verbatim}
\usepackage{graphicx}
\usepackage{cite}
\usepackage{multirow}
\usepackage{graphicx}
\usepackage{xcolor}
\usepackage{gensymb}
\usepackage{amssymb}
\begin{document}

\title{A Miniaturized Dynamic Array for Antenna-Level Physical Layer Security}

\author{Sheng Huang,~\IEEEmembership{Member,~IEEE}, Jacob R. Randall,~\IEEEmembership{Student Member,~IEEE}, Cory Hilton,~\IEEEmembership{Student Member,~IEEE}, Jeffrey A. Nanzer, \IEEEmembership{Senior Member,~IEEE}
        % <-this % stops a space

\thanks{Manuscript received June 13, 2025. \textit{(Corresponding author: Jeffrey A. Nanzer.)}
} 
\thanks{The authors are with the Department of Electrical and Computer Engineering, Michigan State University, East Lansing, MI 48824 USA (e-mail:
	huang287@msu.edu; randa130@msu.edu; hiltonc2@msu.edu; nanzer@msu.edu).}
}
% The paper headers
\markboth{Journal of \LaTeX\ Class Files,~Vol.~14, No.~8, August~2021}%
{Shell \MakeLowercase{\textit{et al.}}: A Sample Article Using IEEEtran.cls for IEEE Journals}

\IEEEpubid{}
% Remember, if you use this you must call \IEEEpubidadjcol in the second
% column for its text to clear the IEEEpubid mark.

\maketitle

\begin{abstract}
A compact dynamic omnidirectional array is proposed for planar physical-layer security through antenna-level directional modulation. Unlike conventional directional-modulation transmitters, the proposed architecture does not rely on phased-array beam synthesis or multiple simultaneous RF chains. Instead, angle-selective information recovery is achieved by switching the excitation paths of a four-element printed meander-line monopole array operating at 5.05~GHz. The resulting state-dependent excitation produces controllable magnitude and phase perturbations in the radiated field, which lead to strongly angle-dependent constellation distortion and bit error rate (BER) behavior. As a result, reliable information recovery is confined to a narrow broadside region in the \(E\)-plane, whereas off-broadside directions exhibit significantly degraded BER. In contrast, the \(H\)-plane remains quasi-static and omnidirectional, providing a full \(360^\circ\) information-recoverable region. The antenna is implemented on a single-layer Rogers RO4350B substrate with a compact footprint of \(0.57\times1.11\lambda_0^2\). A four-path switching network based on commercial RF components is developed to validate the concept experimentally. Communication measurements using 16-QAM at 5.05~GHz demonstrate BER-defined \(E\)-plane information beamwidths (BER\(\leq\)\(10^{-3}\)) of \(30^\circ\) to \(36^\circ\) for the calibrated switching modes, while no bit errors are observed in the measured \(H\)-plane over the full angular range and the corresponding signal-to-noise ratio (SNR) remains above approximately 33 dB. In addition, feed-phase offsets are used to steer the BER-defined information-recoverable sector, demonstrating information-beam steering with the same antenna-level switching mechanism. These results confirm that compact antenna-level directional modulation can provide angularly selective information recovery in one principal plane while preserving omnidirectional coverage in the orthogonal plane.

\end{abstract}


\section{Introduction}
\IEEEPARstart{D}{irectional} modulation (DM) has evolved from a phased-array security concept into a broader class of dynamic electromagnetic transmitters in which information recovery depends on the observation angle. In DM systems, the complex radiated field is engineered so that the intended receiver observes valid communications signals, whereas unintended directions experience symbol distortion and degraded communication quality \cite{5159486,5422702,6645431}. Because this mechanism acts directly at the antenna aperture, it is particularly attractive as a physical-layer security (PLS) strategy for compact wireless hardware where conventional cryptographic protection may be undesirable as the sole line of defense \cite{5751298,7467419,8509094}.
\begin{figure}[t]
	\begin{center}
		\includegraphics[width=3.4in]{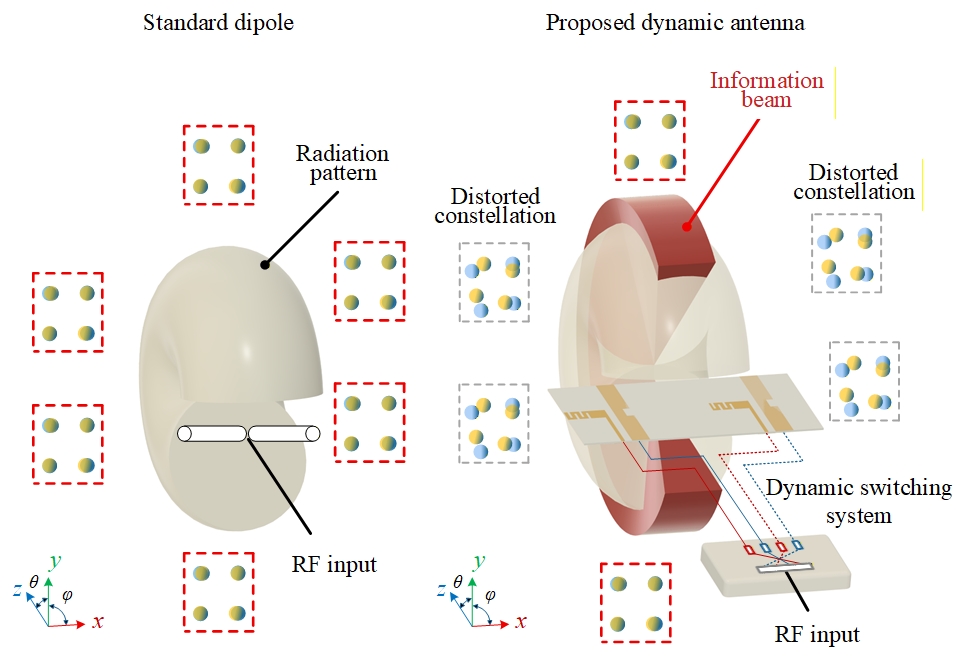}%
	\end{center}
	\caption{Conceptual comparison between a typical omnidirectional radiator and the proposed dynamic omnidirectional array for antenna-level physical-layer security. A conventional radiator provides a wide information-recoverable region over angle. In contrast, the proposed antenna uses a single RF input and a switching-controlled radiator topology to directly impose angle-dependent magnitude and phase perturbations on the radiated field, thereby forming an information beam in the intended direction while causing constellation distortion in unintended directions. Different from conventional phased-array-based secure transmitters, the proposed approach realizes dynamic secure radiation through a compact four-state antenna structure without requiring multiple RF chains or continuous phase-shifter networks.}
	\label{overview}
\end{figure}
Early DM realizations mainly relied on phased arrays and beam-steerable reconfigurable apertures \cite{5159486,5422702,5439878}. Later studies expanded the concept using 4-D arrays \cite{zhu2014dm4d}, dynamic distributed time-modulated arrays \cite{zeng2023timemoddm}, near-field direct antenna modulation \cite{4684619}, distributed antenna array dynamics \cite{9665259}, compact multiport DM antennas \cite{parron2021multiportdm}, and single-element dynamic antennas \cite{10161710}. These works substantially broadened the design space of secure radiation. However, most of them still depend on array-level synchronization, multiport feed coordination, or structurally specialized radiators, making compact planar implementation difficult when omnidirectional behavior is also required.

This paper approaches the problem from a different viewpoint by interpreting the antenna as a duplicated dynamic dipole-array structure rather than only as a compact switched radiator. Specifically, two mirrored radiators are combined to form an equivalent dipole-like cell, and two such cells are then duplicated to create a four-radiator aperture. This interpretation provides a physically transparent route for understanding how switched excitation states perturb both magnitude and phase, how an information beam is formed, and why the secure region remains confined mainly in one principal plane. In this sense, the work is not only an antenna realization, but also a structured array-based interpretation of compact dynamic DM.
The design is implemented with printed meander-line monopoles, whose folded current paths enable compact electrical length within a planar footprint \cite{99054,6171817}. Such radiators are well suited for low-profile integrated implementations and have been widely used in compact wireless and RFID systems \cite{1504825,9128053,5299014,8758307}. By embedding the mirrored-cell concept into a meander-line platform, the proposed antenna achieves a compact overall size of $0.57 \times 1.11 \times 0.0085\,\lambda_0^3$ at $5~\mathrm{GHz}$ while preserving omnidirectional behavior in the $H$-plane.
Motivated by recent progress in compact dynamic secure antennas, including energy-efficient electrically small DM antennas \cite{9674846}, high-efficiency phase-center control through spatial amplitude modulation \cite{10286341}, and steerable secure dynamic phased arrays for V2X links \cite{10979360}, the present work develops a fully planar four-state architecture driven by a single RF input. Switching the feed-path power ratios introduces state-dependent perturbations into the E-plane field while keeping the H-plane nearly unchanged, thereby producing a narrow information-recoverable sector together with broad omnidirectional coverage elsewhere. The mirrored duplicated-cell interpretation also leads naturally to the theoretical model developed in Section II, where the structure is analyzed as a two-cell dynamic dipole array.
\begin{figure*}[t]
	\centering
	\includegraphics[width=7in]{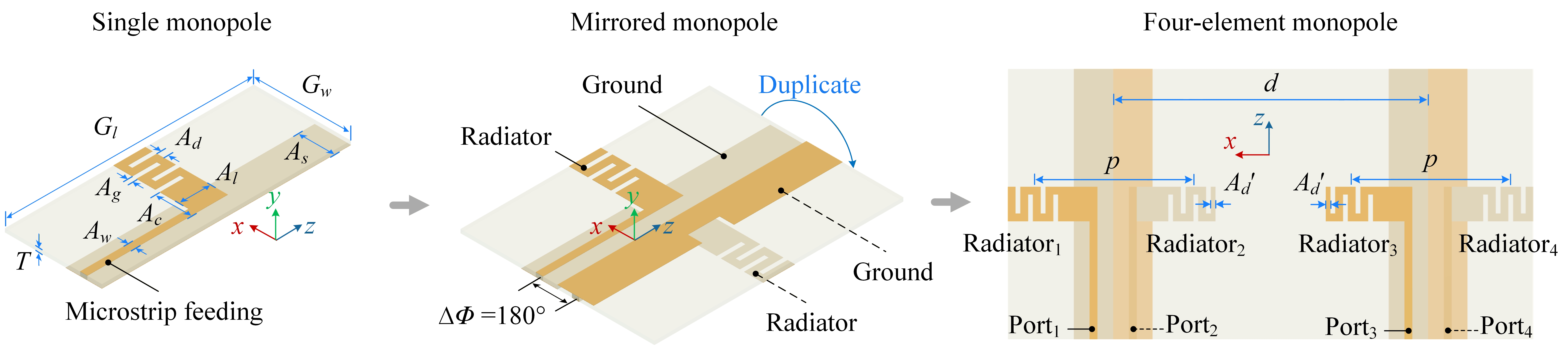}%
	\hfil
	\caption{Geometry and design evolution of the proposed planar meander-line antenna on Rogers RO4350B for operation at 5 GHz. The left panel shows the single-element structure with microstrip feeding and geometrical parameters. The middle panel illustrates the duplicated layout and the anti-phase relationship between the two radiating regions, forming an equivalent dipole-like cell. The right panel shows the resulting four-element configuration, where \(p\) and \(d\) denote the intra-cell and inter-cell spacings, respectively.}
\end{figure*}
\begin{table}[t]
	\caption{Geometrical Parameters of the Proposed Antenna Structure (Unit: mm)\label{tab:table1}}
	\centering
	\renewcommand{\arraystretch}{1.2}
	\setlength{\tabcolsep}{4pt}
	\begin{tabular}{cccccccccccc}
		\hline\hline
		\(G_l\) & \(G_w\) & \(T\)& \(d\) & \(p\) & \(A_d\) & \(A_s\) & \(A_g\) & \(A_w\)  & \(A_d'\) & \(A_l\)& \(A_c\)\\
		\hline
		34.3 & 13.4 & 0.508 & 48 & 20& 1 & 5 & 0.6 & 1  & 0.6 & 3.4&5\\
		\hline\hline
	\end{tabular}
\end{table}

The main contributions of this paper are summarized as follows. First, a compact mirrored four-radiator antenna is formulated and interpreted as a duplicated dynamic dipole-array architecture for secure transmission. Second, the mirrored-cell viewpoint provides a clear physical explanation of the state-dependent field perturbations and the resulting information beam formation. Third, the proposed antenna realizes a hardware-efficient DM platform with a single RF input and a planar switching-controlled topology. Finally, the concept is verified through full-wave simulation, communication analysis, and experimental characterization.
The rest of the paper is arranged around this design-to-validation flow. Section~II develops the mirrored meander-line element, extends it to the four-radiator aperture, and derives the switching-state model used to describe the antenna-level modulation. Section~III uses this model to evaluate the angular communication response, with emphasis on observed BER and the influence of spacing and excitation imbalance. Section~IV reports the prototype implementation, RF switching network, and measured communication performance. The main findings are summarized in Section~V.

\section{The Design of Planar Meander Line Antenna}

\begin{figure*}[t]
	\centering
	\subfloat[]{\includegraphics[width=2.35in]{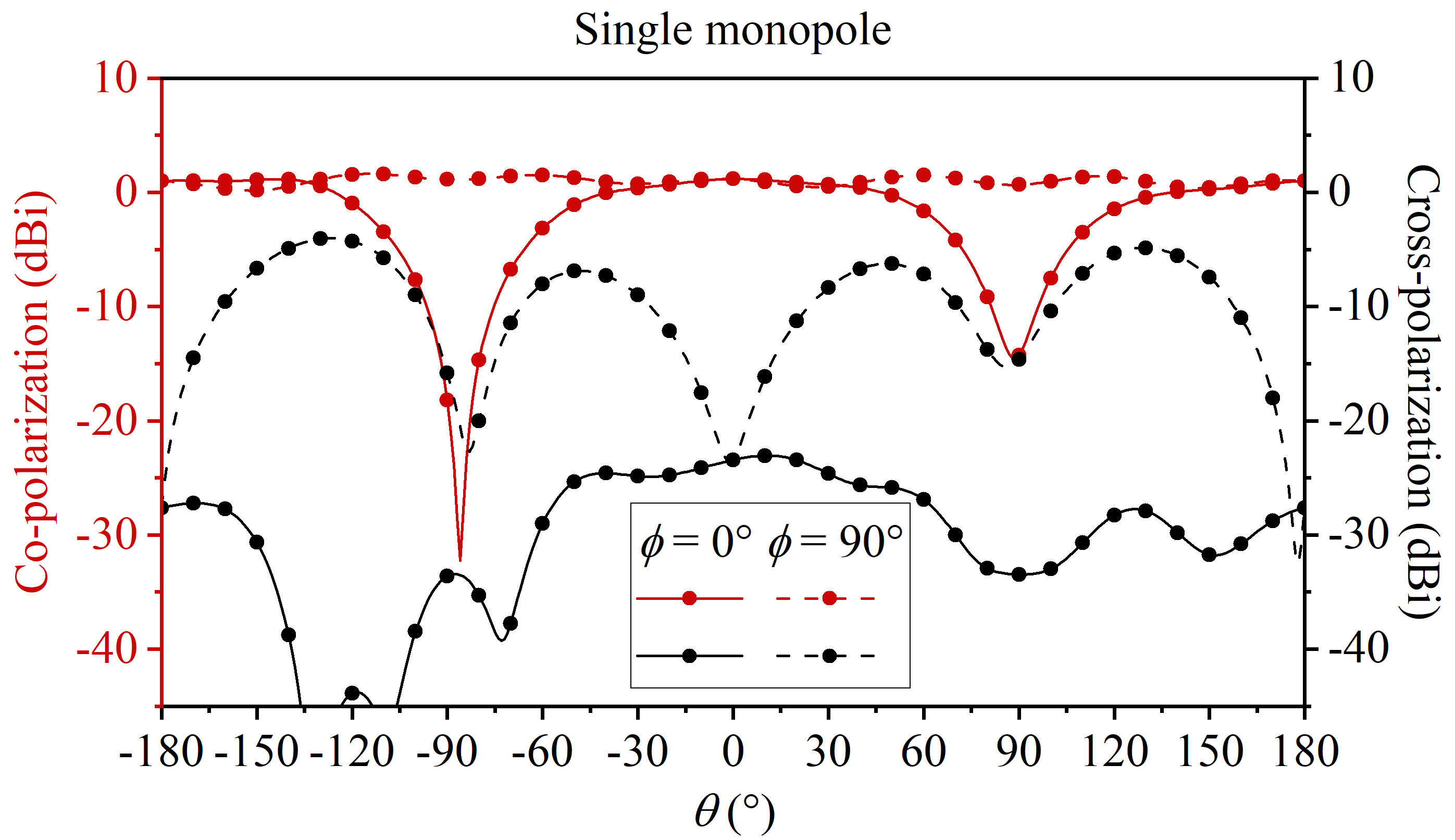}}\hfil
	\subfloat[]{\includegraphics[width=2.35in]{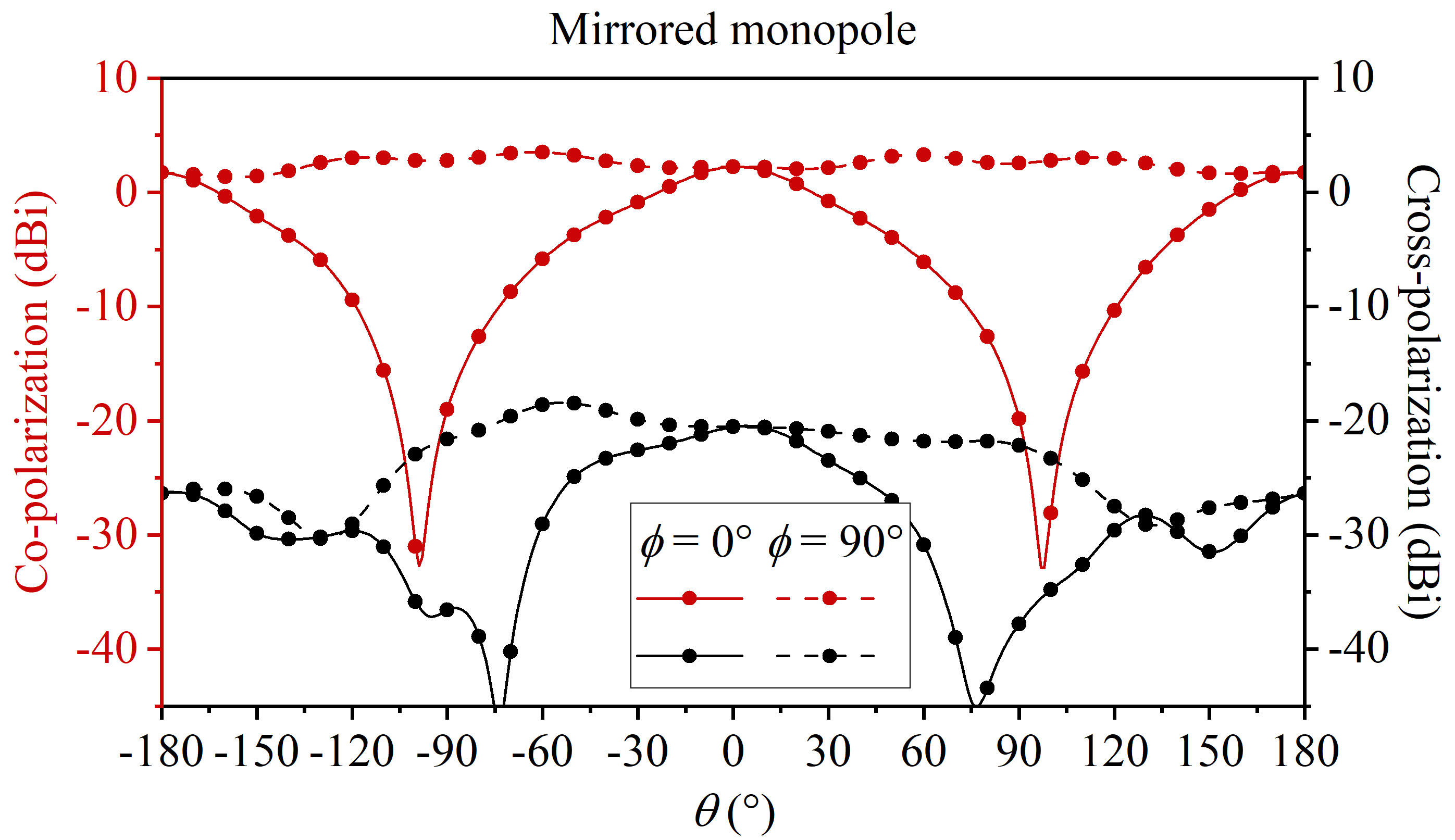}%
	}
	\hfil
	\subfloat[]{\includegraphics[width=2.35in]{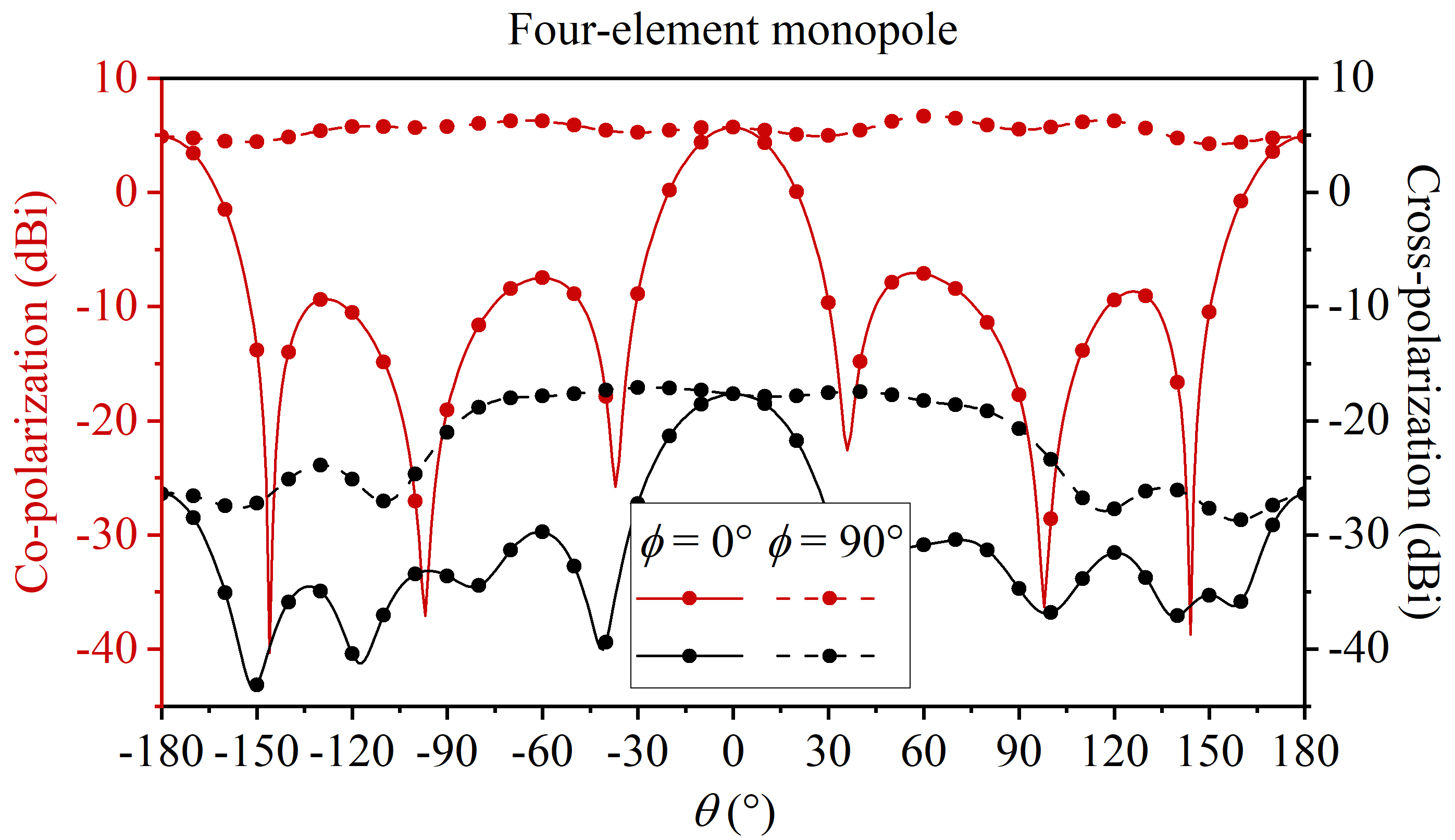}%
	}\hfil
	\caption{Simulated co- and cross-polarized realized gains of the proposed antenna during the structural evolution at 5 GHz. (a) Single monopole. (b) Mirrored monopole. (c) Four-element monopole. For each configuration, the realized gains are shown for the \(\phi=0^\circ\) and \(\phi=90^\circ\) planes. The simulated array patterns are with uniform power excitation on all ports.}\label{single tuning}
\end{figure*}

\subsection{Single Meander Line Monopole}

The geometry evolution of the proposed antenna is shown in Fig.~2. The design begins with a single planar meander-line monopole realized on a Rogers RO4350B substrate \((\varepsilon_r=3.48,\ \tan\delta=0.0037,\ T=0.508~\mathrm{mm})\). The single radiator is fed by a microstrip line of width \(A_w\), while its meandered and capacitive-loading sections are parameterized by \(A_d\), \(A_g\), \(A_c\), and \(A_l\). The overall substrate dimensions are \(G_l\) and \(G_w\). By mirroring the single monopole, the two-radiator configuration in Fig.~2 is obtained. The mirrored layout enforces an anti-phase relation \((\Delta\phi=180^\circ)\) between the two radiating regions, so that the combined structure behaves as an equivalent dipole-like cell. This feature is the basis of the dynamic switching behavior developed later. The final four-element configuration is generated by duplicating this dipole-like cell along the array axis. As indicated in Fig.~2, \(p\) denotes the intra-cell spacing, \(d\) denotes the inter-cell spacing, and \(A_d'\) denotes the narrowed width of radiators 2 and 3 for frequency tuning. Accordingly, the proposed antenna can be regarded as a duplicated two-cell array, in which the cell-level anti-phase radiation and the array-level spacing jointly determine the overall dynamic response. The optimized dimensions for the \(5~\mathrm{GHz}\) design are listed in Table~\ref{tab:table1}.

The initial meander-line geometry is selected according to the self-resonant design approach reported in~\cite{99054,10379428}, while the capacitive-loading concept follows compact planar RFID antenna implementations such as~\cite{4020418}. In the equivalent loaded meander-line model, the folded section contributes an effective inductance \(L_m\) and a distributed capacitance \(C_m\), while the capacitive patch introduces an additional loading capacitance \(C_p\) between the top printed patch and the bottom ground plane through the RO4350B substrate. The resulting resonance condition is approximated as
\begin{equation}
	f_0 \approx
	\frac{1}{2\pi\sqrt{L_m\left(C_m+C_p\right)}} ,
\end{equation}
where the three equivalent parameters are estimated from the printed geometry as
\begin{subequations}
	\begin{align}
	L_m &\approx
	\frac{120 N A_l\sqrt{\varepsilon_{\mathrm{eff}}}}{c}
	\ln\left(\frac{A_d+A_g}{A_d}\right),\\
	C_m &\approx
	\frac{(2N-1)A_l\sqrt{\varepsilon_{\mathrm{eff}}}}
	{2\pi\cdot469c},\\
	C_p &\approx
	\eta_p\varepsilon_0\varepsilon_r
	\frac{A_lA_c}{T}.
	\end{align}
\end{subequations}
Here, \(N\) is the number of meander sections, \(c\) is the speed of light, \(\varepsilon_{\mathrm{eff}}\) is the effective dielectric constant of the substrate, and \(\eta_p\) is an effective capacitive-patch coupling factor that accounts for fringing fields, finite-ground effects, and the nonuniform field distribution around the patch and feed transition. Using the dimensions in Table~\ref{tab:table1}, \(\varepsilon_r=3.48\), and \(\eta_p\approx0.4\), this loaded estimate gives \(N\approx2\) for resonance near \(5.05~\mathrm{GHz}\). The final dimensions are then refined through full-wave optimization.

The corresponding radiation evolution is presented in Fig.~3, which compares the co- and cross-polarized realized gains of the single monopole, mirrored monopole, and four-element monopole at \(5~\mathrm{GHz}\) in the \(\phi=0^\circ\) and \(\phi=90^\circ\) planes. As the structure evolves from a single radiator to the duplicated four-element configuration, the broadside realized gain increases from 1.17 dBi to 2.24 dBi and finally to 5.68 dBi with uniform current excitation. This confirms that the mirrored configuration enhances the effective radiation aperture, while the final four-element arrangement provides further constructive radiation enhancement in the broadside direction. Meanwhile, the overall patterns remain consistent with monopole-type broadside radiation, which supports the use of the duplicated architecture as the physical basis for the subsequent dynamic array design.

\subsection{Theoretical Analysis of the Two-Cell Dynamic Dipole Array}

The proposed four-radiator structure can be interpreted as a two-cell dynamic dipole array. 
Each cell consists of two mirrored radiators separated by an internal center-to-center spacing \(p\). 
The two radiators in each cell are always driven with a \(180^\circ\) phase difference, so that each cell behaves as an equivalent dipole-like radiating unit. 
The complete structure is then formed by duplicating this dynamic dipole cell along the array axis, with a center-to-center spacing \(d\) between the two cells.

In this work, the \(+z\)-axis, i.e., \(\theta=0^\circ\), is selected as the reference communication direction. Since the duplicated array is arranged along the \(x\)-direction, we define the internal phase variable
\begin{equation}
	\psi=\frac{kp}{2}\sin\theta\cos\phi ,
\end{equation}
and the inter-cell phase variable
\begin{equation}
	\gamma=kd\sin\theta\cos\phi ,
\end{equation}
where \(k=2\pi/\lambda_0\) is the free-space wavenumber. 
Here, \(\psi\) denotes the phase accumulation within each dipole-like cell, whereas \(\gamma\) denotes the phase accumulation between the two duplicated cells.

For a single dynamic dipole cell, two switching states are defined as
\begin{subequations}\label{eq:single_cell_states}
	\begin{align}
		S_1:\quad I_{1,\mathrm{L}} &= \sqrt{\alpha}I_0, &
		I_{1,\mathrm{R}} &= -I_0, \label{eq:single_cell_states_a}\\
		S_2:\quad I_{2,\mathrm{L}} &= I_0, &
		I_{2,\mathrm{R}} &= -\sqrt{\alpha}I_0. \label{eq:single_cell_states_b}
	\end{align}
\end{subequations}
where \(\alpha\) is the power ratio and \(I_0\) is the reference current amplitude.
The equivalent current response of one dipole-like cell is therefore written as
\begin{subequations}\label{eq:IS_expand}
	\begin{equation}\label{eq:IS_expand_a}
		I_{S_1}
		=
		I_0\Big[(\sqrt{\alpha}+1)\cos\psi
		-j(\sqrt{\alpha}-1)\sin\psi\Big]
	\end{equation}
	\begin{equation}\label{eq:IS_expand_b}
		I_{S_2}
		=
		I_0\Big[(\sqrt{\alpha}+1)\cos\psi
		+j(\sqrt{\alpha}-1)\sin\psi\Big]
	\end{equation}
\end{subequations}
It is observed that \(I_{S_1}\) and \(I_{S_2}\) share the same real term but have opposite quadrature terms. 
Therefore, the two states of a single cell form a phase-reversed pair. 
Define
\begin{subequations}\label{eq:Ic_Q}
	\begin{equation}\label{eq:Ic_Q_a}
		I_{\mathrm{c}}=I_0(\sqrt{\alpha}+1)\cos\psi
	\end{equation}
	\begin{equation}\label{eq:Ic_Q_b}
		\mathcal{Q}=I_0(\sqrt{\alpha}-1)\sin\psi
	\end{equation}
\end{subequations}
Then the two cell states can be rewritten compactly as
\begin{subequations}\label{eq:IS_compact}
	\begin{equation}\label{eq:IS_compact_a}
		I_{S_1}=I_{\mathrm{c}}-j\mathcal{Q},
	\end{equation}
	\begin{equation}\label{eq:IS_compact_b}
		I_{S_2}=I_{\mathrm{c}}+j\mathcal{Q}.
	\end{equation}
\end{subequations}
Hence, the switching of a single dynamic dipole cell is fundamentally a phase-pattern reversal process, where the perturbation term appears in quadrature with respect to the common current component.

The complete antenna consists of two duplicated dipole-like cells centered at \(-d/2\) and \(+d/2\), respectively. 
Four switching states are defined as
\begin{subequations}
	\begin{align}
		A &= (S_1,S_1), \\
		B &= (S_2,S_2), \\
		C &= (S_1,S_2), \\
		D &= (S_2,S_1).
	\end{align}
\end{subequations}
The corresponding equivalent array currents are
\begin{subequations}
	\begin{align}
		I_A
		&=
		I_{S_1}e^{-j\gamma/2}+I_{S_1}e^{j\gamma/2}, \\
		I_B
		&=
		I_{S_2}e^{-j\gamma/2}+I_{S_2}e^{j\gamma/2}, \\
		I_C
		&=
		I_{S_1}e^{-j\gamma/2}+I_{S_2}e^{j\gamma/2}, \\
		I_D
		&=
		I_{S_2}e^{-j\gamma/2}+I_{S_1}e^{j\gamma/2}.
	\end{align}
\end{subequations}
Substituting \(I_{S_1}=I_{\mathrm{c}}-j\mathcal{Q}\) and \(I_{S_2}=I_{\mathrm{c}}+j\mathcal{Q}\) yields
\begin{subequations}
	\begin{align}
		I_A
		&= 2(I_{\mathrm{c}}-j\mathcal{Q})\cos\frac{\gamma}{2}, \\
		I_B
		&= 2(I_{\mathrm{c}}+j\mathcal{Q})\cos\frac{\gamma}{2}, \\
		I_C
		&= 2I_{\mathrm{c}}\cos\frac{\gamma}{2}
		+ 2\mathcal{Q}\sin\frac{\gamma}{2}, \\
		I_D
		&= 2I_{\mathrm{c}}\cos\frac{\gamma}{2}
		- 2\mathcal{Q}\sin\frac{\gamma}{2}.
	\end{align}
\end{subequations}
These expressions directly reveal the switching mechanisms considered in this work. 

First, for the \(A\)--\(B\) switching pair,
\begin{subequations}\label{eq:AB_pair}
	\begin{equation}\label{eq:AB_pair_a}
		I_{\mathrm{avg},AB}
		=
		\frac{I_A+I_B}{2}
		=
		2I_{\mathrm{c}}\cos\frac{\gamma}{2}
	\end{equation}
	\begin{equation}\label{eq:AB_pair_b}
		I_{\Delta,AB}
		=
		\frac{I_B-I_A}{2}
		=
		j\,2\mathcal{Q}\cos\frac{\gamma}{2}.
	\end{equation}
\end{subequations}
Therefore, the \(A\)--\(B\) switching pair produces a purely quadrature perturbation with respect to the common current term. 
This means that the \(A\)--\(B\) pair mainly introduces differential phase, which is the array-level manifestation of the phase-pattern reversal inherited from the single dynamic dipole cell. 

Second, for the \(C\)--\(D\) switching pair,
\begin{subequations}\label{eq:CD_pair}
	\begin{equation}\label{eq:CD_pair_a}
		I_{\mathrm{avg},CD}
		=
		\frac{I_C+I_D}{2}
		=
		2I_{\mathrm{c}}\cos\frac{\gamma}{2}
	\end{equation}
	\begin{equation}\label{eq:CD_pair_b}
		I_{\Delta,CD}
		=
		\frac{I_C-I_D}{2}
		=
		2\mathcal{Q}\sin\frac{\gamma}{2}.
	\end{equation}
\end{subequations}
In this case, the perturbation term is purely in phase with the common current component. 
Hence, the \(C\)--\(D\) pair mainly introduces differential magnitude, rather than differential phase. 
Physically, this means that the current magnitude is redistributed between the two array states, while the phase center remains nearly unchanged. 

For compact notation, the reference, phase-perturbation, and magnitude-perturbation components are defined as
\begin{subequations}\label{eq:ABCD_components}
	\begin{align}
		I_{\mathrm{ref}} &= 2I_{\mathrm{c}}\cos\frac{\gamma}{2}, \label{eq:ABCD_components_a}\\
		I_{\phi} &= 2\mathcal{Q}\cos\frac{\gamma}{2}, \label{eq:ABCD_components_b}\\
		I_{m} &= 2\mathcal{Q}\sin\frac{\gamma}{2}. \label{eq:ABCD_components_c}
	\end{align}
\end{subequations}

Third, for the three-state \(A\)--\(C\)--\(D\) switching sequence, the average channel is
\begin{equation}\label{eq:ACD_avg}
	I_{\mathrm{avg},ACD}
	=
	\frac{I_A+I_C+I_D}{3}
	=
	I_{\mathrm{ref}}-\frac{j}{3}I_{\phi},
\end{equation}
and the state-dependent perturbations with respect to this average channel are
\begin{subequations}\label{eq:ACD_delta}
	\begin{align}
		\Delta I_A
		&=
		I_A-I_{\mathrm{avg},ACD}
		=
		-\frac{2j}{3}I_{\phi}, \label{eq:ACD_delta_a}\\
		\Delta I_C
		&=
		I_C-I_{\mathrm{avg},ACD}
		=
		I_m+\frac{j}{3}I_{\phi}, \label{eq:ACD_delta_c}\\
		\Delta I_D
		&=
		I_D-I_{\mathrm{avg},ACD}
		=
		-I_m+\frac{j}{3}I_{\phi}. \label{eq:ACD_delta_d}
	\end{align}
\end{subequations}
Thus, unlike the purely pairwise \(A\)--\(B\) and \(C\)--\(D\) cases, the \(A\)--\(C\)--\(D\) sequence retains the full magnitude-perturbation channel \(I_m\) while introducing a weaker residual phase-perturbation component through \(I_{\phi}\). This provides a theoretical basis for the three-state dispersion and BER results examined in Section~III.

Fourth, for the full \(A\)--\(B\)--\(C\)--\(D\) switching sequence, the four states can be rewritten as
\begin{subequations}\label{eq:ABCD_states}
	\begin{align}
		I_A &= I_{\mathrm{ref}} - jI_{\phi}, \label{eq:ABCD_states_a}\\
		I_B &= I_{\mathrm{ref}} + jI_{\phi}, \label{eq:ABCD_states_b}\\
		I_C &= I_{\mathrm{ref}} + I_{m}, \label{eq:ABCD_states_c}\\
		I_D &= I_{\mathrm{ref}} - I_{m}, \label{eq:ABCD_states_d}
	\end{align}
\end{subequations}
It follows that the full four-state switching simultaneously excites two orthogonal perturbation channels:
\begin{itemize}
	\item a phase perturbation channel \(I_{\phi}\), inherited from \(A\)--\(B\) switching;
	\item a magnitude perturbation channel \(I_{m}\), inherited from \(C\)--\(D\) switching.
\end{itemize}
Therefore, the full \(A\)--\(B\)--\(C\)--\(D\) switching realizes combined phase--magnitude dynamic modulation. Moreover, the total perturbation envelope of the full four-state set is
\begin{equation}
	\sqrt{|I_{\phi}|^2+|I_{m}|^2}
	=
	2|\mathcal{Q}|
	=
	2I_0(\sqrt{\alpha}-1)|\sin\psi|,
\end{equation}
which indicates that the switching imbalance at the cell level determines the available perturbation strength, while the inter-cell phase \(\gamma\) determines how this perturbation is partitioned between differential phase and differential magnitude.

At the reference communication direction, i.e., the \(+z\)-axis \((\theta=0^\circ)\), one has \(\sin\theta\cos\phi=0\), which yields \(\psi=\gamma=\mathcal{Q}=0\). Consequently,
\begin{equation}
	I_{\Delta,AB}=I_{\Delta,CD}=0.
\end{equation}
This means that both the differential phase perturbation and the differential magnitude perturbation vanish along the reference communication direction. 
Therefore, the \(A\)--\(B\), \(C\)--\(D\), \(A\)--\(C\)--\(D\), and full \(A\)--\(B\)--\(C\)--\(D\) switching cases can all support low-distortion communication around the \(+z\)-axis, while off-axis directions experience angle-dependent phase or magnitude perturbations.
\begin{figure*}[t]
	\centering
	\subfloat[]{\includegraphics[width=3.2in]{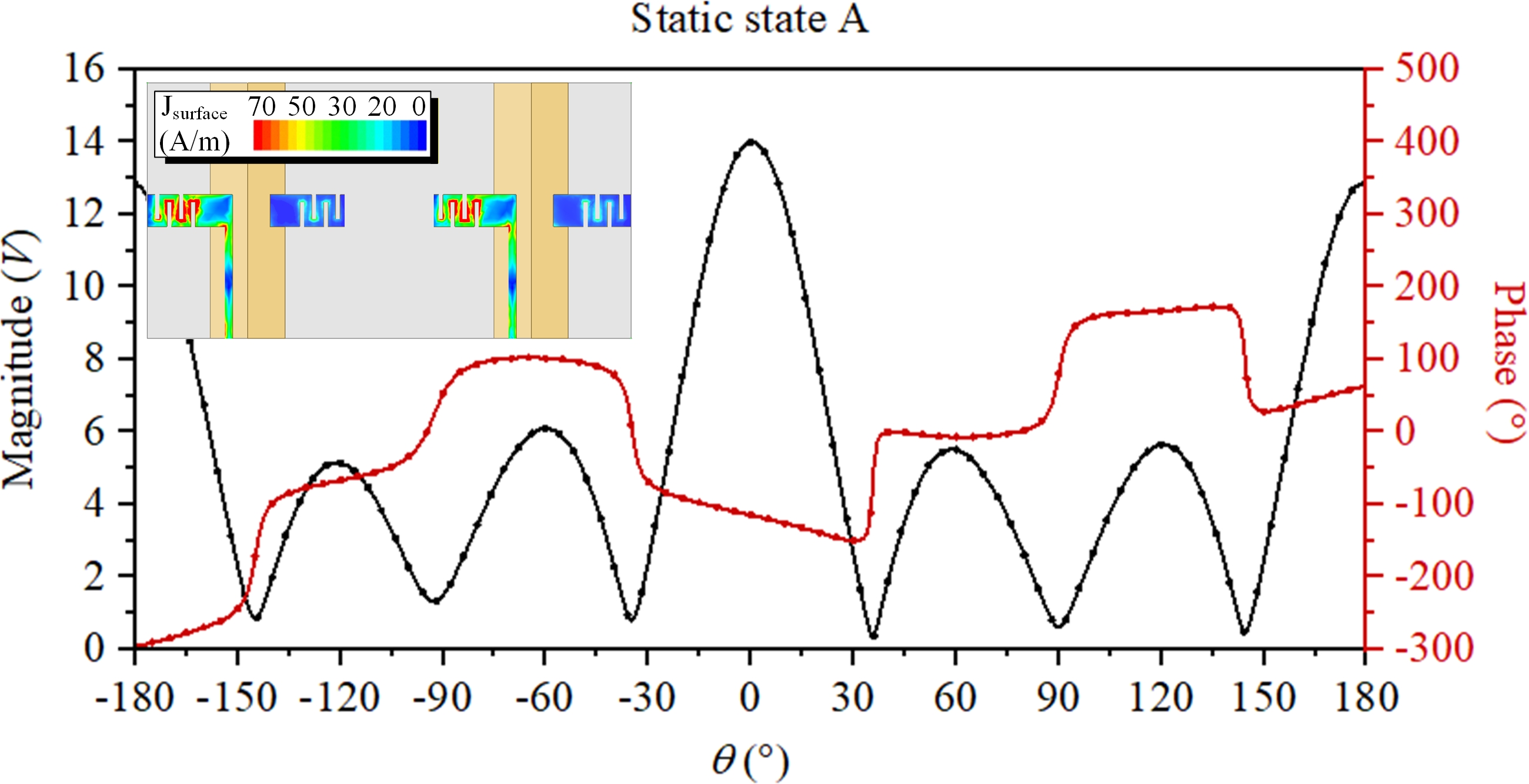}}\hspace{0.04\textwidth}
	\subfloat[]{\includegraphics[width=3.2in]{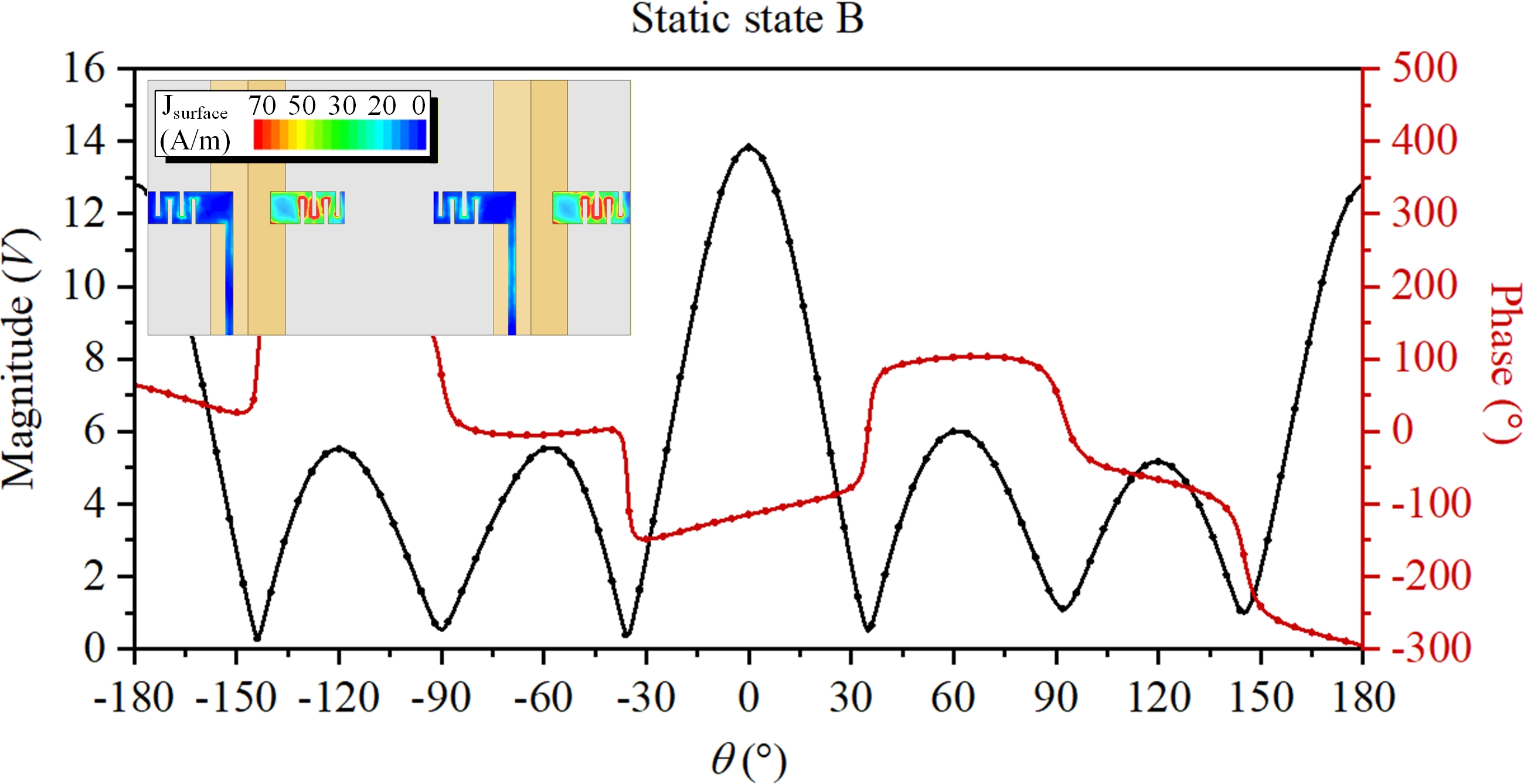}}\\
	\subfloat[]{\includegraphics[width=3.2in]{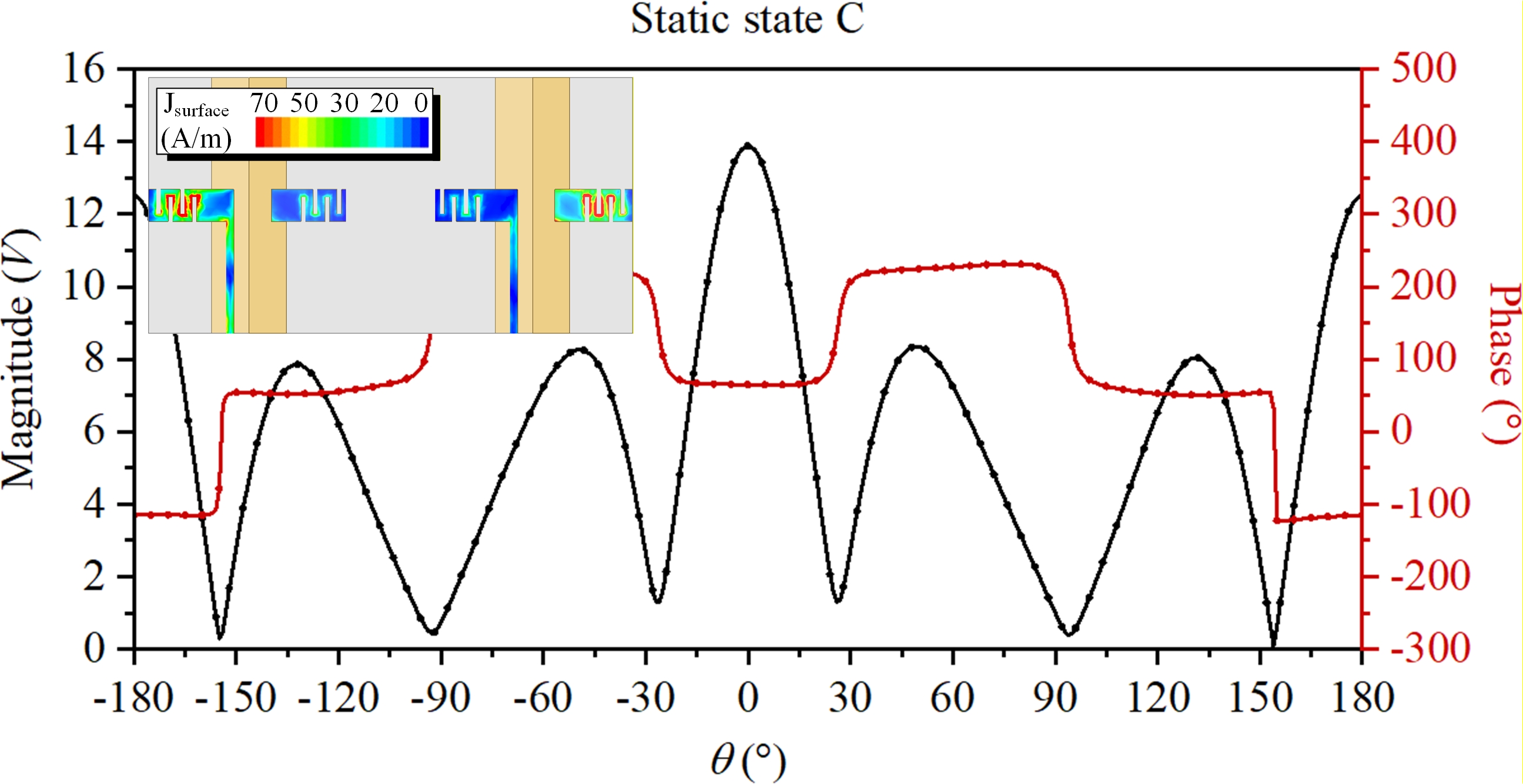}}\hspace{0.04\textwidth}
	\subfloat[]{\includegraphics[width=3.2in]{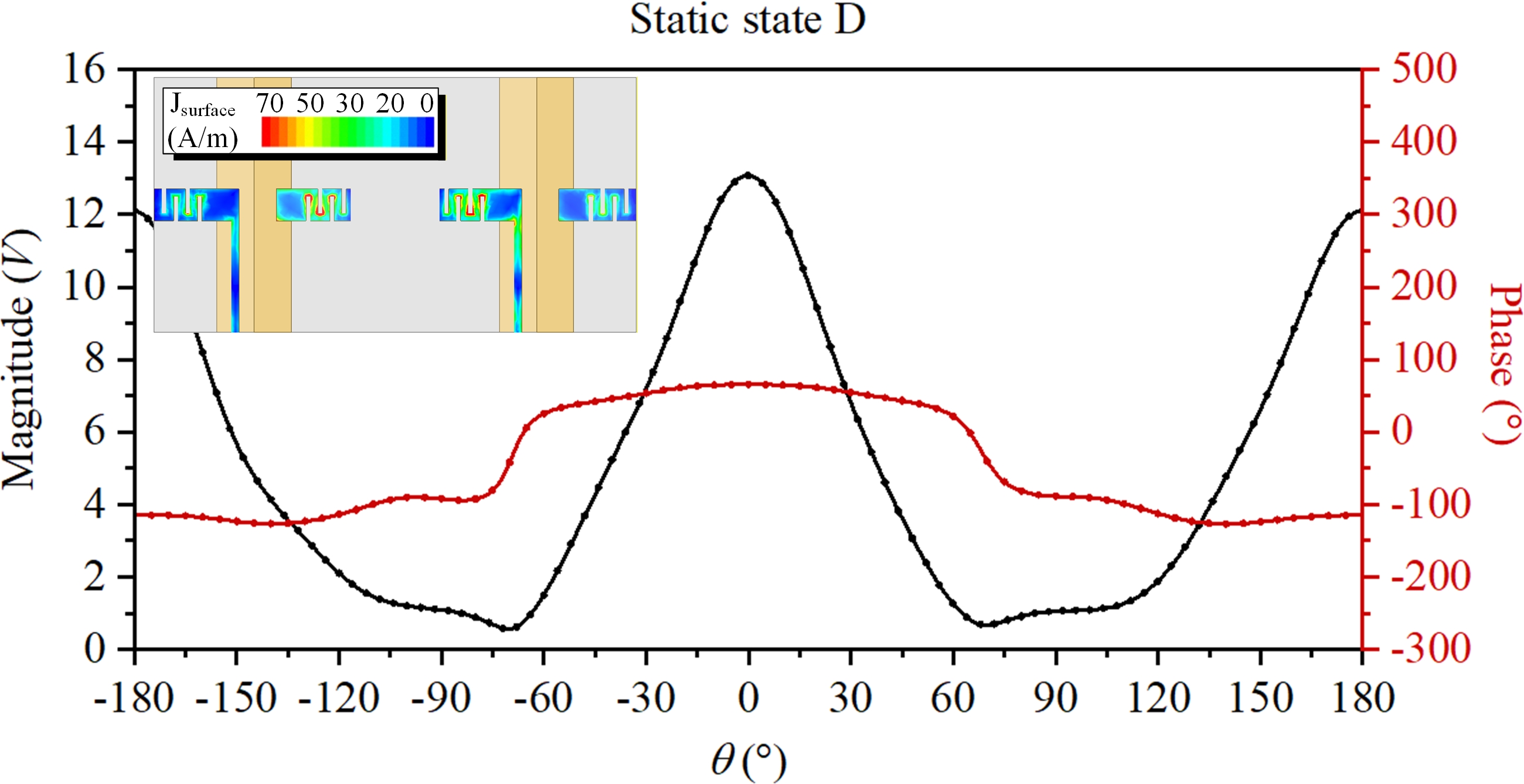}}
	\caption{HFSS-simulated surface current distributions and the corresponding magnitude and phase patterns in the \(x\)-\(z\) plane for the proposed dynamic four-element array with \(\alpha = 9~\mathrm{dB}\) and an inter-cell spacing of \(d = 48~\mathrm{mm}\) at 5 GHz under four static switching states: (a) State A, (b) State B, (c) State C, and (d) State D.}
	\label{fig:four_states_current_patterns}
\end{figure*}
For communication analysis, the average received powers of the \(A\)--\(B\), \(C\)--\(D\), \(A\)--\(C\)--\(D\), and full four-state switching cases can be written as
\begin{subequations}\label{eq:avg_power_cases}
	\begin{align}
		\bar{P}_{AB}(\theta,\phi)
		&\propto
		|I_{\mathrm{avg},AB}(\theta,\phi)|^2
		+
		|I_{\Delta,AB}(\theta,\phi)|^2, \label{eq:avg_power_cases_a}\\
		\bar{P}_{CD}(\theta,\phi)
		&\propto
		|I_{\mathrm{avg},CD}(\theta,\phi)|^2
		+
		|I_{\Delta,CD}(\theta,\phi)|^2, \label{eq:avg_power_cases_b}\\
		\bar{P}_{ACD}(\theta,\phi)
		&\propto
		|I_{\mathrm{avg},ACD}(\theta,\phi)|^2 \notag\\
		&\quad+
		\frac{1}{3}\left(
		|\Delta I_A|^2
		+
		|\Delta I_C|^2
		+
		|\Delta I_D|^2
		\right), \label{eq:avg_power_cases_c}\\
		\bar{P}_{ABCD}(\theta,\phi)
		&\propto
		|I_{\mathrm{ref}}(\theta,\phi)|^2
		+
		|I_{\phi}(\theta,\phi)|^2
		+
		|I_{m}(\theta,\phi)|^2. \label{eq:avg_power_cases_d}
	\end{align}
\end{subequations}
For the \(A\)--\(C\)--\(D\) case, the effective perturbation power is governed by the three-state variance in \eqref{eq:avg_power_cases_c}. 
For the full four-state switching case, a useful normalized measure is the information-beam efficiency, defined as the fraction of the time-averaged dynamic field power carried by the recoverable reference component,
\begin{equation}
	\eta_{\mathrm{IB}}(\theta,\phi)
	=
	\frac{|I_{\mathrm{ref}}(\theta,\phi)|^2}
	{|I_{\mathrm{ref}}(\theta,\phi)|^2+|I_{\phi}(\theta,\phi)|^2+|I_{m}(\theta,\phi)|^2},
\end{equation}
where the denominator represents the sum of the recoverable reference power and the switching-induced perturbation power. 
For the two-state cases, the same interpretation is applied by replacing the reference component with the corresponding average channel, \(I_{\mathrm{avg},AB}\) or \(I_{\mathrm{avg},CD}\), and the perturbation power with the pairwise differential channel, \(I_{\Delta,AB}\) or \(I_{\Delta,CD}\).
When the perturbation terms remain much smaller than the reference current term, the received symbols cluster tightly around the ideal constellation points. 
However, once the perturbation terms become comparable to the reference term, angle-dependent symbol displacement occurs, which manifests as constellation distortion.
Accordingly, the constellation distortion is qualitatively related to
\begin{equation}
	\text{constellation distortion}
	\;\propto\;
	\frac{\sqrt{|I_{\phi}(\theta,\phi)|^2+|I_{m}(\theta,\phi)|^2}}
	{|I_{\mathrm{ref}}(\theta,\phi)|}.
\end{equation}

The above formulation establishes a hierarchical interpretation of the proposed antenna. 
At the cell level, each proposed unit behaves as a dynamic dipole-like element with two anti-phase currents and two phase-reversed states. 
At the array level, duplicating the dipole-like cell introduces the additional phase variable \(\gamma\), which determines whether the cell-level perturbation appears as differential phase (\(A\)--\(B\) switching), differential magnitude (\(C\)--\(D\) switching), or combined phase--magnitude modulation (\(A\)--\(C\)--\(D\) and full \(A\)--\(B\)--\(C\)--\(D\) switching). 
Consequently, the angular distributions of radiation distortion and BER can be systematically controlled through the duplicated-array configuration.

To further validate the four-state interpretation, Fig.~\ref{fig:four_states_current_patterns} shows the HFSS-simulated surface current distributions and the corresponding magnitude and phase patterns in the \(x\)-\(z\) plane for \(d = 0.8\lambda_0\). As indicated by \eqref{eq:ABCD_states_a} and \eqref{eq:ABCD_states_b}, States \(A\) and \(B\) differ mainly through the \(\pm jI_{\phi}\) term, and therefore exhibit relatively similar magnitude envelopes but more noticeable phase variation, confirming that the \(A\)--\(B\) pair is predominantly governed by the phase-perturbation channel \(I_{\phi}\) in \eqref{eq:ABCD_components_b}. By contrast, States \(C\) and \(D\) follow the \(\pm I_m\) relation in \eqref{eq:ABCD_states_c} and \eqref{eq:ABCD_states_d}, and show more evident magnitude redistribution, which is consistent with the magnitude-perturbation channel \(I_m\) in \eqref{eq:ABCD_components_c}. The simulated broadside realized gains of States \(A\), \(B\), \(C\), and \(D\) are 4.83, 4.79, 5.07, and 4.54 dBi, respectively. The near-equality between States \(A\) and \(B\) further supports the phase-dominant nature of the \(A\)--\(B\) pair, whereas the larger gain deviation between States \(C\) and \(D\) indicates stronger broadside magnitude redistribution. Meanwhile, the residual variations observed in both magnitude and phase confirm that the practical antenna departs from the ideal orthogonal perturbation model, mainly due to mutual coupling, finite radiator size, and switching-dependent current redistribution.

\subsection{Effect of Inter-Cell Spacing and Mutual Coupling on Information Beam and BER}

In the ideal uncoupled case, the information beam distribution is determined by the inter-cell spacing \(d\) through the array phase term \(\gamma\), which redistributes the differential phase and differential magnitude channels through \(\cos(\gamma/2)\) and \(\sin(\gamma/2)\). In practice, however, for achieving compact design, reducing \(d\) also reduces the separation between the nearest radiators \(R_2\) and \(R_3\), thereby increasing their mutual coupling. As a result, the actual information beam is affected not only by the intended array phase term, but also by coupling-induced channel mixing. To describe this effect, a first-order complex coupling coefficient is introduced as
\begin{equation}
	\kappa(d)=\kappa_r(d)+j\kappa_i(d).
\end{equation}
Instead of treating mutual coupling as a uniform complex scaling, it is modeled here as a first-order perturbation that modifies the reference channel and breaks the orthogonality between the phase- and magnitude-perturbation channels. To this end, three spacing-dependent perturbation coefficients, namely \(\eta_r(d)\), \(\eta_{\phi}(d)\), and \(\eta_m(d)\), are introduced to characterize the coupling-induced deviations of the reference, phase, and magnitude terms, respectively. The ideal and effective terms are then written as
\begin{subequations}\label{eq:ideal_effective_components}
	\begin{align}
		I_{\mathrm{ref}}^{(0)} &= 2I_{\mathrm{c}}\cos\frac{\gamma}{2}, &
		I_{\mathrm{ref}} &= \left(1+\eta_r(d)\right)I_{\mathrm{ref}}^{(0)}, \label{eq:ideal_effective_components_a}\\
		I_{\phi}^{(0)} &= 2\mathcal{Q}\cos\frac{\gamma}{2}, &
		\widetilde{I}_{\phi} &= \left(j+\eta_{\phi}(d)\right)I_{\phi}^{(0)}, \label{eq:ideal_effective_components_b}\\
		I_{m}^{(0)} &= 2\mathcal{Q}\sin\frac{\gamma}{2}, &
		\widetilde{I}_{m} &= \left(1+\eta_m(d)\right)I_{m}^{(0)}. \label{eq:ideal_effective_components_c}
	\end{align}
\end{subequations}
where \(I_{\mathrm{ref}}^{(0)}\), \(I_{\phi}^{(0)}\), and \(I_m^{(0)}\) denote the ideal uncoupled reference, phase-perturbation, and magnitude-perturbation terms, respectively, while \(I_{\mathrm{ref}}\), \(\widetilde{I}_{\phi}\), and \(\widetilde{I}_{m}\) denote their coupled counterparts. Here, \(\eta_r(d)\), \(\eta_{\phi}(d)\), and \(\eta_m(d)\) are assumed to be small complex perturbation coefficients. More specifically, under the first-order approximation, one may write
\begin{subequations}\label{eq:eta_kappa_relation}
	\begin{align}
		\eta_r(d) &\approx \kappa_r(d), \label{eq:eta_kappa_relation_a}\\
		\eta_{\phi}(d) &\approx j\kappa_r(d)-\kappa_i(d), \label{eq:eta_kappa_relation_b}\\
		\eta_m(d) &\approx \kappa_r(d)+j\kappa_i(d), \label{eq:eta_kappa_relation_c}
	\end{align}
\end{subequations}
where the imaginary part \(\kappa_i(d)\) in the reference channel mainly introduces a common phase offset and is therefore absorbed into the overall phase reference. These relations indicate that both the phase- and magnitude-perturbation channels are modified by the real part of the coupling coefficient and contaminated by the imaginary part.

Accordingly, the coupled four states preserve the same compact form, with the ideal terms replaced by their effective counterparts. For the \(A\)--\(B\) and \(C\)--\(D\) pairs, the average and differential currents become
\begin{subequations}
	\begin{align}
		I_{\mathrm{avg},AB} &= I_{\mathrm{ref}}, &
		I_{\Delta,AB} &= \widetilde I_{\phi}, \\
		I_{\mathrm{avg},CD} &= I_{\mathrm{ref}}, &
		I_{\Delta,CD} &= \widetilde I_{m}.
	\end{align}
\end{subequations}
Thus, under mutual coupling, the \(A\)--\(B\) pair is no longer purely differential phase, and the \(C\)--\(D\) pair is no longer purely differential magnitude, since \(\widetilde{I}_{\phi}\) and \(\widetilde{I}_{m}\) generally contain both in-phase and quadrature components.

Accordingly, the actual information beam can be expressed as
\begin{equation}
	I_{\mathrm{beam}}(\theta,\phi;d)
	=
	I_{\mathrm{ideal}}(\theta,\phi;d)
	+
	\Delta I_{\mathrm{cpl}}(\theta,\phi;d),
\end{equation}
where \(d\) affects the beam through both the array-phase redistribution \(\gamma\) and the coupling-induced channel mixing.
The information-beam efficiency for the full four-state switching case is then defined as
\begin{equation}
	\eta_{\mathrm{IB}}(\theta,\phi;d)
	=
	\frac{|I_{\mathrm{ref}}|^2}
	{|I_{\mathrm{ref}}|^2+|\widetilde{I}_{\phi}|^2+|\widetilde{I}_{m}|^2},
	\label{eq:IB_eff_revised}
\end{equation}
where \(I_{\mathrm{ref}}\), \(\widetilde{I}_{\phi}\), and \(\widetilde{I}_{m}\) are functions of \((\theta,\phi;d)\). This quantity measures the normalized recoverable-field power fraction within the time-averaged dynamic radiation response. For the \(A\)--\(B\) and \(C\)--\(D\) switching pairs, the recoverable component is the pairwise average channel and the undesired dynamic component is the corresponding differential channel, giving
\begin{subequations}\label{eq:IB_eff_pairs}
	\begin{align}
		\eta_{\mathrm{IB},AB}(\theta,\phi;d)
		&=
		\frac{|I_{\mathrm{avg},AB}|^2}
		{|I_{\mathrm{avg},AB}|^2+|I_{\Delta,AB}|^2}, \label{eq:IB_eff_pairs_a}\\
		\eta_{\mathrm{IB},CD}(\theta,\phi;d)
		&=
		\frac{|I_{\mathrm{avg},CD}|^2}
		{|I_{\mathrm{avg},CD}|^2+|I_{\Delta,CD}|^2}. \label{eq:IB_eff_pairs_b}
	\end{align}
\end{subequations}
Since the perturbation coefficients \(\eta_{\phi}(d)\) and \(\eta_m(d)\) are assumed to increase with the mutual-coupling strength, their magnitudes generally become larger as \(d\) decreases. Consequently, the leakage terms embedded in \(\widetilde{I}_{\phi}\) and \(\widetilde{I}_{m}\) increase, which enlarges the perturbation contribution in the denominator of \eqref{eq:IB_eff_revised} and thus lowers the information-beam efficiency, even if the ideal beam-shaping effect of \(\gamma\) remains favorable.

Because larger switching-induced perturbations displace the received constellation more strongly from its calibrated reference, a lower information-beam efficiency is expected to correspond to a higher observed BER. Therefore, the element spacing \(d\) must be chosen to preserve the desired information-beam shaping while limiting coupling-induced leakage. This tradeoff leads to an optimal spacing that balances array-phase control and mutual-coupling suppression.
\begin{figure*}[!t]
	\centering
	\subfloat[]{\includegraphics[width=2.26in]{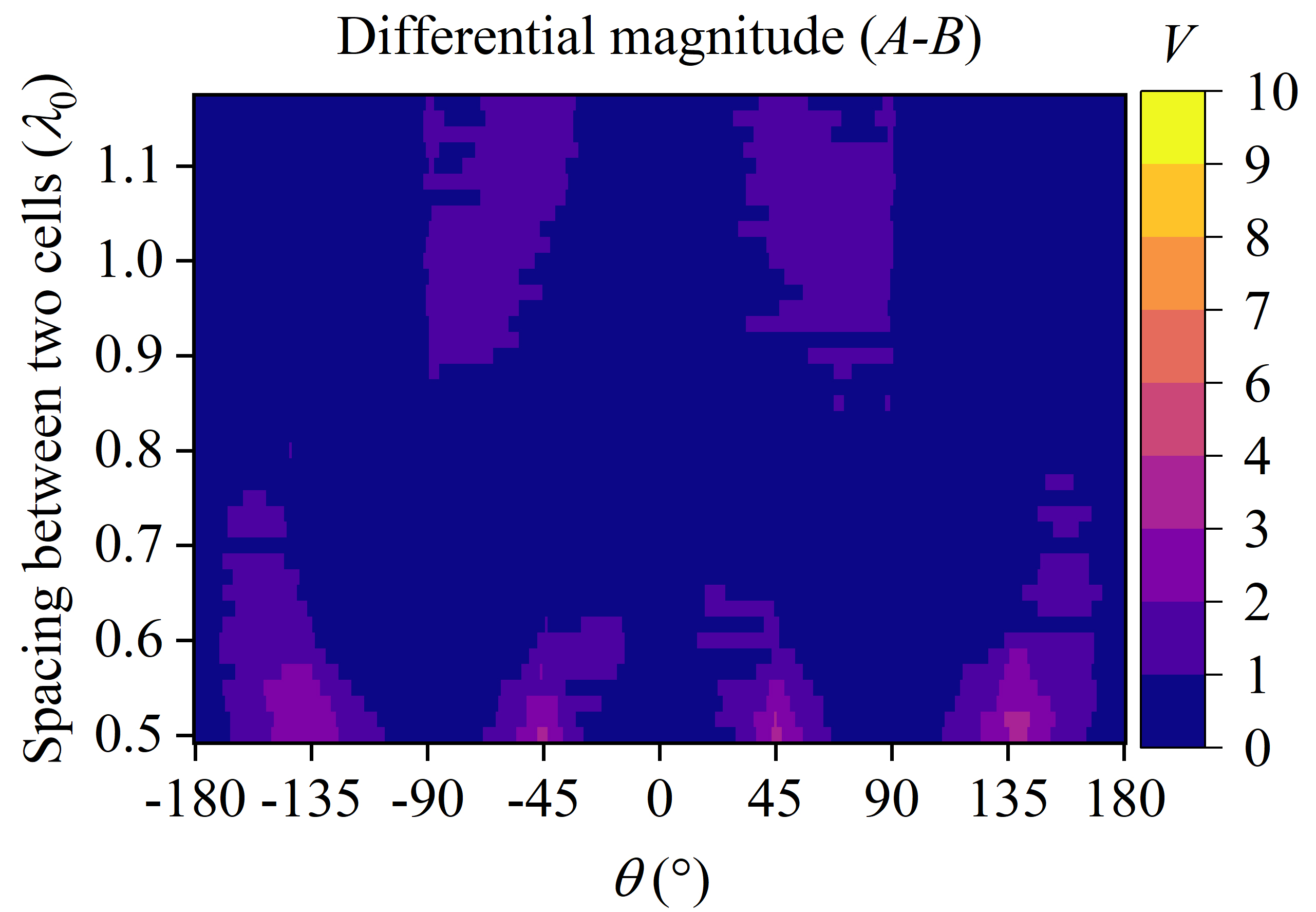}}\hfil
	\subfloat[]{\includegraphics[width=2.33in]{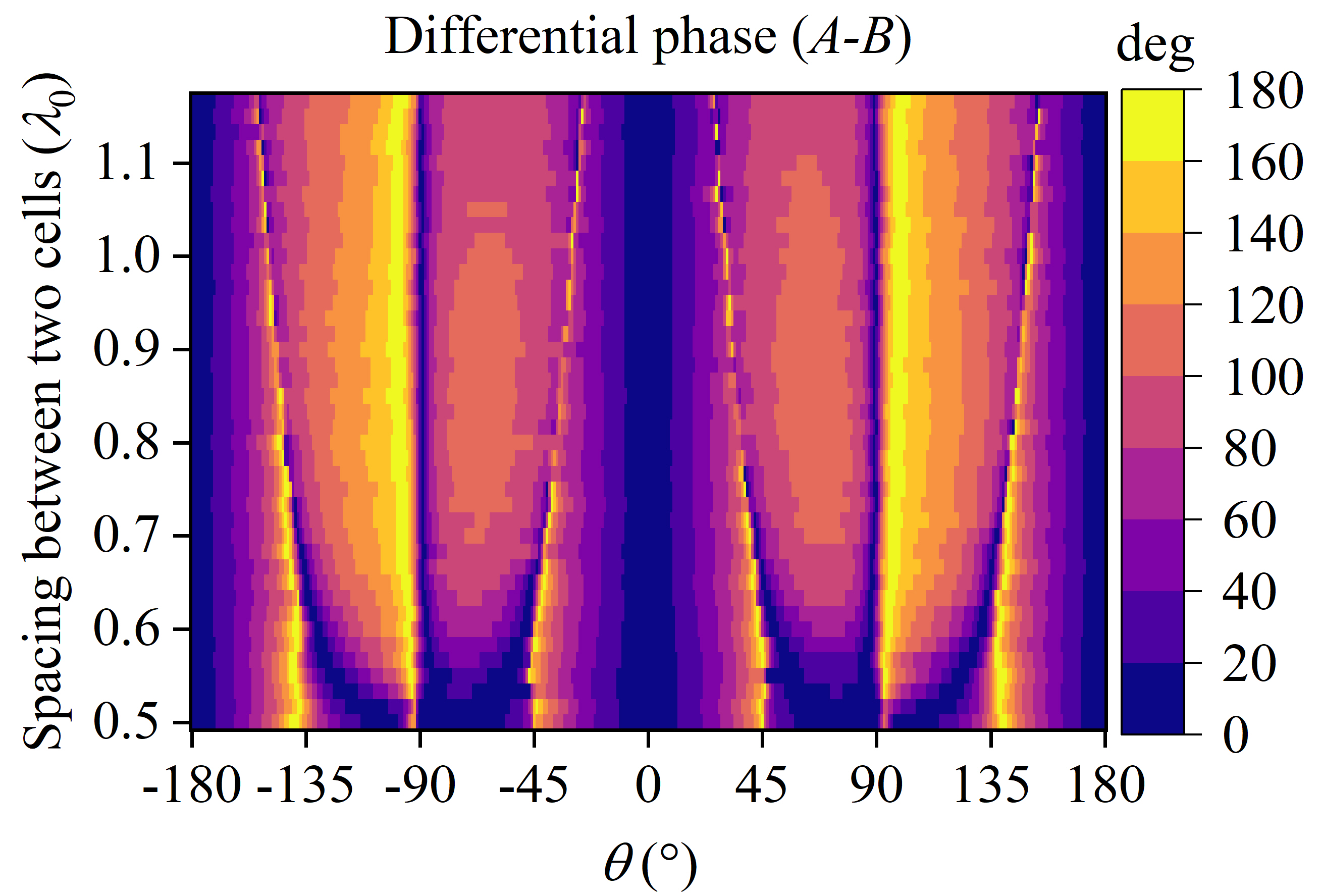}}\hfil
	\subfloat[]{\includegraphics[width=2.4in]{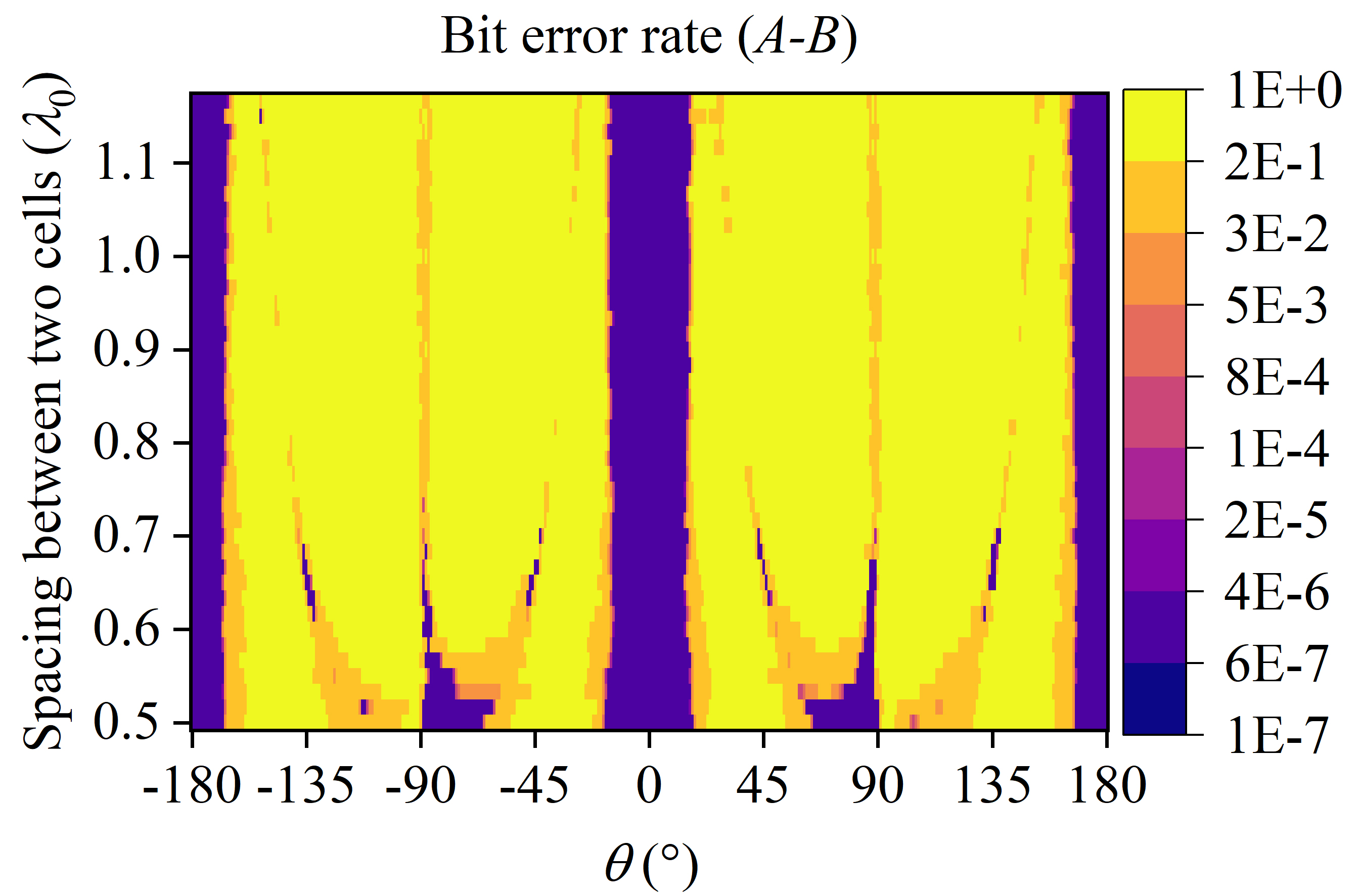}}\hfil\\
	\subfloat[]{\includegraphics[width=2.26in]{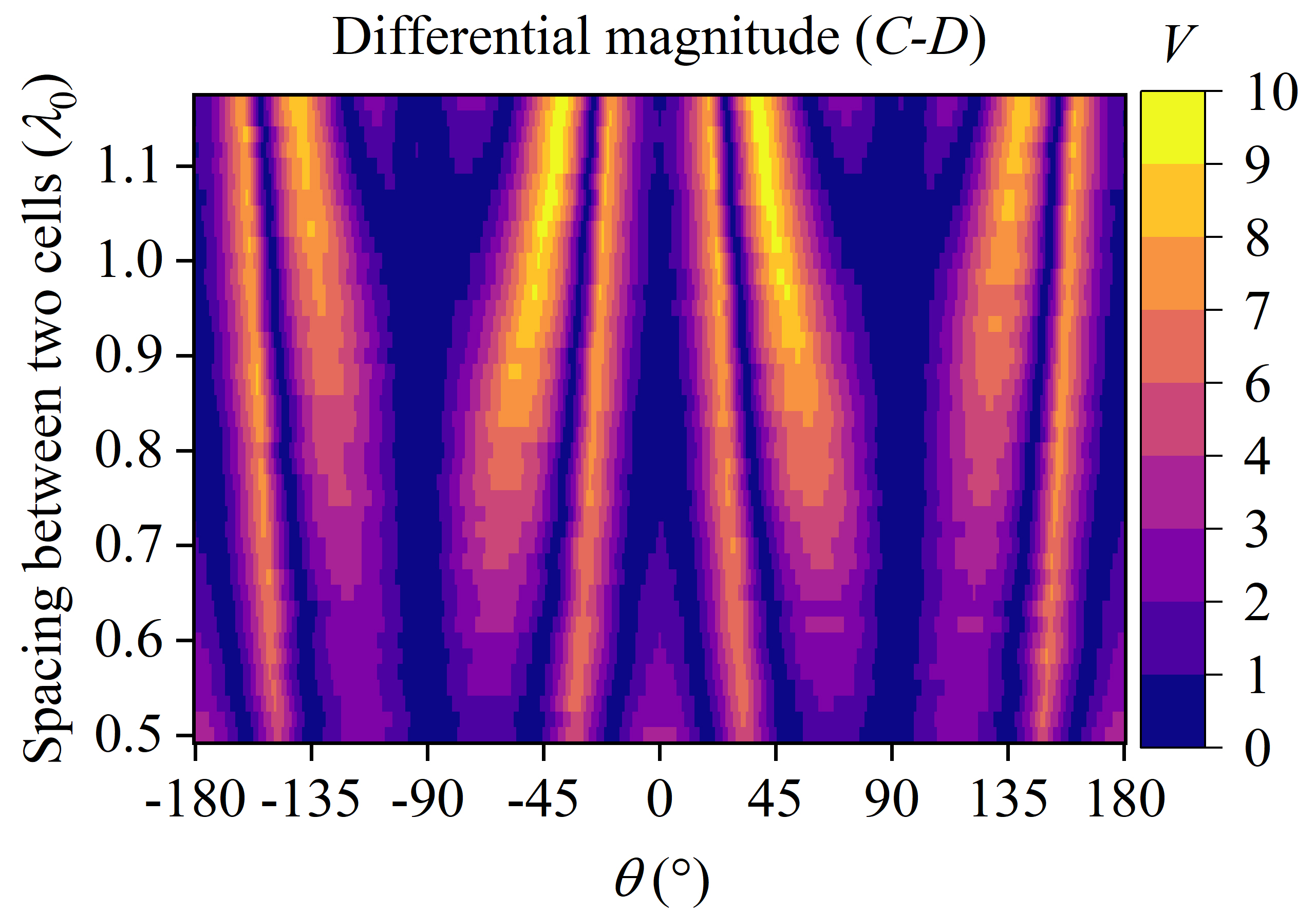}}\hfil
	\subfloat[]{\includegraphics[width=2.33in]{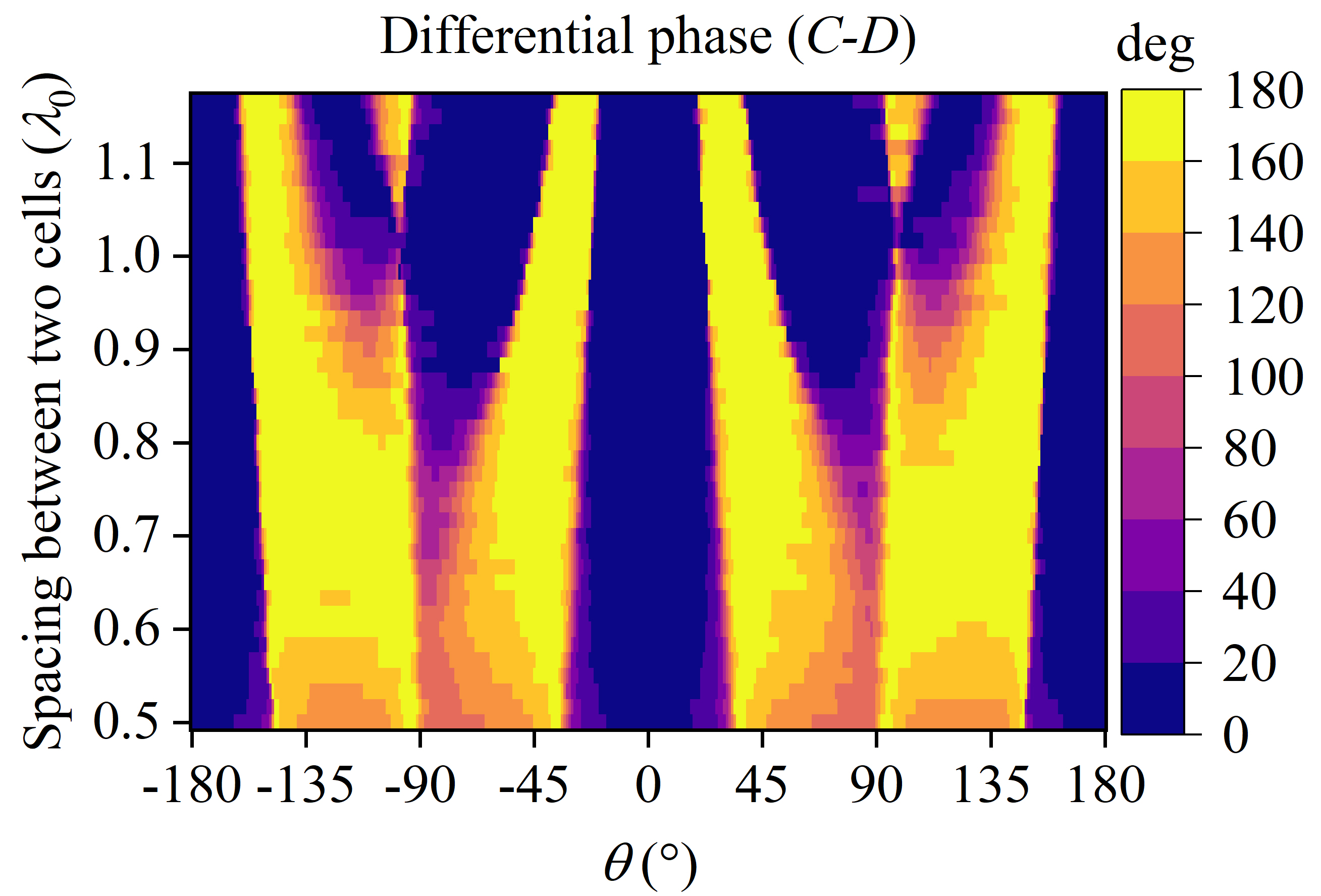}}\hfil
	\subfloat[]{\includegraphics[width=2.4in]{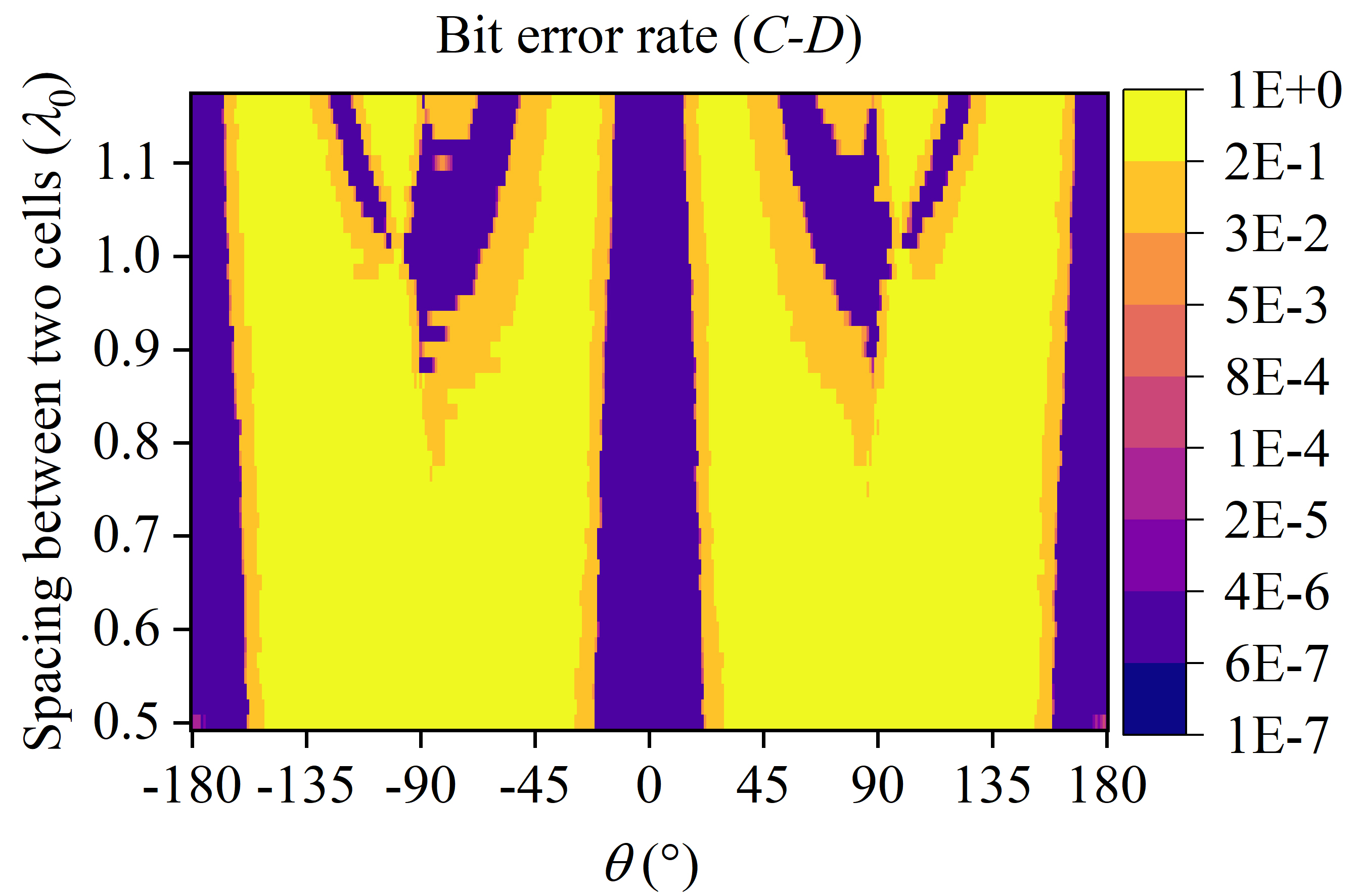}}\hfil\\
	\subfloat[]{\includegraphics[width=2.26in]{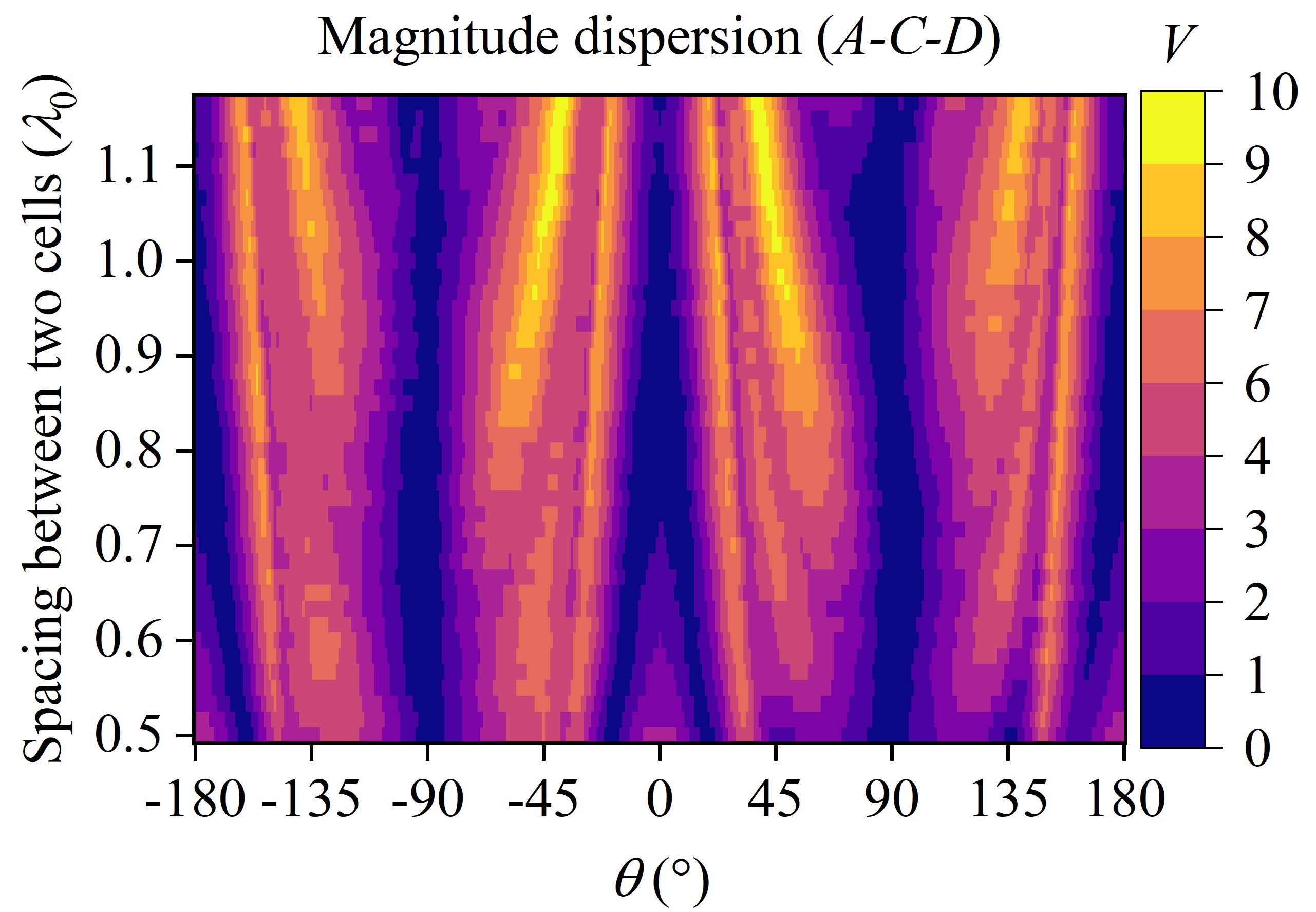}}\hfil
	\subfloat[]{\includegraphics[width=2.33in]{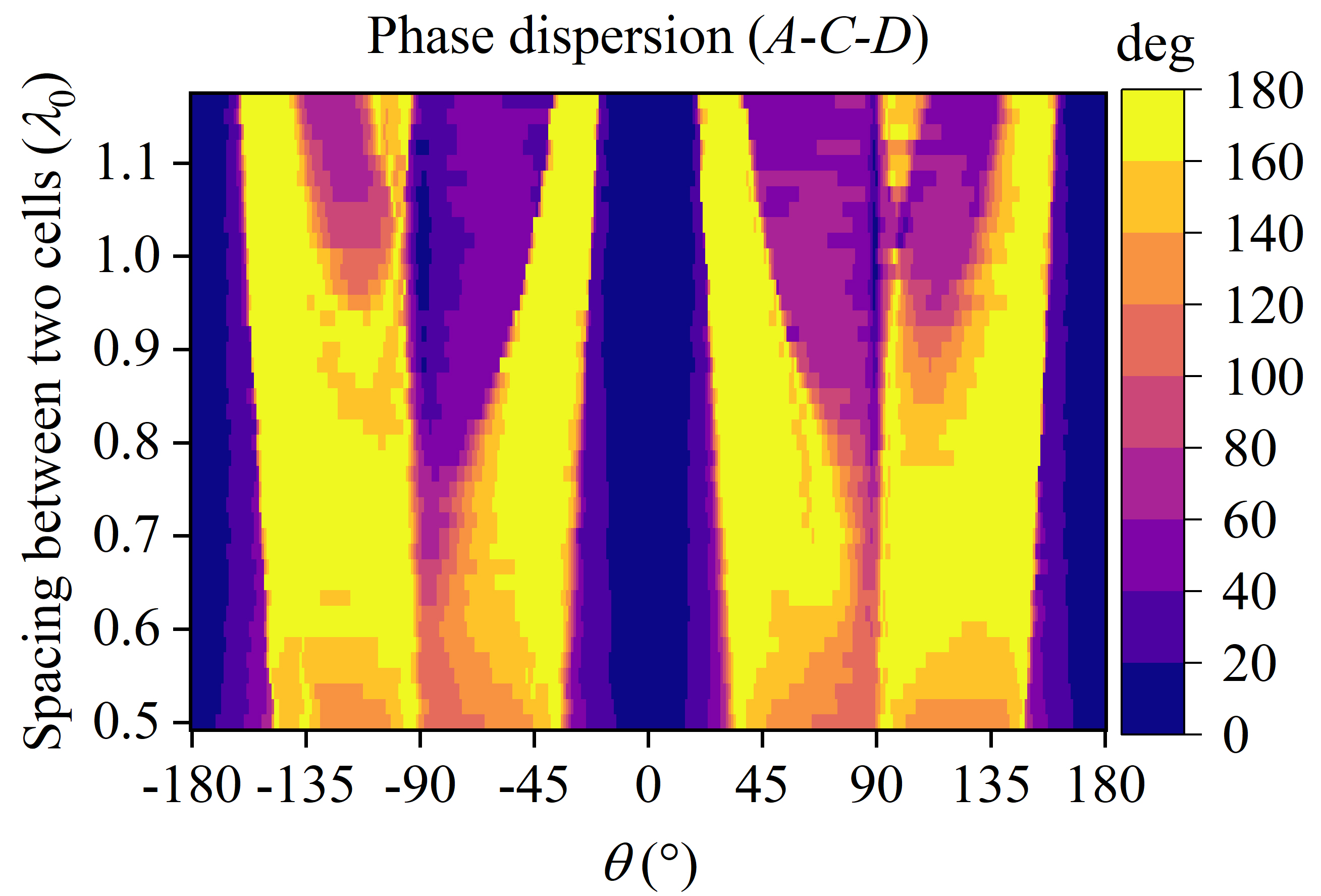}}\hfil
	\subfloat[]{\includegraphics[width=2.4in]{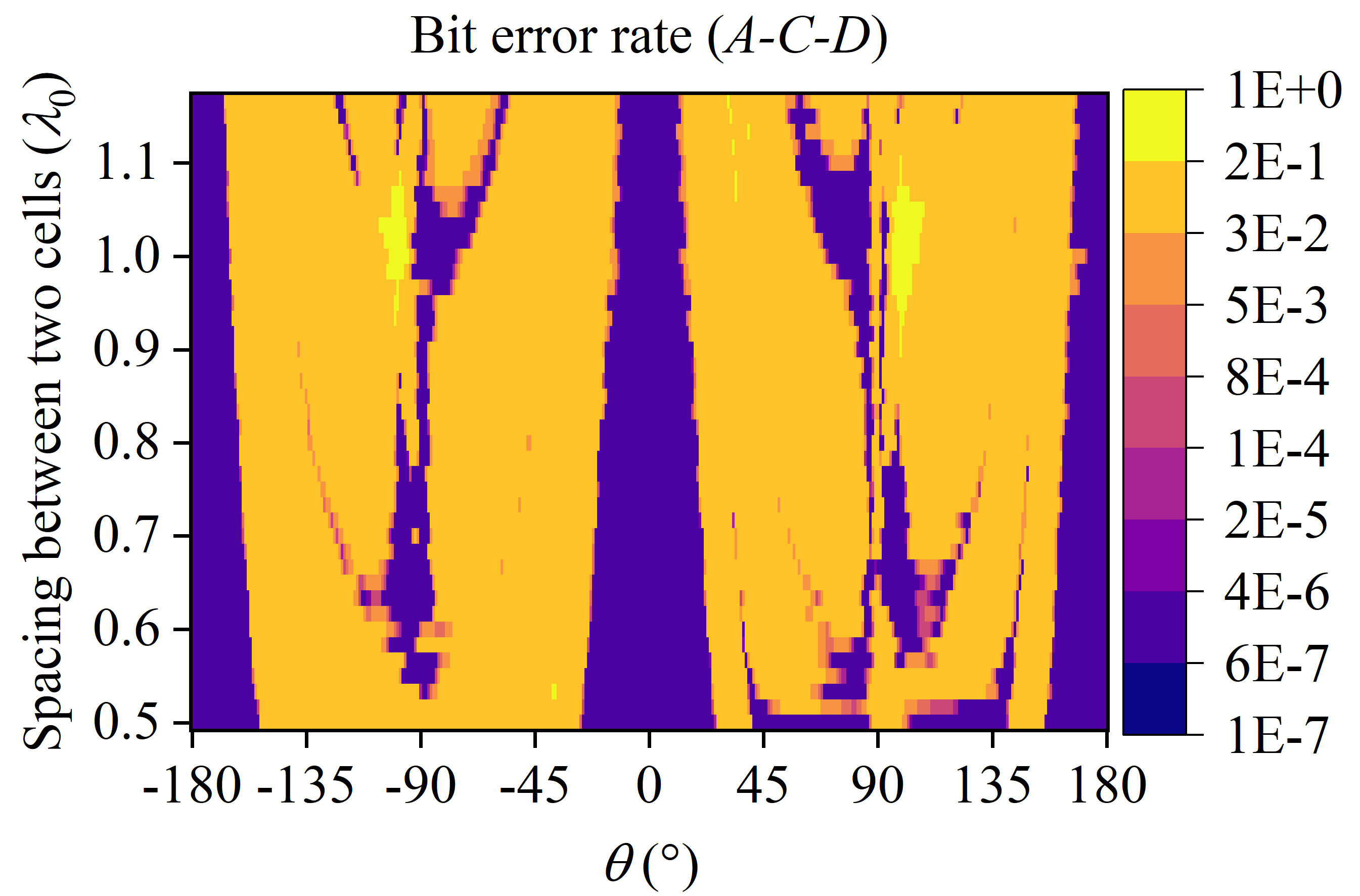}}\hfil\\
	\subfloat[]{\includegraphics[width=2.26in]{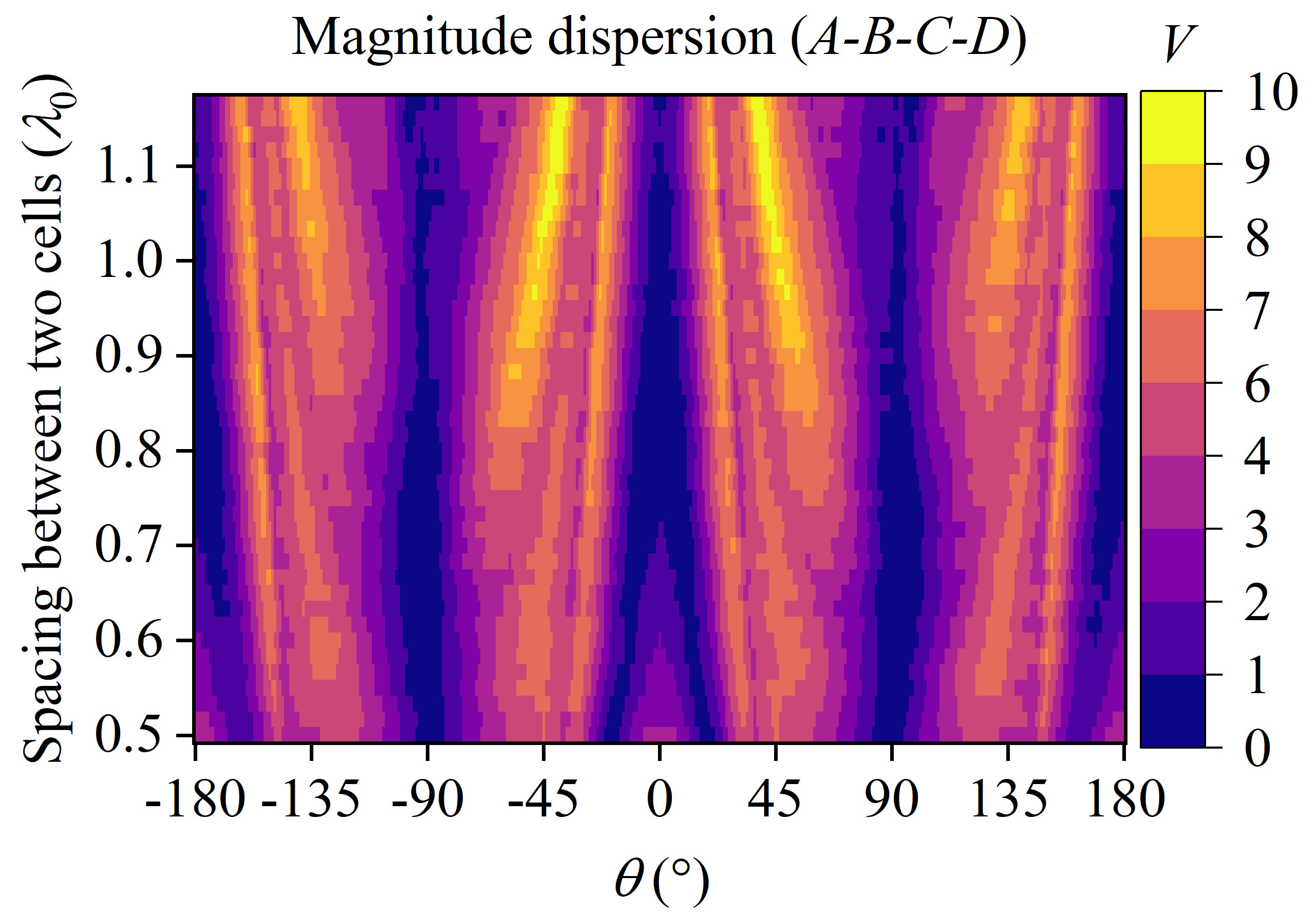}}\hfil
	\subfloat[]{\includegraphics[width=2.33in]{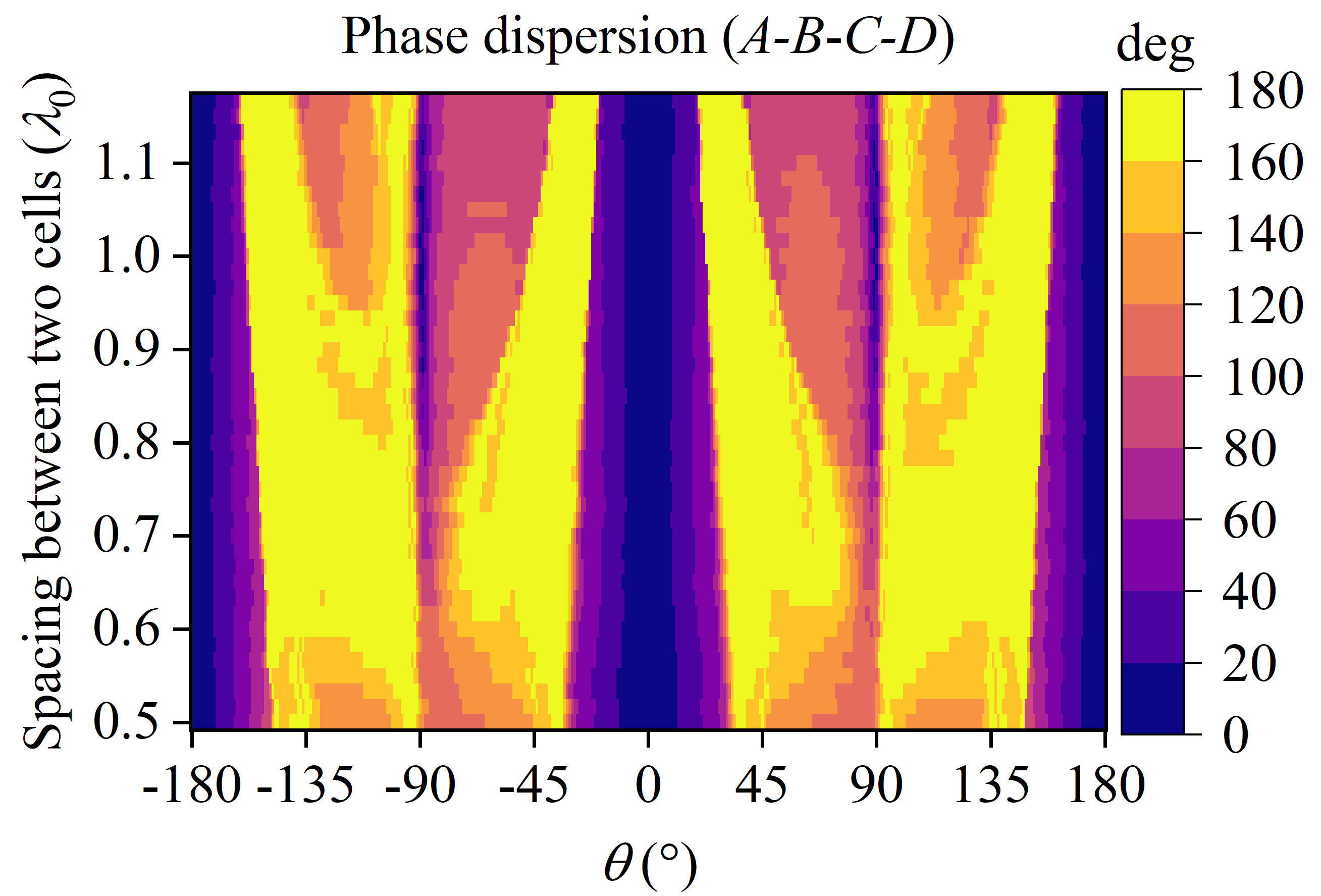}}\hfil
	\subfloat[]{\includegraphics[width=2.4in]{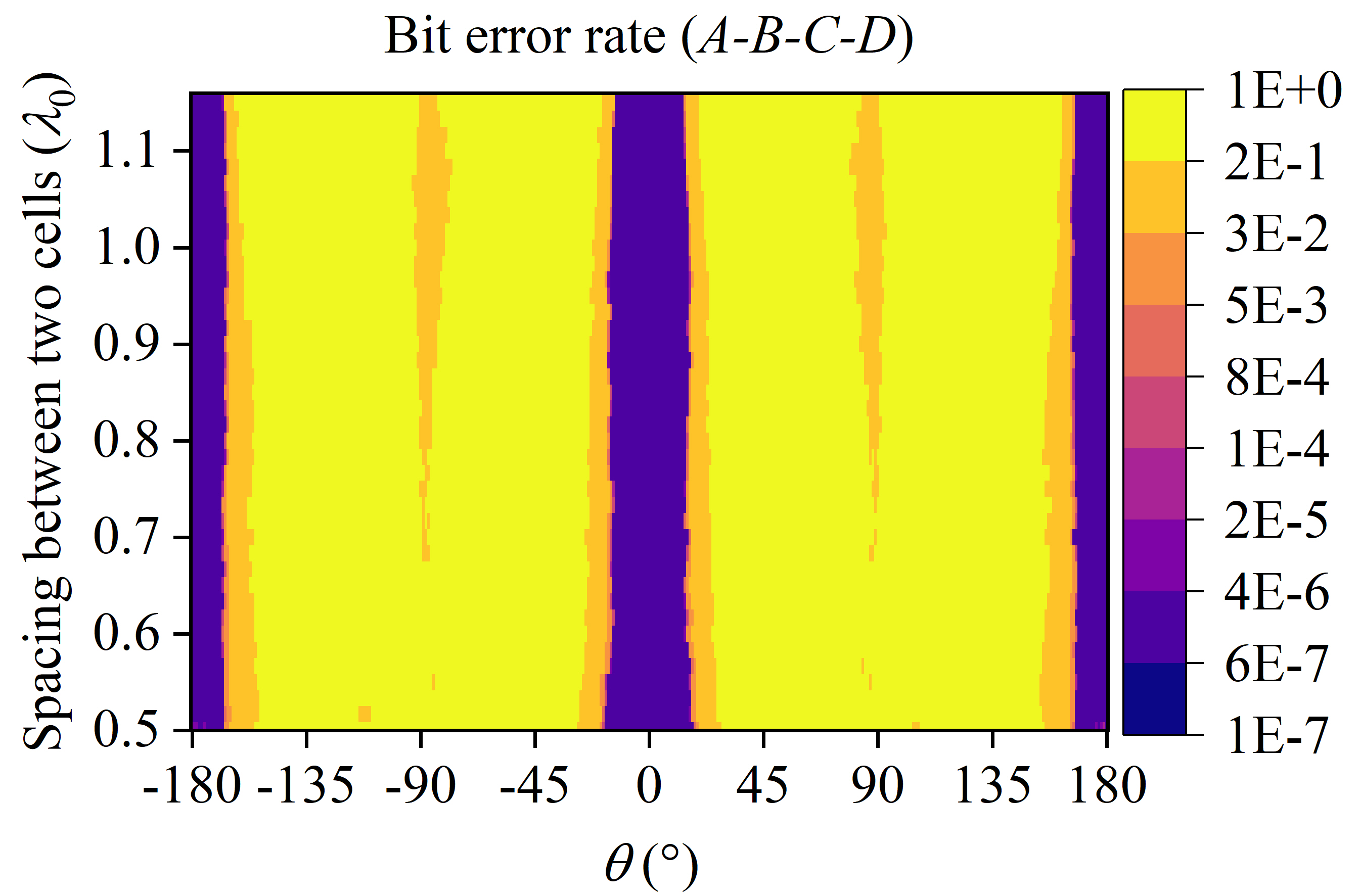}}\hfil
	\caption{Simulated radiated field variation and BER of the proposed four-state dynamic array versus the inter-cell spacing $d$ and observation angle $\theta$. (a)--(c) correspond to the $A$--$B$ switching case and show the absolute differential magnitude $|\Delta V_{AB}|$, the absolute differential phase $|\Delta \phi_{AB}|$, and the corresponding 16-QAM BER, respectively. (d)--(f) correspond to the $C$--$D$ switching case and show $|\Delta V_{CD}|$, $|\Delta \phi_{CD}|$, and the corresponding BER, respectively. (g)--(i) correspond to the three-state $A$--$C$--$D$ switching case and show $D_{\mathrm{mag}}(\theta,d)$, $D_{\phi}(\theta,d)$, and the corresponding BER, respectively. (j)--(l) correspond to the full $A$--$B$--$C$--$D$ switching case and show $D_{\mathrm{mag}}(\theta,d)$, $D_{\phi}(\theta,d)$, and the corresponding BER, respectively.}\label{fig:discrepancy_ber_vs_d}
\end{figure*}
\begin{figure*}[!t]
	\centering
	\subfloat[]{\includegraphics[width=2.26in]{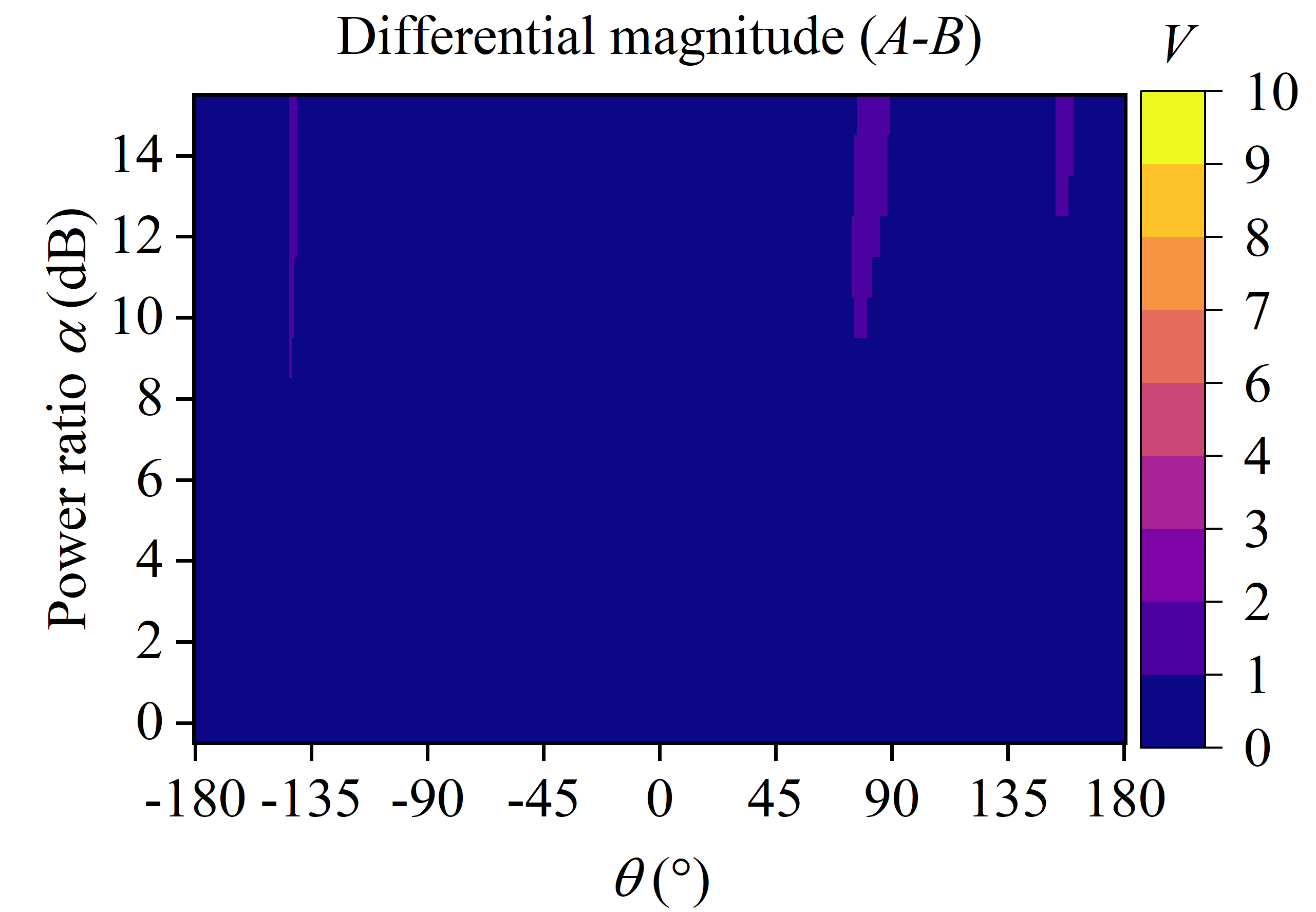}}\hfil
	\subfloat[]{\includegraphics[width=2.33in]{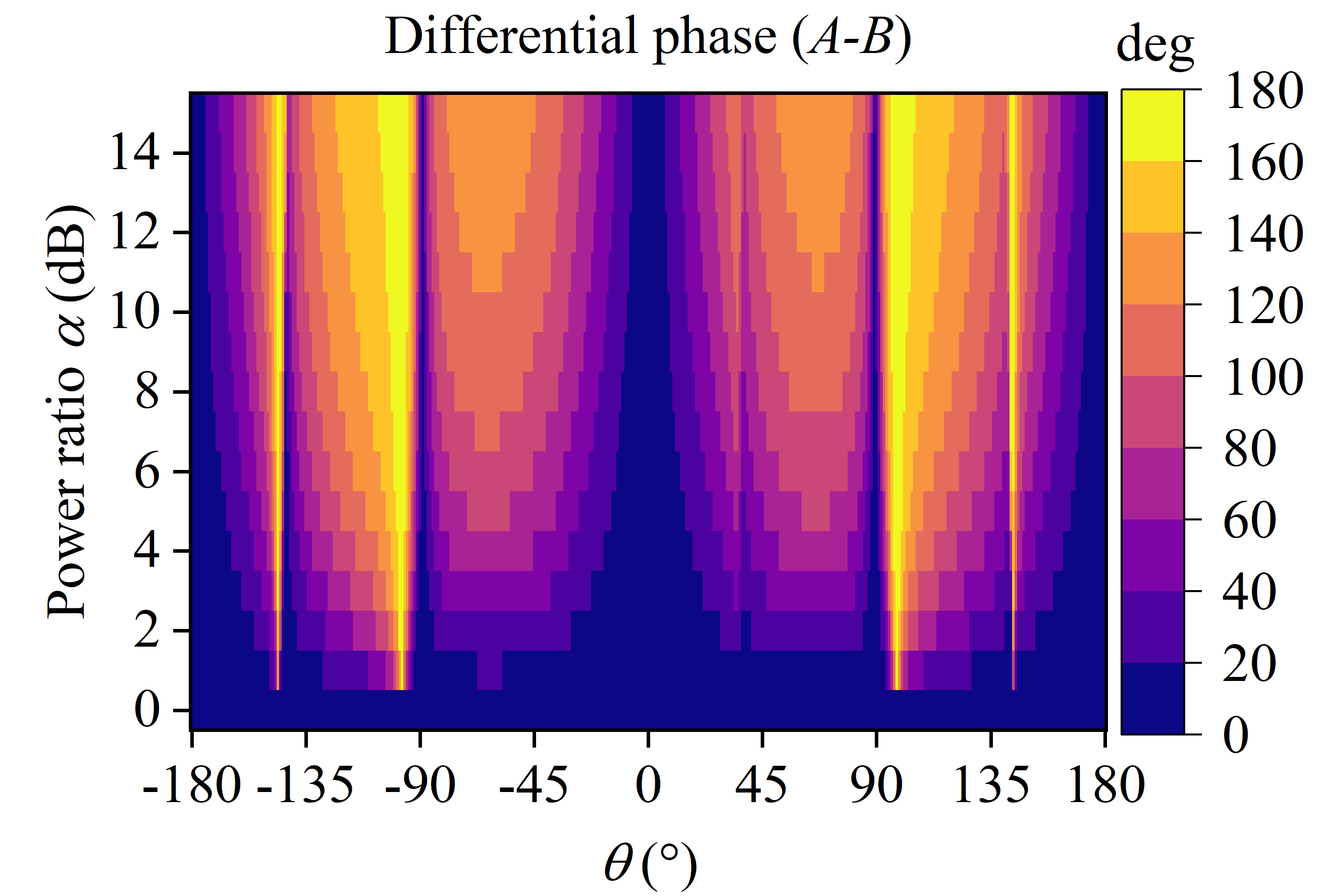}}\hfil
	\subfloat[]{\includegraphics[width=2.4in]{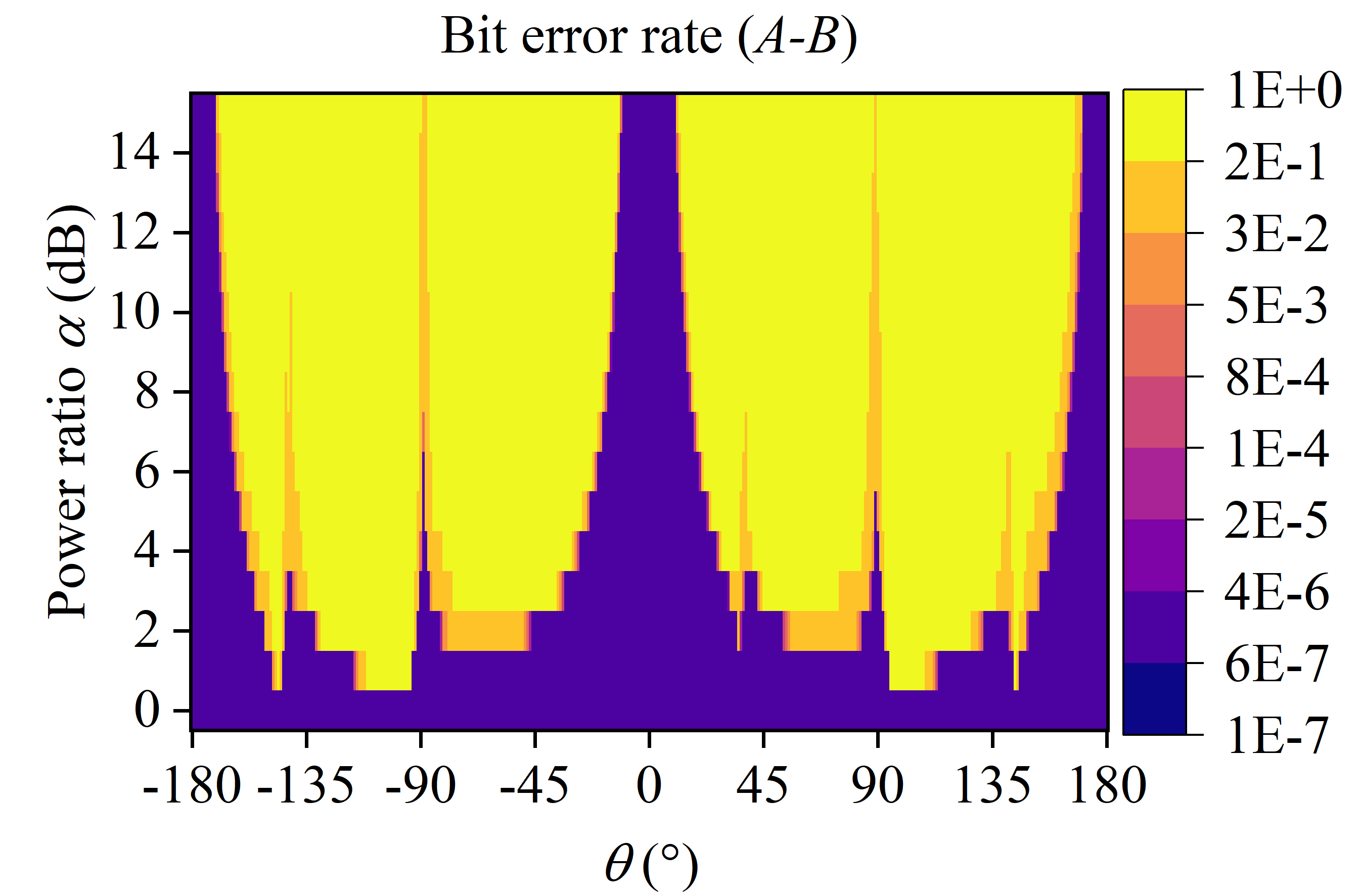}}\hfil\\
	\subfloat[]{\includegraphics[width=2.26in]{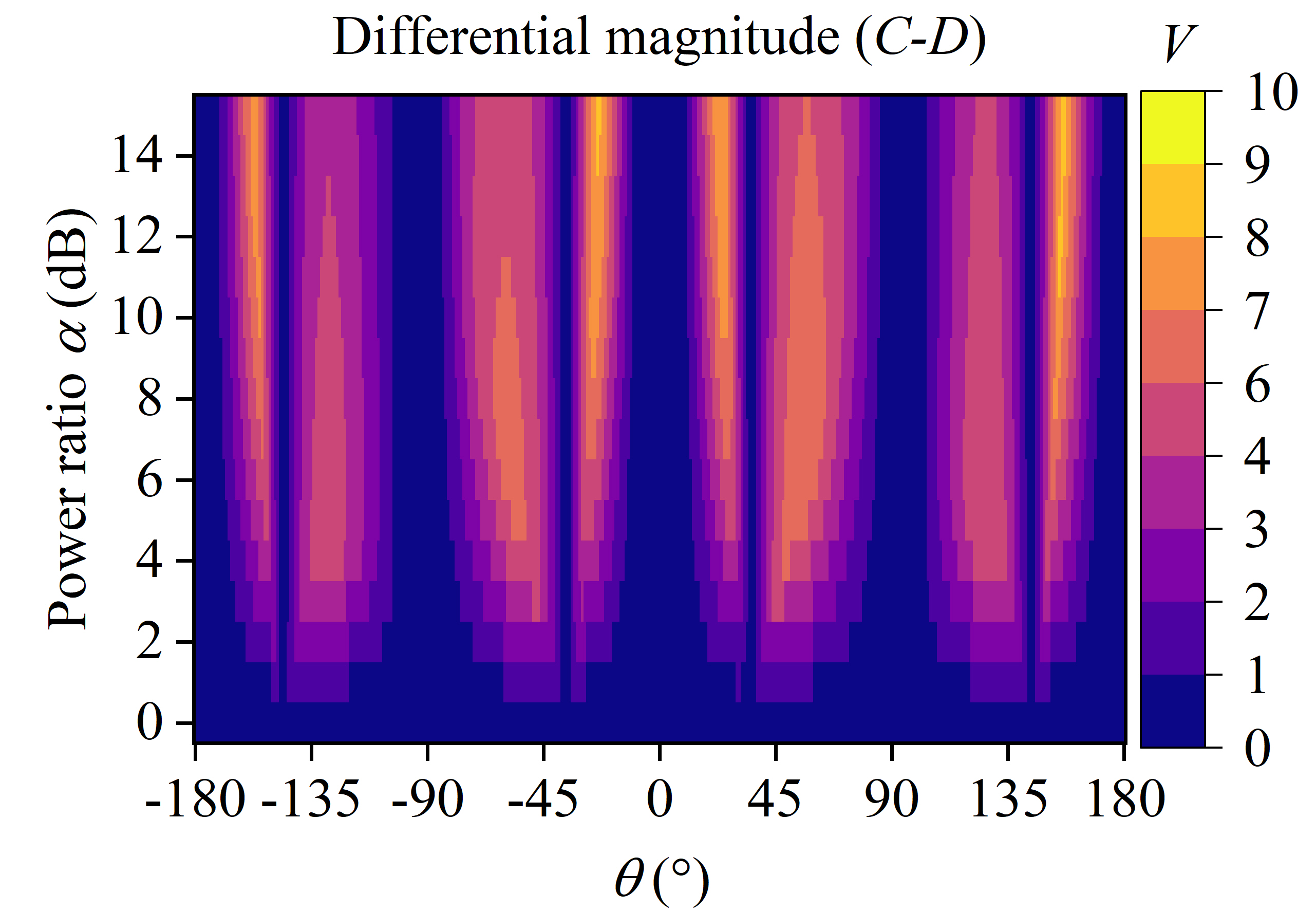}}\hfil
	\subfloat[]{\includegraphics[width=2.33in]{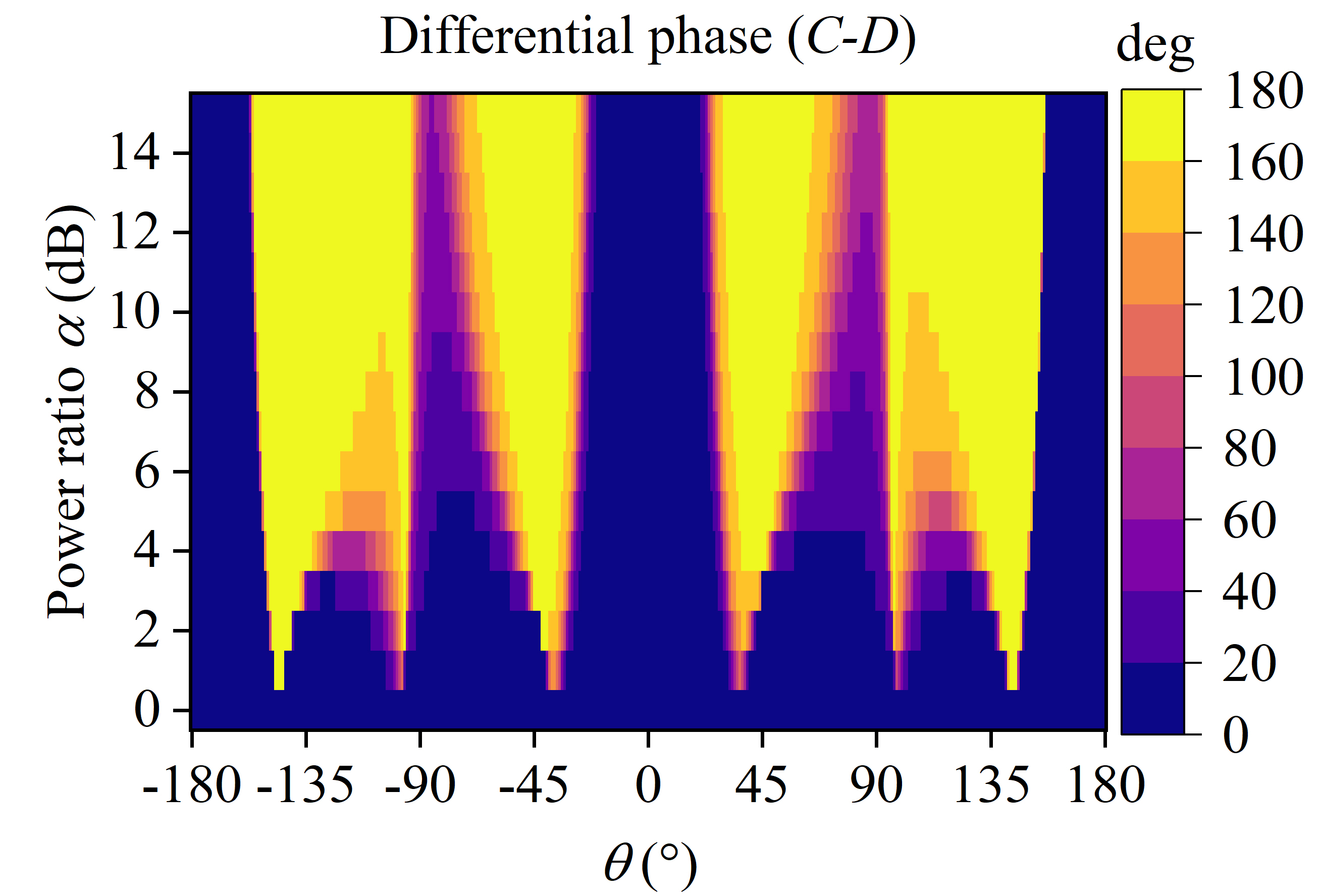}}\hfil
	\subfloat[]{\includegraphics[width=2.4in]{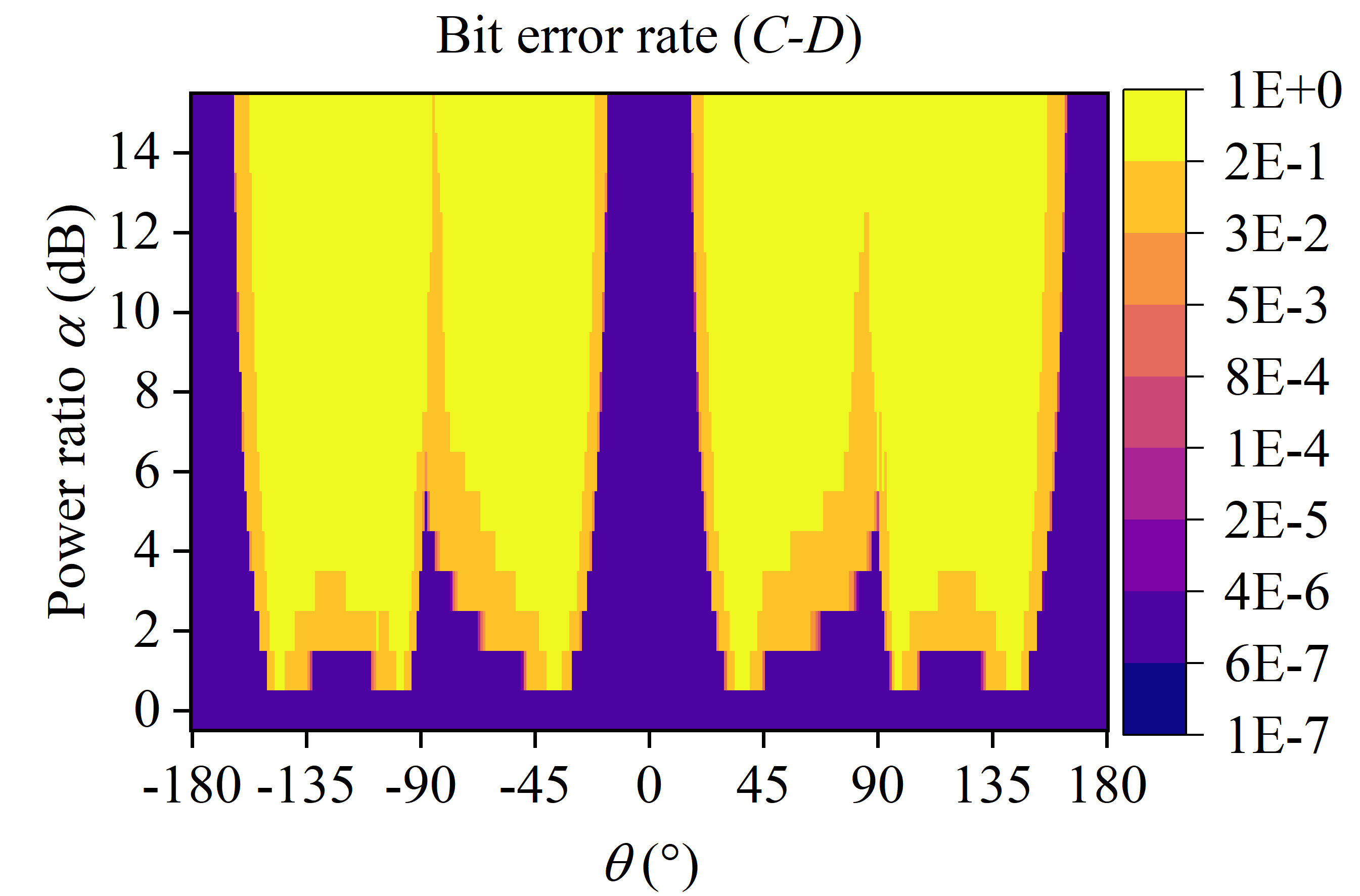}}\hfil\\
	\subfloat[]{\includegraphics[width=2.26in]{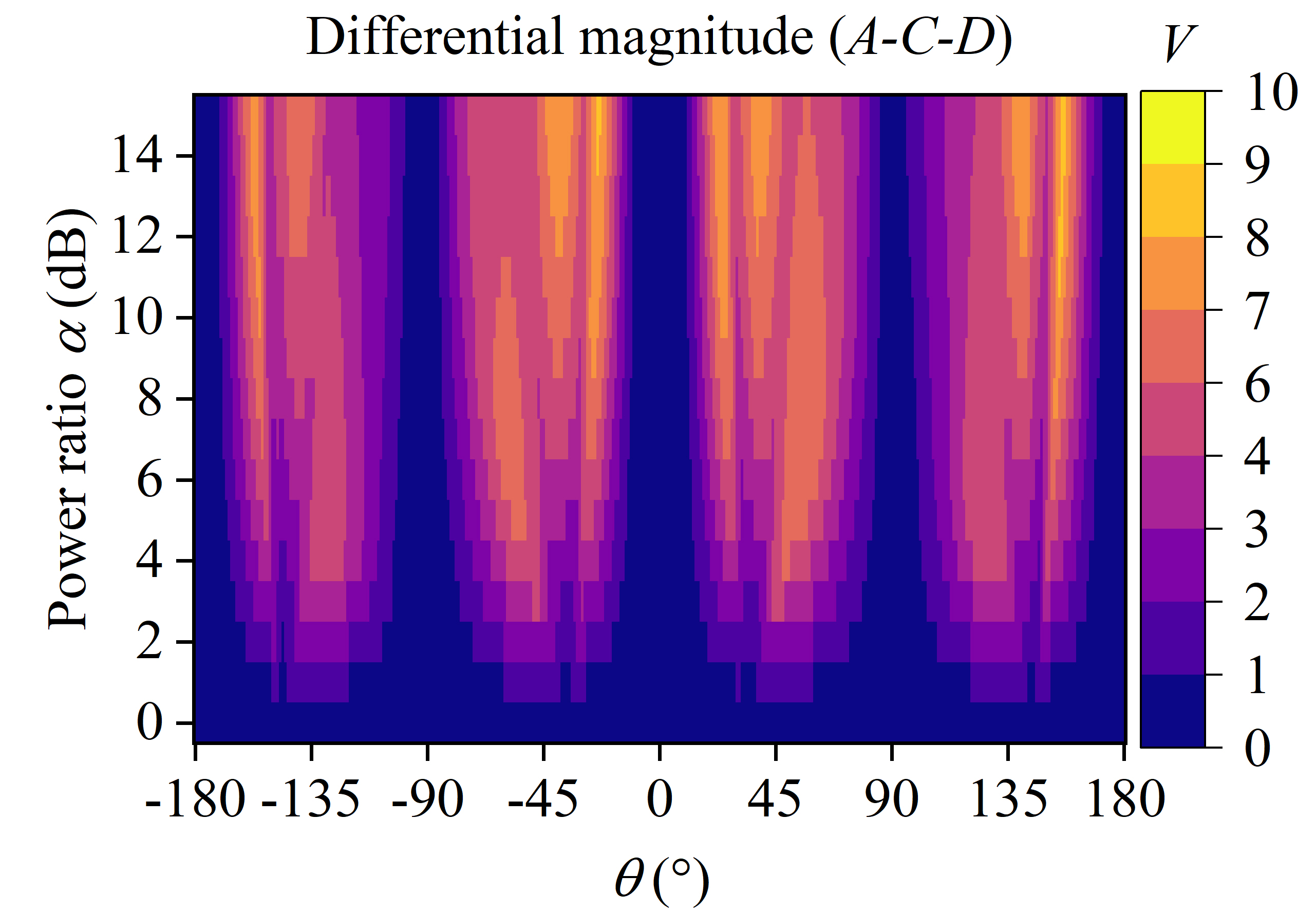}}\hfil
	\subfloat[]{\includegraphics[width=2.33in]{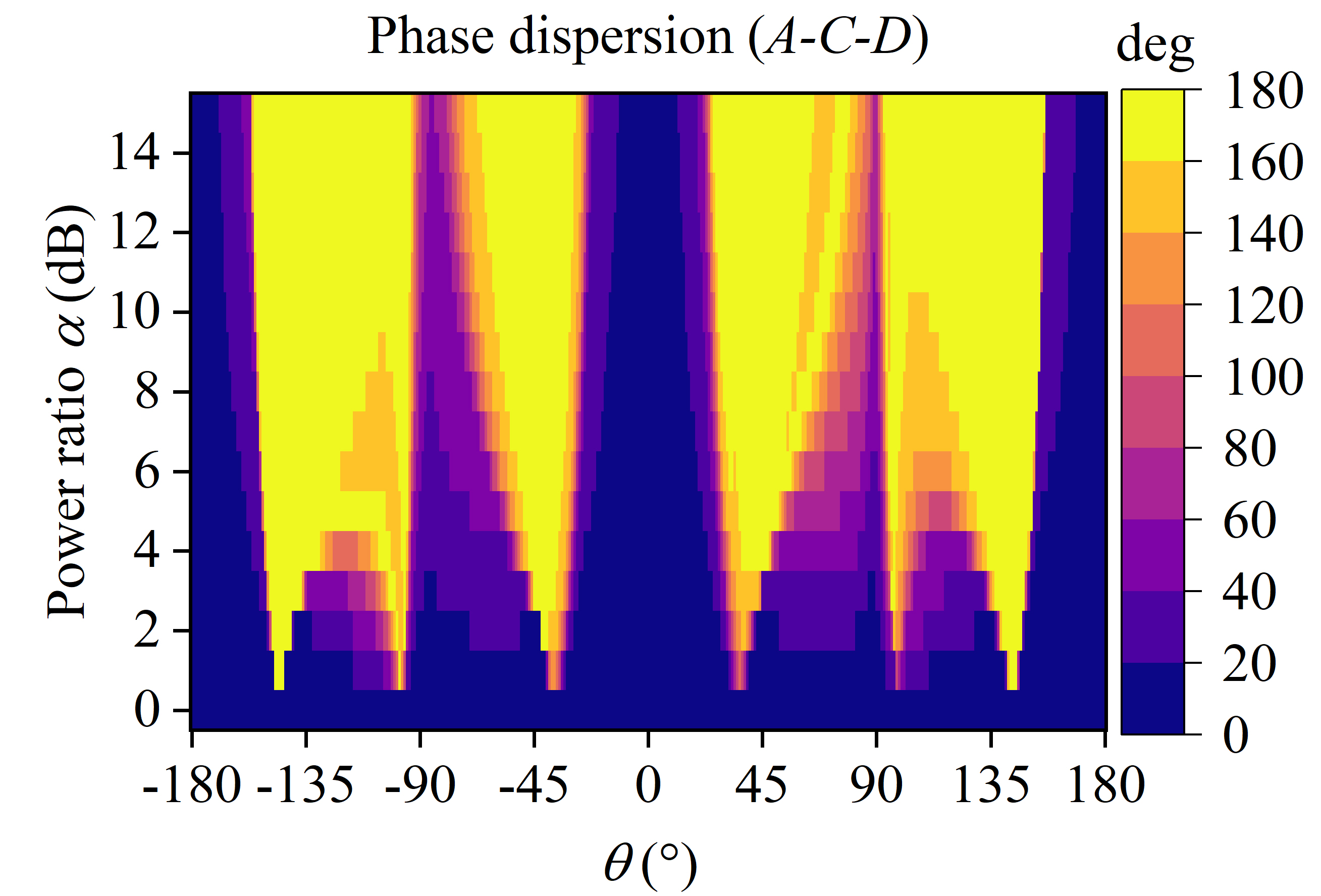}}\hfil
	\subfloat[]{\includegraphics[width=2.4in]{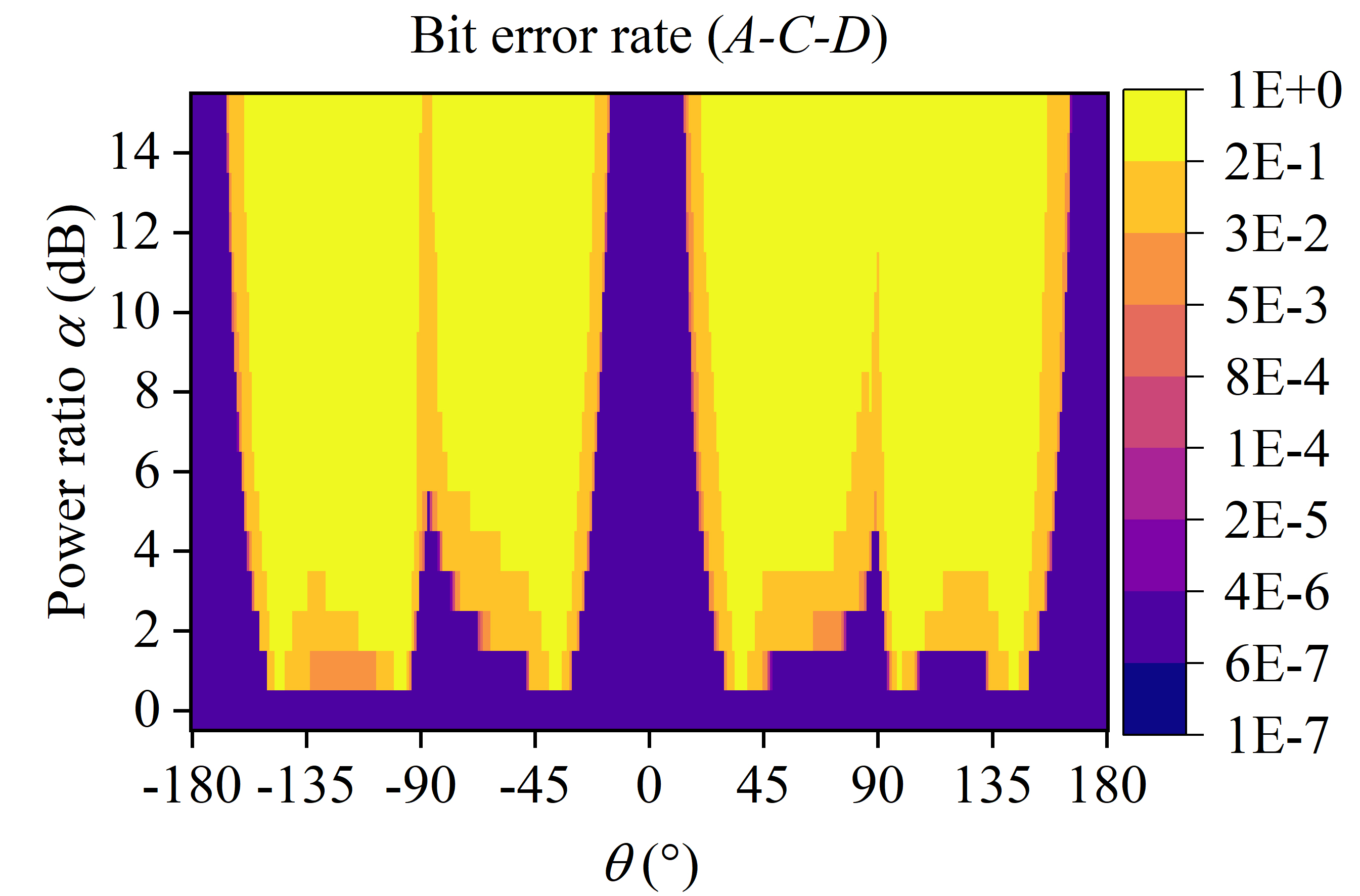}}\hfil\\
	\subfloat[]{\includegraphics[width=2.26in]{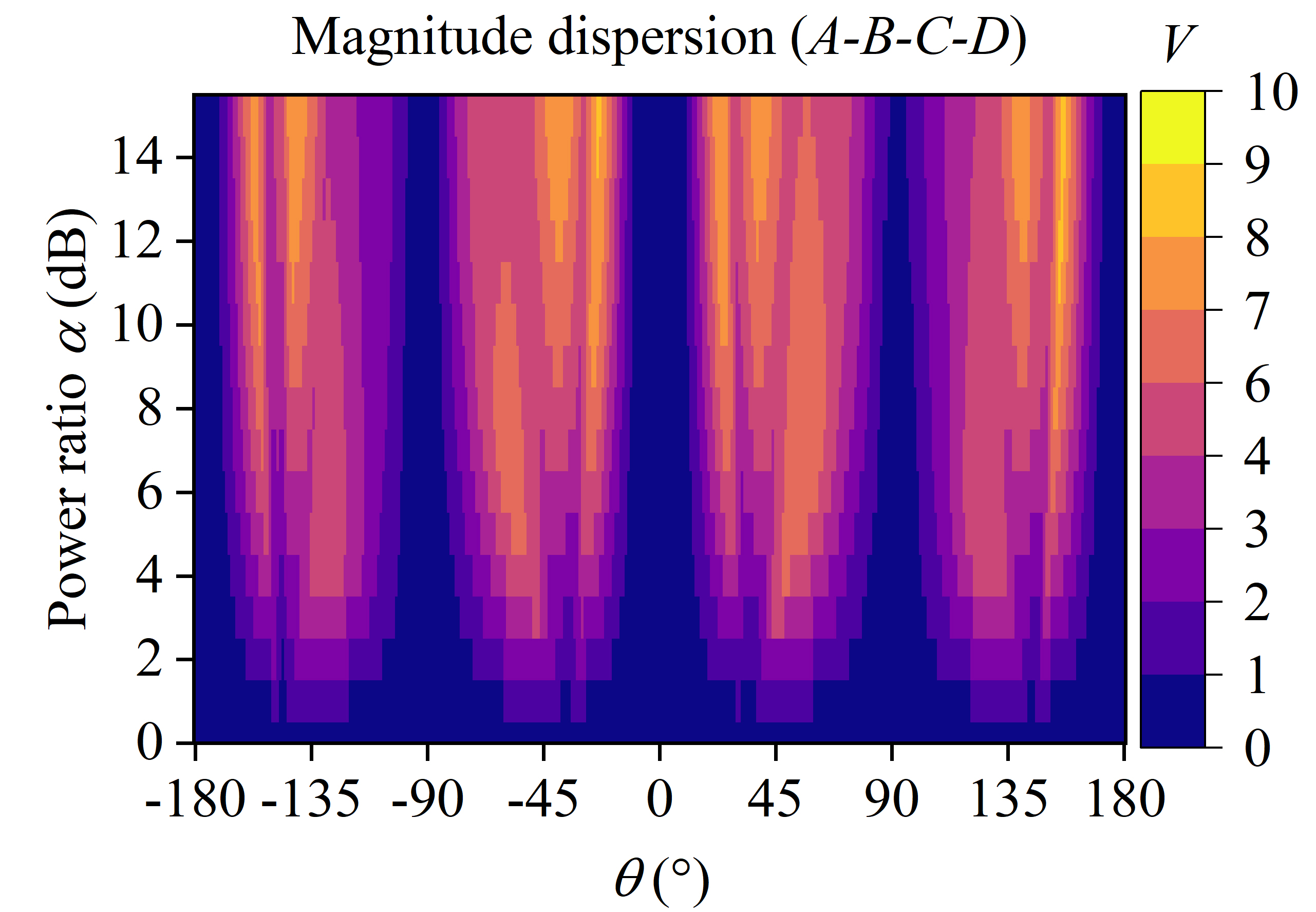}}\hfil
	\subfloat[]{\includegraphics[width=2.33in]{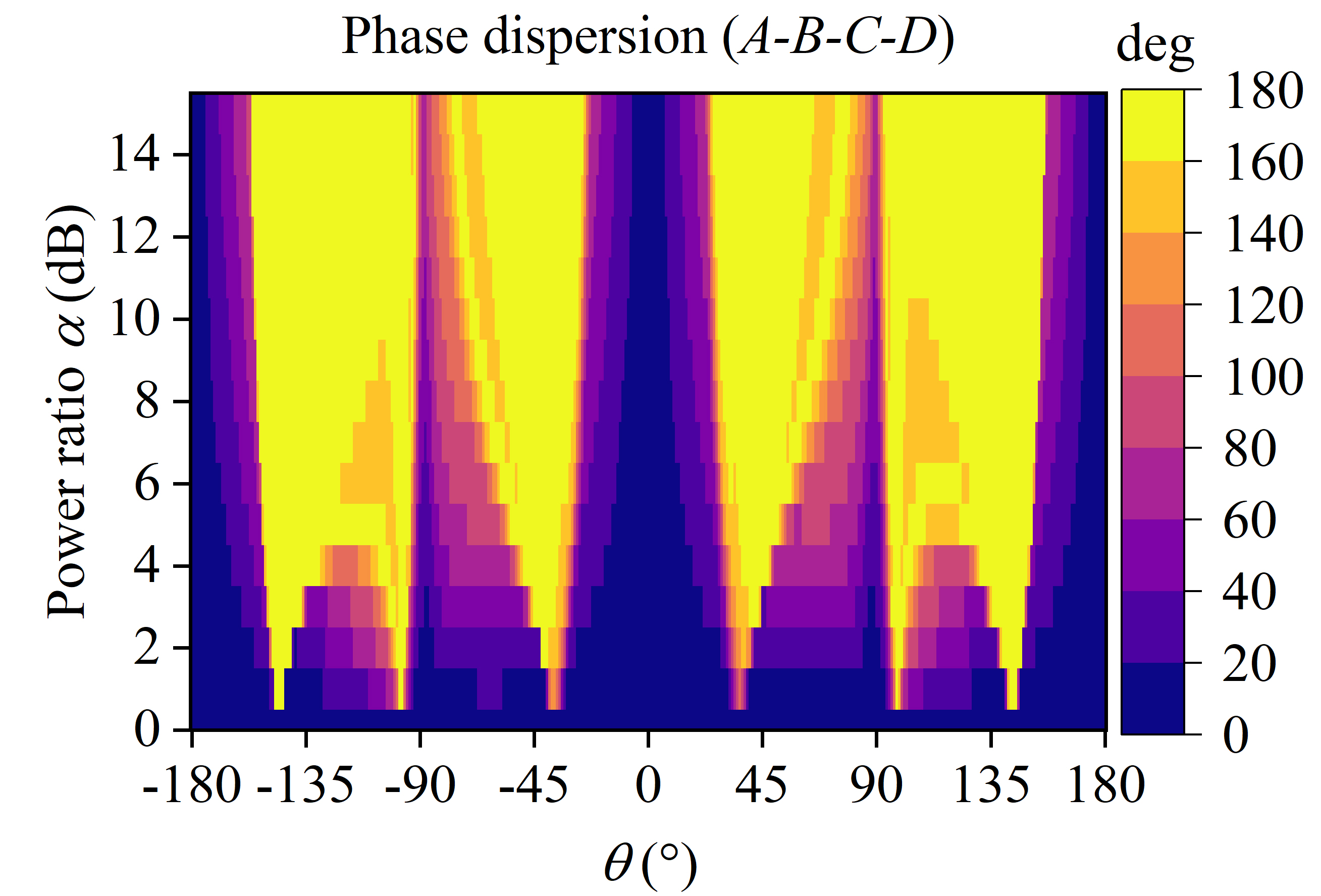}}\hfil
	\subfloat[]{\includegraphics[width=2.4in]{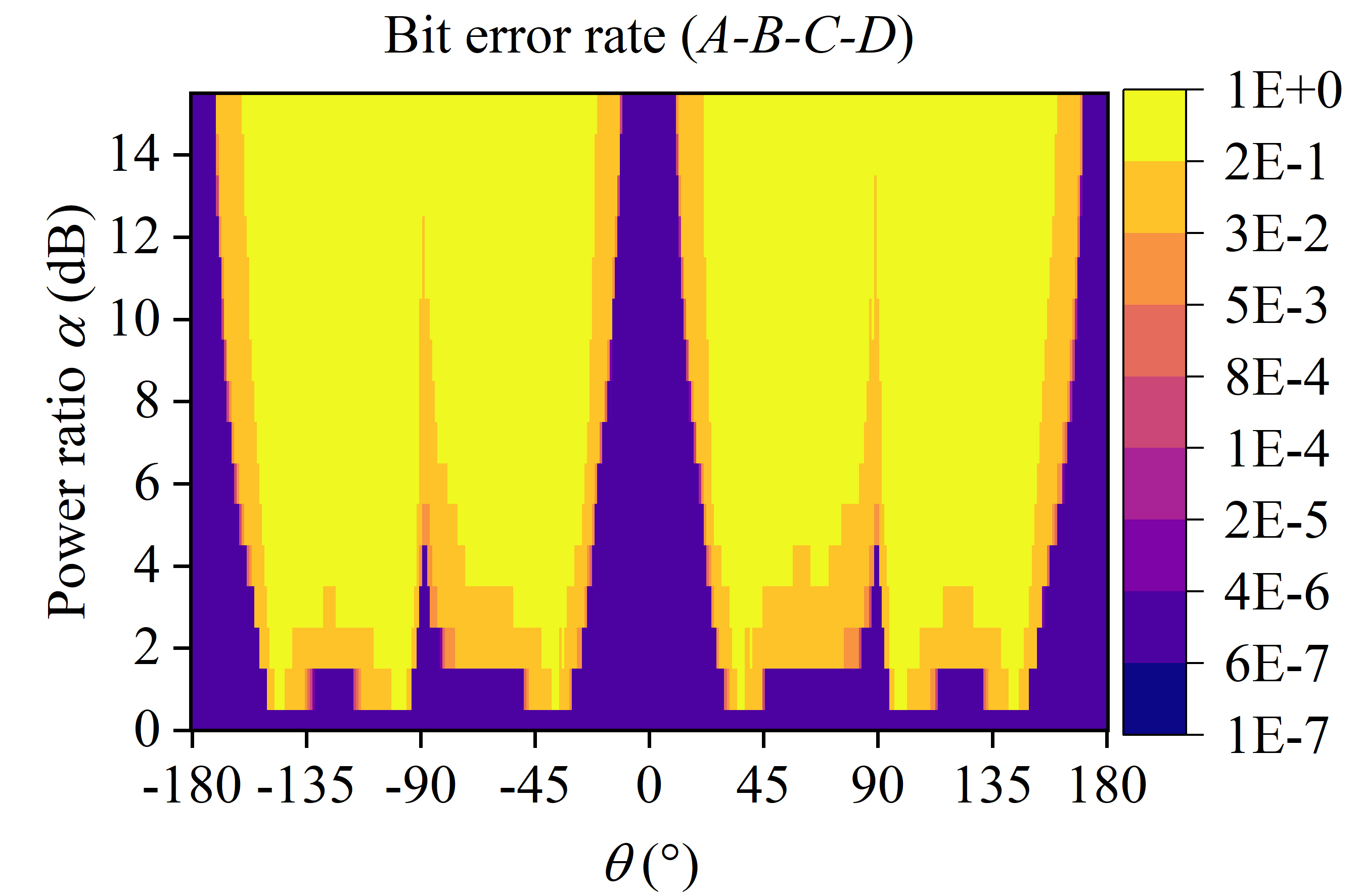}}\hfil
	\caption{Simulated radiated field variation and BER of the proposed four-state dynamic array versus power ratio \(\alpha\) and observation angle \(\theta\). (a)--(c) correspond to the \(A\)--\(B\) switching case and show the absolute differential magnitude \(|\Delta V_{AB}|\), the absolute differential phase \(|\Delta \phi_{AB}|\), and the corresponding 16-QAM BER, respectively. (d)--(f) correspond to the \(C\)--\(D\) switching case and show \(|\Delta V_{CD}|\), \(|\Delta \phi_{CD}|\), and the corresponding BER, respectively. (g)--(i) correspond to the three-state \(A\)--\(C\)--\(D\) switching case and show the magnitude dispersion \(D_{\mathrm{mag}}(\theta,\alpha)\), the phase dispersion \(D_{\phi}(\theta,\alpha)\), and the corresponding BER, respectively. (j)--(l) correspond to the full \(A\)--\(B\)--\(C\)--\(D\) switching case and show \(D_{\mathrm{mag}}(\theta,\alpha)\), \(D_{\phi}(\theta,\alpha)\), and the corresponding BER, respectively.}\label{fig:discrepancy_ber_vs_alpha}
\end{figure*}Fig.~\ref{fig:discrepancy_ber_vs_d} further validates the spacing-dependent interpretation by showing the simulated radiated field variation and BER versus the inter-cell spacing $d$ and observation angle $\theta$. For the two-state cases, the plotted quantities are the absolute differential magnitude $|\Delta V_{ij}|$ and the absolute differential phase $|\Delta \phi_{ij}|$, where $(i,j)\in\{(A,B),(C,D)\}$. For the multi-state cases, $D_{\mathrm{mag}}(\theta,d)$ denotes the magnitude dispersion, i.e., the difference between the maximum and minimum field magnitudes among the active switching states, whereas $D_{\phi}(\theta,d)$ denotes the phase dispersion, i.e., the maximum absolute wrapped phase difference among all active state pairs. For the BER evaluation, a 48-kb pseudorandom bit sequence is Gray-coded and modulated using 16-QAM, and the communication channel is directly constructed from the simulated complex far-field responses exported from HFSS \cite{10286341}. The switching states are alternated on a symbol-by-symbol basis, and equalization is performed using the average channel over the active switching states. The SNR control parameter is set to 40 dB so that the observed BER is dominated primarily by the directional modulation effect rather than by insufficient SNR. It is observed that larger radiated field variation, particularly larger phase discrepancy or phase variation, generally results in more severe constellation distortion and hence higher BER, consistent with (20)--(27), while a relatively stable low-BER region is preserved around broadside.

\subsection{Effect of the Power Ratio on Information Beam and BER}

Fig.~\ref{fig:discrepancy_ber_vs_alpha} summarizes the simulated differential field characteristics and the corresponding 16-QAM BER of the proposed array as functions of the power ratio \(\alpha\) and observation angle \(\theta\). The results are consistent with the theoretical model developed above. In particular, the power ratio directly modifies the perturbation term \(\mathcal{Q}=I_0(\sqrt{\alpha}-1)\sin\psi\), and therefore changes the phase- and magnitude-perturbation channels \(I_{\phi}=2\mathcal{Q}\cos(\gamma/2)\) and \(I_m=2\mathcal{Q}\sin(\gamma/2)\). Since the information-beam efficiency is determined by the relative strengths of the reference and perturbation terms, increasing \(\alpha\) generally enhances the radiated field variation away from broadside and further confines the information-recoverable region. This trend is also consistent with the amplitude-ratio-controlled information beam behavior reported in \cite{10286341}, where increasing amplitude asymmetry narrows the recoverable angular region.
\begin{figure*}[t]
	\centering
	\subfloat[]{%
		\includegraphics[width=5.8in]{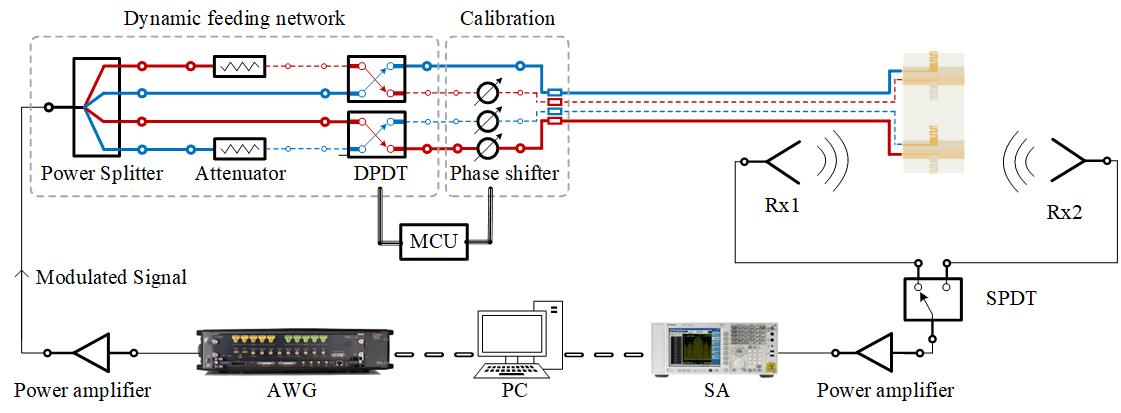}
	}\hfil
	\subfloat[]{%
		\includegraphics[width=3.42in]{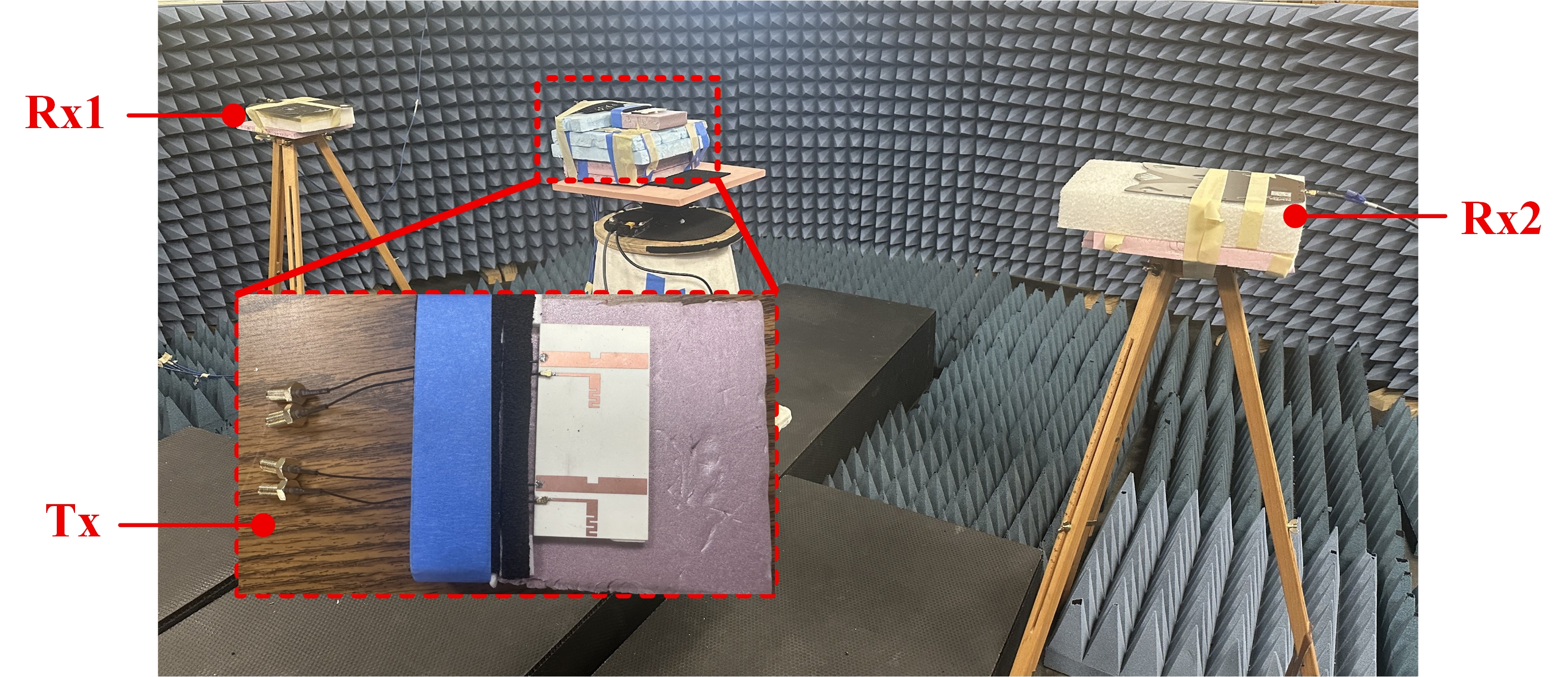}
	}\hfil
	\subfloat[]{%
		\includegraphics[width=3.55in]{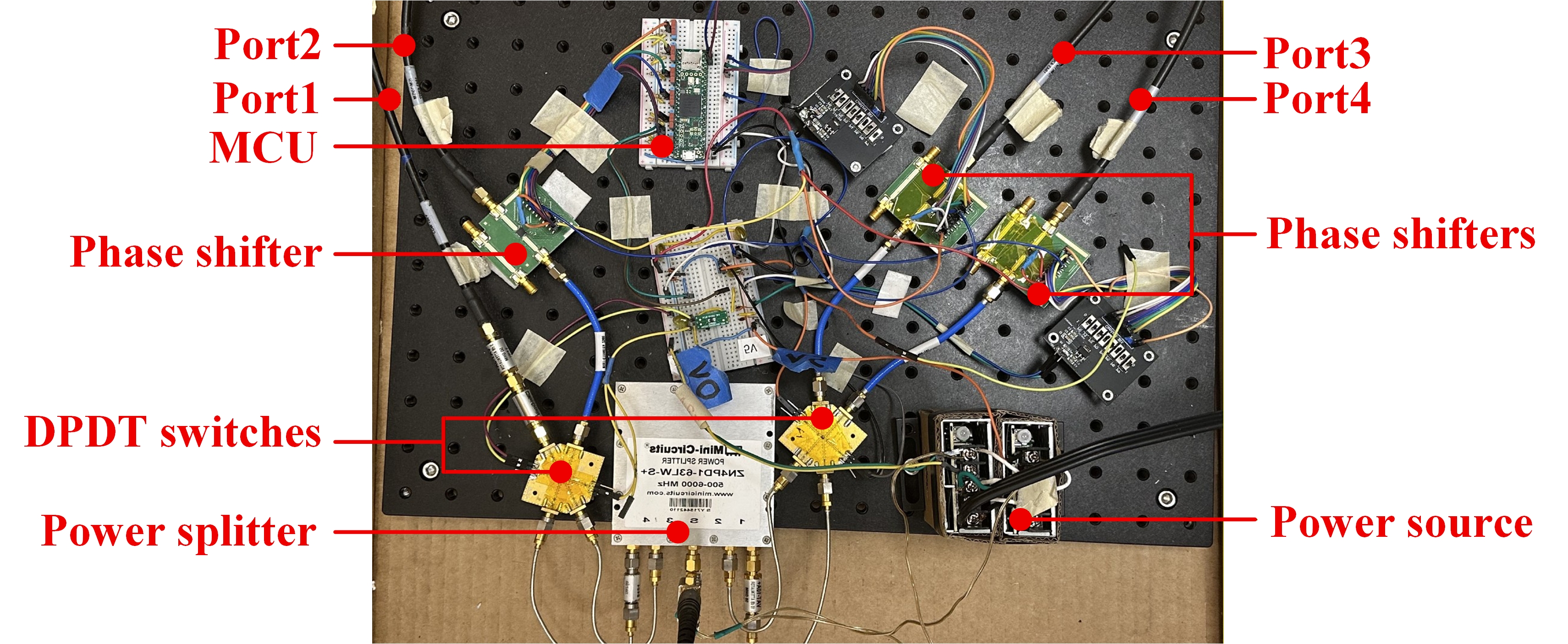}
	}
	
	\caption{Experimental platform for BER characterization of the proposed dynamic omnidirectional array. (a) Block diagram of the communication measurement system. (b) Photograph of the over-the-air measurement setup, where the proposed dynamic array is used at the transmitter and a broadband Vivaldi antenna is used at the receiver. (c) Photograph of the dynamic feeding and calibration network, including the power splitter, DPDT switches, phase shifters, microcontroller unit (MCU), and power supply.}
	\label{fig:measurement_setup}
\end{figure*}
For the \(A\)--\(B\) switching pair, \(|\Delta V_{AB}|\) remains close to zero over most angles and power ratios, whereas \(|\Delta\phi_{AB}|\) is dominant. This agrees with the theoretical result that the \(A\)--\(B\) case is mainly governed by the phase-perturbation channel \(I_{\phi}\). Consequently, the corresponding BER map exhibits a narrow low-BER region around broadside, while off-broadside directions experience strong phase distortion and therefore severe BER degradation.

By contrast, the \(C\)--\(D\) switching pair exhibits much stronger differential magnitude variation, confirming that this case is predominantly governed by the magnitude-perturbation channel \(I_m\). Meanwhile, the differential phase is also non-negligible, indicating practical phase leakage caused by nonideal field mixing and mutual coupling. As a result, the BER distribution of the \(C\)--\(D\) case remains strongly angle dependent, but its spatial pattern differs from that of the \(A\)--\(B\) case because the dominant perturbation mechanism is now magnitude redistribution rather than phase reversal.

\begin{figure}[t]
	\centering
	\includegraphics[width=3.45in]{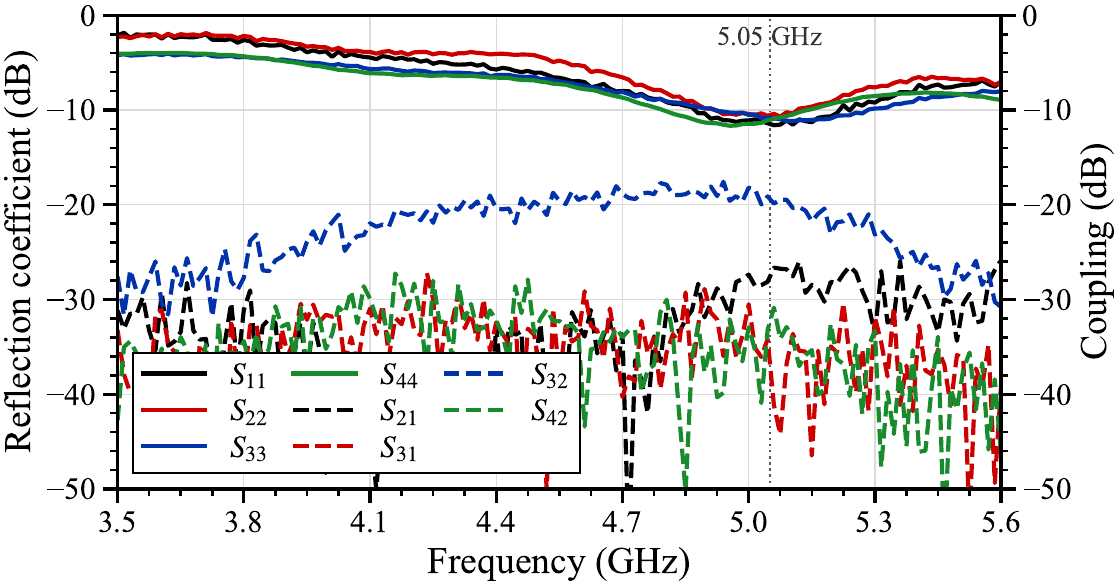}
	\caption{Measured \(S\)-parameter response of the fabricated four-element array. The solid traces show the reflection coefficients \(S_{11}\), \(S_{22}\), \(S_{33}\), and \(S_{44}\) on the left axis, while the dashed traces show the selected inter-port coupling coefficients on the right axis. During each two-port VNA measurement, the unused antenna ports were terminated with matched 50~\(\Omega\) loads.}
	\label{measured_s11}
\end{figure}
For the full \(A\)--\(B\)--\(C\)--\(D\) switching case, both the magnitude dispersion \(D_{\mathrm{mag}}(\theta,\alpha)\) and the phase dispersion \(D_{\phi}(\theta,\alpha)\) become significant, showing that the two perturbation channels are simultaneously excited. This combined effect produces the strongest directional BER selectivity among the three cases. Overall, Fig.~\ref{fig:discrepancy_ber_vs_alpha} confirms that \(\alpha\) serves as a direct control parameter for the tradeoff between perturbation strength and information beam confinement. Unlike the two-state phase-center-dynamic mechanism in \cite{10286341}, the proposed four-state antenna enables the same control variable to regulate not only phase-dominant perturbation in the \(A\)--\(B\) case, but also magnitude-dominant and combined phase--magnitude perturbation in the \(C\)--\(D\) and full four-state cases.

\section{System Design and Measurement}

\subsection{Communication Performance Measurement}
The experimental setup used to characterize the BER performance of the proposed array is shown in Fig.~\ref{fig:measurement_setup}. The four-element array was fabricated on a printed circuit substrate using a standard chemical-etching process. Each element was connected through an MHF4-RP coaxial jumper cable (dc--8~GHz) and an I-PEX-compatible surface-mounted receptacle. All RF interconnections were designed for 50~\(\Omega\) operation. 
\begin{figure*}[!t]
	\centering
	\subfloat[]{\includegraphics[width=3.32in]{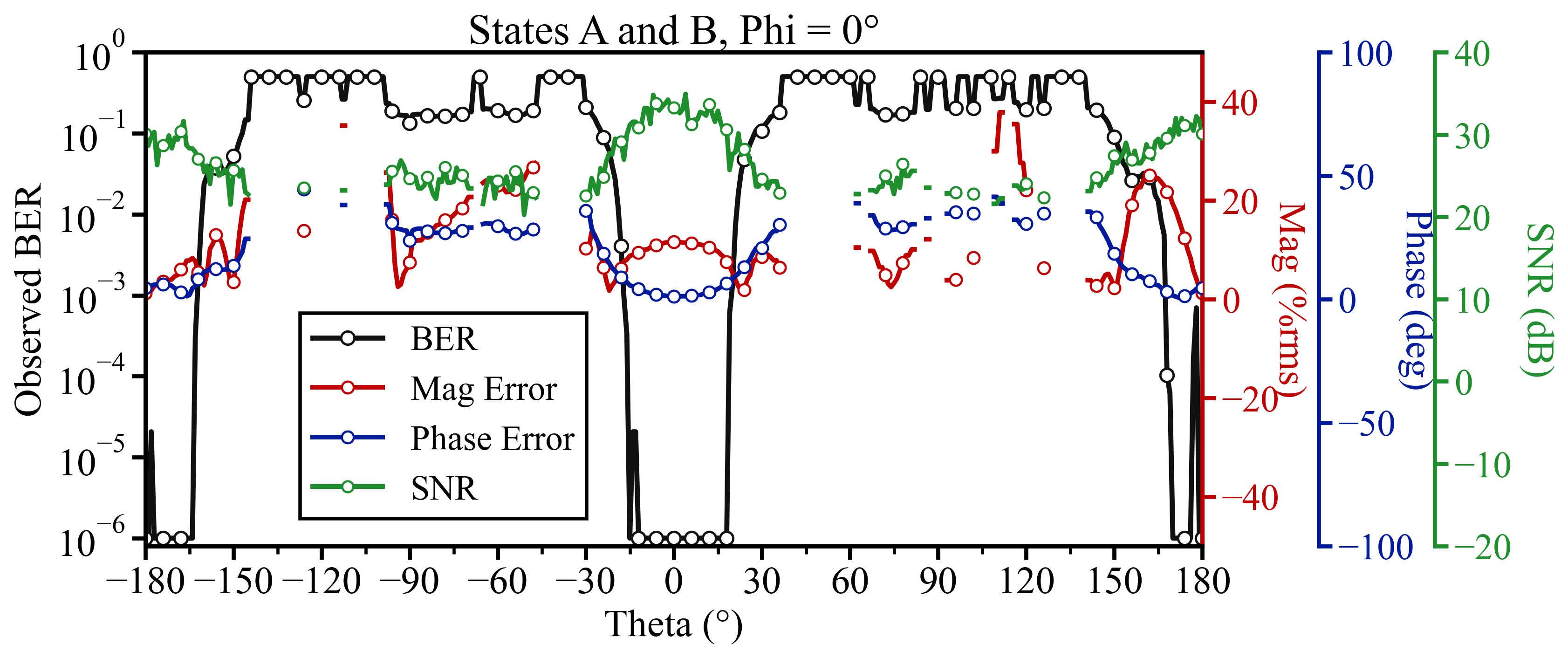}}\hfil
	\subfloat[]{\includegraphics[width=3.32in]{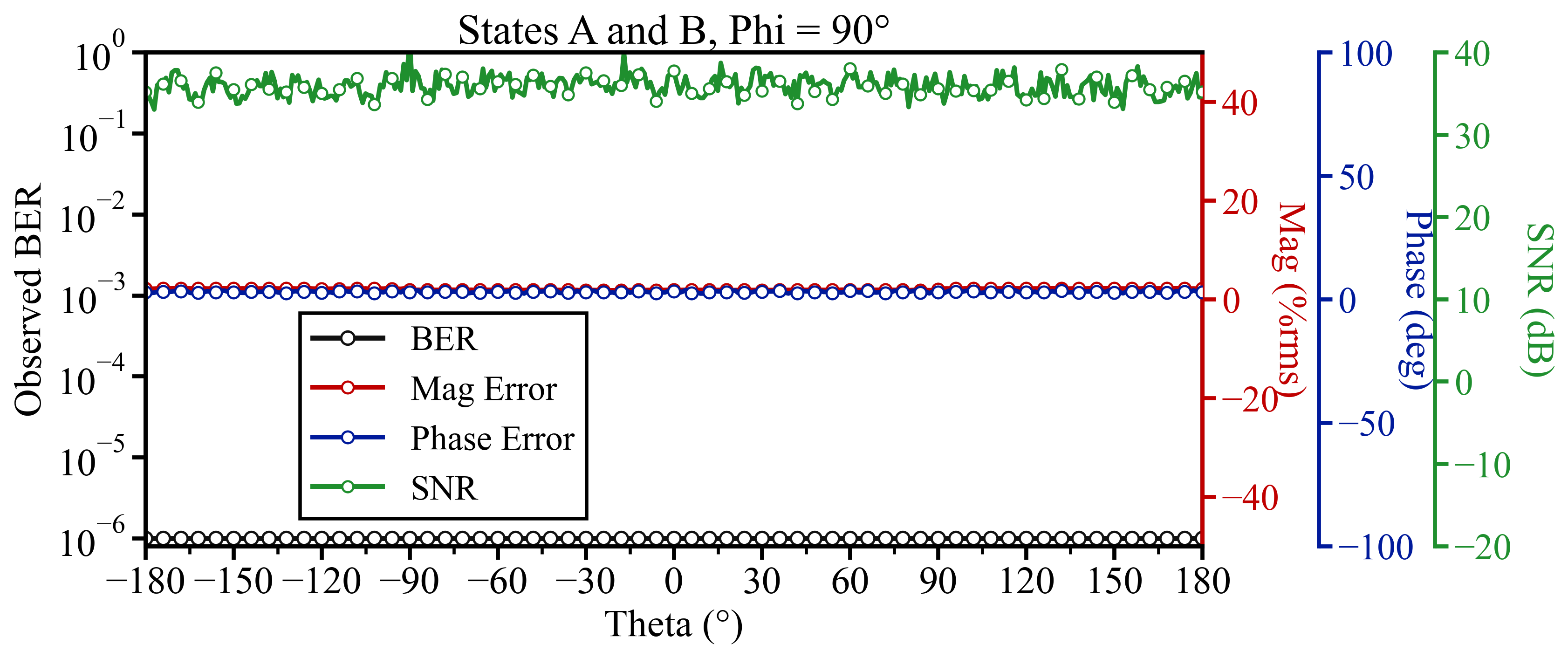}}\\
	\subfloat[]{\includegraphics[width=3.32in]{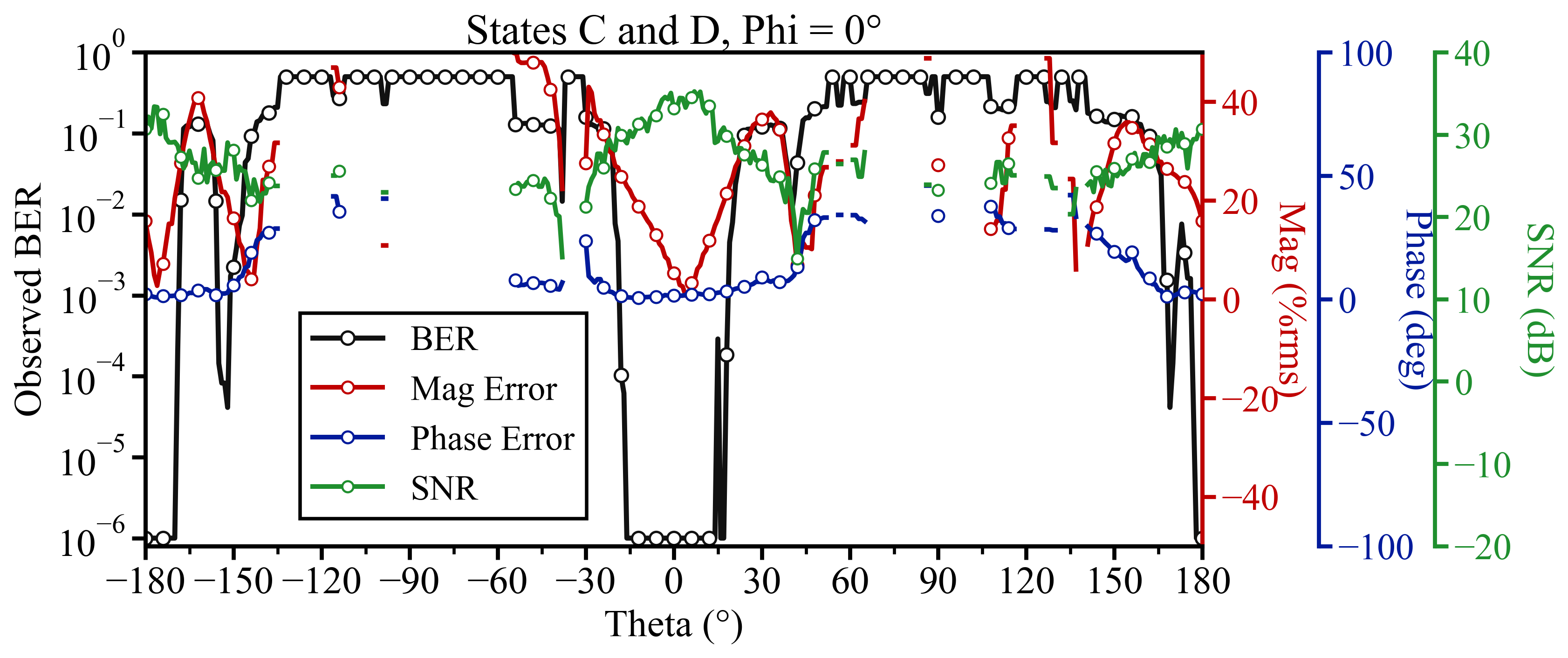}}\hfil
	\subfloat[]{\includegraphics[width=3.32in]{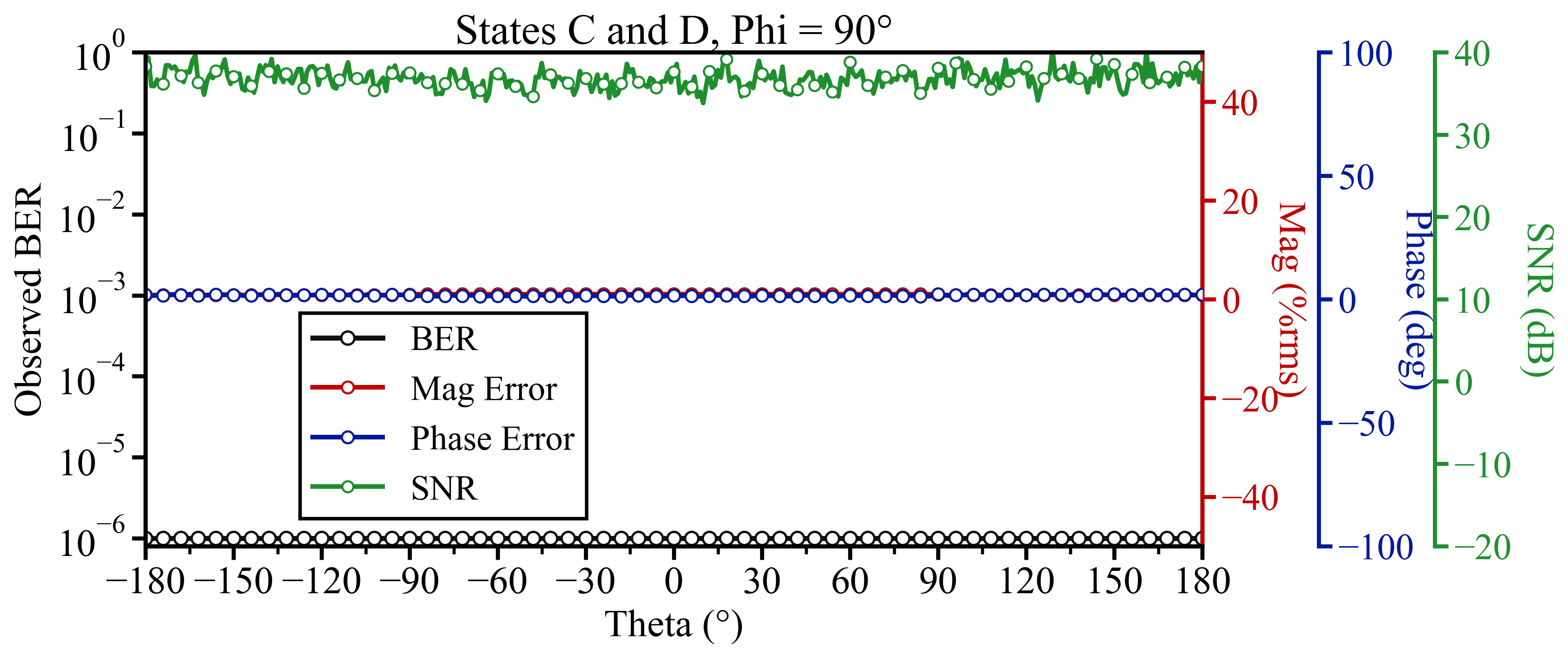}}\\
	\subfloat[]{\includegraphics[width=3.32in]{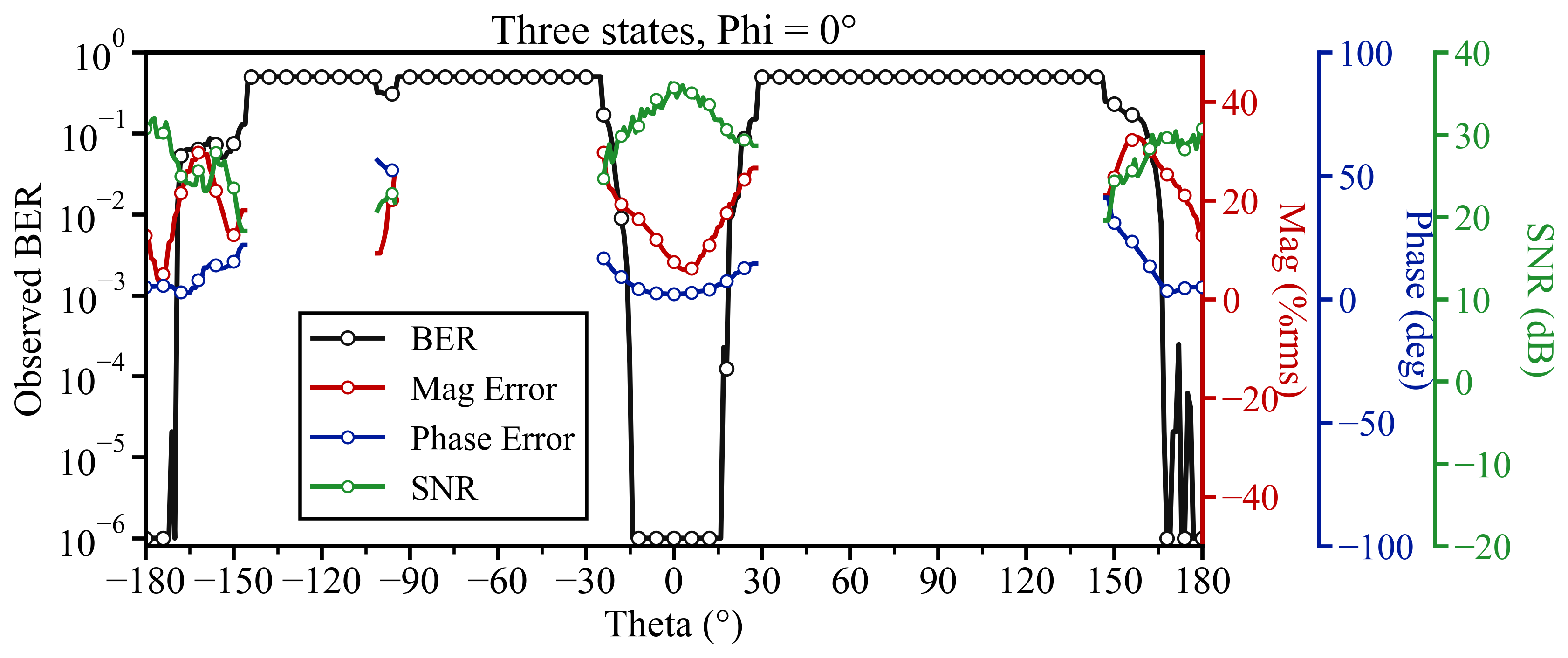}}\hfil
	\subfloat[]{\includegraphics[width=3.32in]{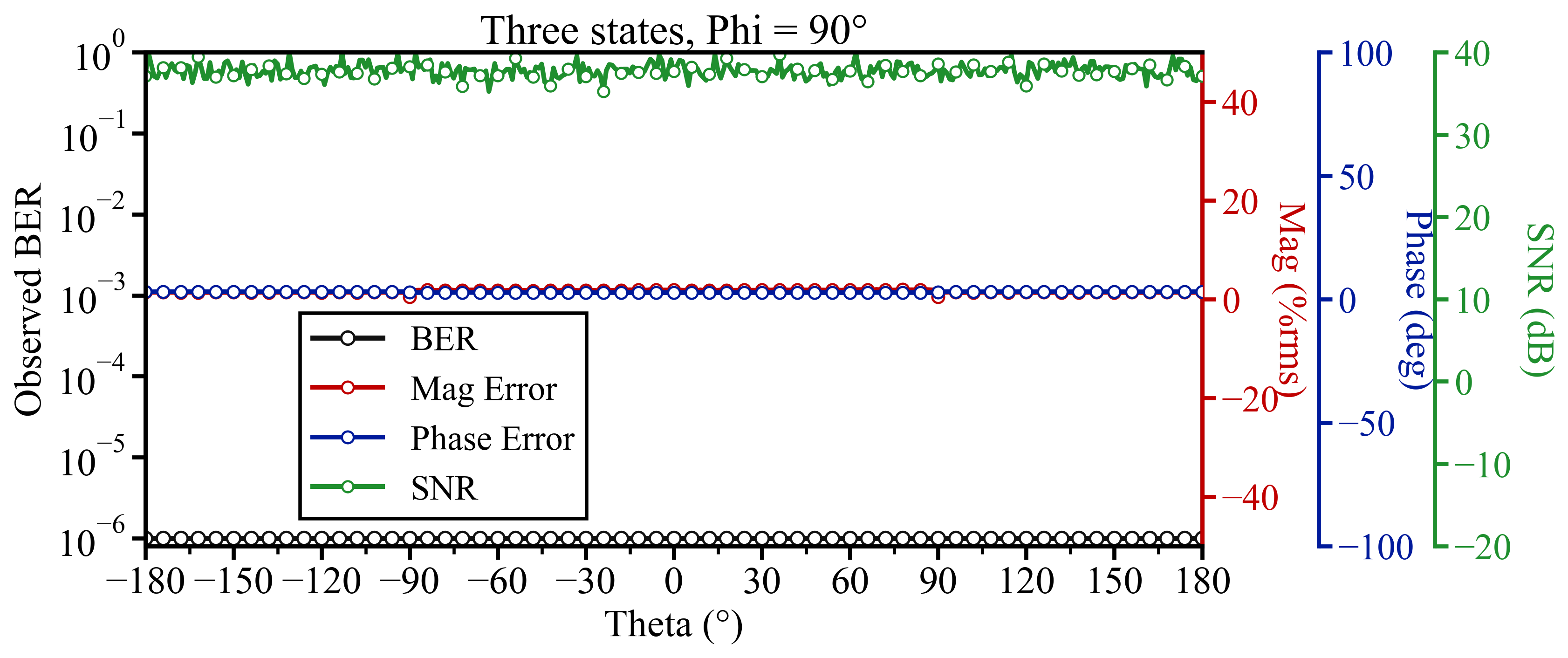}}\\
	\subfloat[]{\includegraphics[width=3.32in]{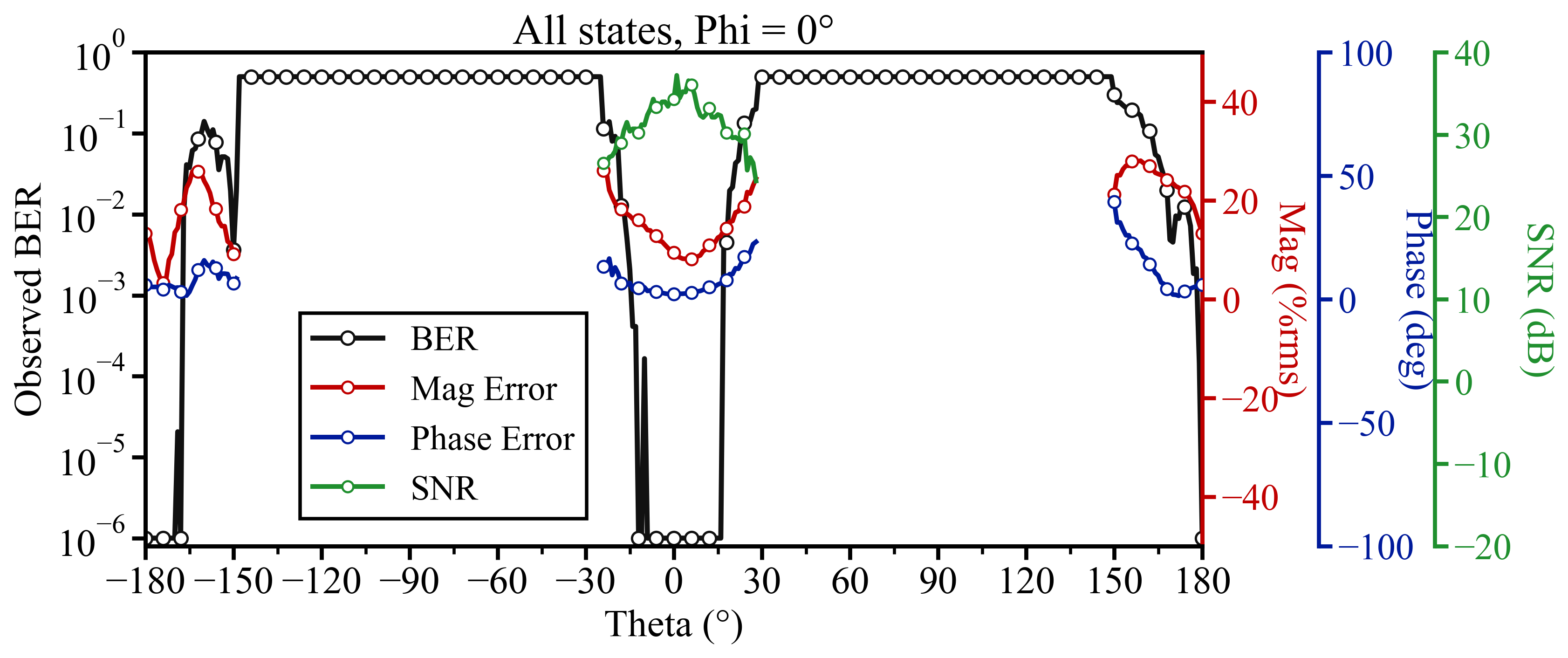}}\hfil
	\subfloat[]{\includegraphics[width=3.32in]{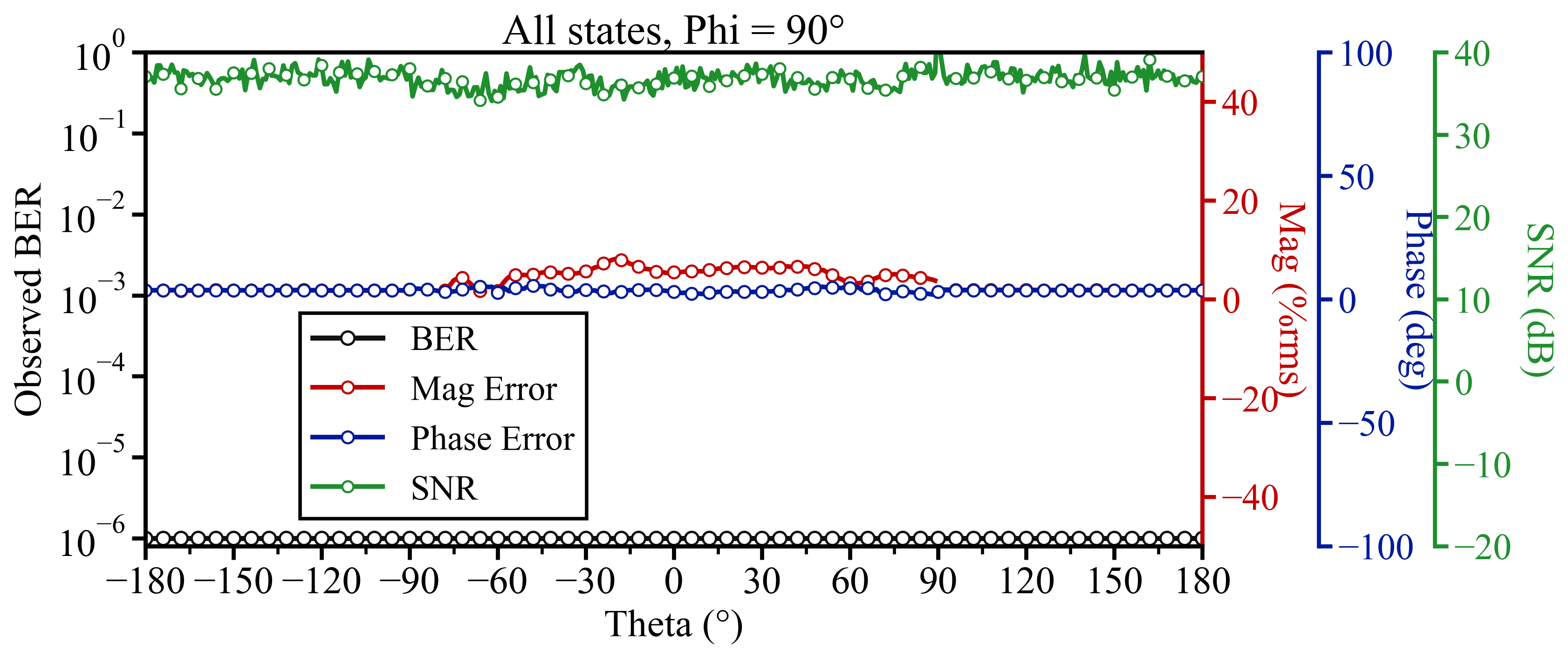}}
	\caption{Measured 16-QAM BER, magnitude error, phase error, and SNR of the proposed dynamic array under calibrated switching operation at 5.05~GHz. (a) and (b) show the \(A\)--\(B\) switching case in the \(E\)- and \(H\)-planes, respectively. (c) and (d) show the \(C\)--\(D\) switching case in the \(E\)- and \(H\)-planes, respectively. (e) and (f) show the three-state switching case in the \(E\)- and \(H\)-planes, respectively. (g) and (h) show the full \(A\)--\(B\)--\(C\)--\(D\) switching case in the \(E\)- and \(H\)-planes, respectively. Samples for which demodulation failed are plotted at BER \(=0.5\), corresponding to angles where the magnitude and phase errors were too large for reliable waveform recovery; the corresponding magnitude and phase errors are left blank because they could not be reliably measured after demodulation failed.}
	\label{fig:measured_ber_switching}
\end{figure*}
For the communication experiment, the array and the switching circuitry were placed at the transmitter. A digitally modulated RF signal at 5.05~GHz was generated by a Keysight M8190A arbitrary waveform generator (AWG) and then amplified to maintain a sufficiently high received SNR, so that the measured BER variation was dominated primarily by the intentionally controlled radiation characteristics rather than by insufficient link budget. At the receiver, two broadband Vivaldi antennas (TSA800, 0.8--6~GHz) were used to capture the transmitted signal for front and back hemispheres, respectively. Their RF paths were controlled by a single-pole-double-through (SPDT) switch (RF Lambda RFSP2TR0218G). To characterize BER as a function of angle, the transmitting array was mounted on a precision rotatable platform, while the Tx--Rx separation was kept in the far-field region. The received signal was captured by a Keysight N9030A signal analyzer (SA) and extract the BER, magnitude error, phase error, and received SNR at each observation angle.

The four-path dynamic feeding network is illustrated in Fig.~\ref{fig:measurement_setup}(c). A broadband four-way power splitter was used to distribute the modulated signal to four RF paths. Two double-pole double-throw (DPDT) RF switches were employed to control the excitation states of the switching pairs, and these switches were digitally driven by a Teensy~4.1 microcontroller unite (MCU) for real-time path reconfiguration. During the BER measurements, the selected switching sequence was cycled at a switching frequency of 2~kHz. Differential attenuation was intentionally introduced among the RF paths to realize spatial amplitude modulation. In addition, digitally controlled phase shifters were inserted in three of the four branches to compensate for phase offsets introduced by cables, attenuators, and other RF components, and also to allow controlled adjustment of the information beam direction.

The measured reflection and coupling responses of the fabricated array are presented in Fig.~\ref{measured_s11}, where all unused ports were terminated with matched 50~\(\Omega\) loads.

As shown in Fig.~\ref{measured_s11}, the four measured reflection coefficients are clustered near the intended operating frequency of 5~GHz, with the representative port responses around \(-8.2\) to \(-11.7~\mathrm{dB}\). The selected inter-element coupling terms remain substantially lower than the reflection responses; at 5.05~GHz, the plotted coupling coefficients are approximately \(-19~\mathrm{dB}\) or below, with several nonadjacent paths below \(-28~\mathrm{dB}\). These measured \(S\)-parameters confirm that the fabricated array and interconnects provide a usable matched four-port platform for the subsequent switching experiment.

Based on the available laboratory setup, the switching network was calibrated at 5.05~GHz to realize an approximate differential power ratio of \(\alpha=12~\mathrm{dB}\), which was adjusted through attenuators and digitally controlled phase shifters while compensating for branch-dependent phase offsets introduced by cables and RF components. Prior to each angular BER measurement, the phase shifters and receiver position were adjusted to minimize the magnitude and phase errors at broadside, thereby establishing a consistent reference condition for subsequent off-broadside evaluation. For signal generation and post-processing, 16-QAM modulation and demodulation were implemented using Keysight IQTools and PathWave vector signal analysis software. The array was operated in transmit mode, and BER was evaluated as a function of observation angle under the calibrated dynamic switching conditions.

The measured communication results are summarized in Fig.~\ref{fig:measured_ber_switching}. In the \(E\)-plane, the BER response exhibits a distinct information-recoverable sector around broadside rather than a conventional gain beam. Using BER \(=10^{-3}\) as the recovery criterion, the measured broadside information beam for the \(A\)--\(B\) switching case spans approximately \(-17^\circ\leq\theta\leq19^\circ\), corresponding to a beamwidth of \(36^\circ\). For the \(C\)--\(D\) switching case, the information beam is approximately \(-18^\circ\leq\theta\leq18^\circ\), also giving a \(36^\circ\) beamwidth. When three states are used in the \(A\)--\(C\)--\(D\) sequence, the recoverable sector is reduced to approximately \(-15^\circ\leq\theta\leq18^\circ\), corresponding to a \(33^\circ\) beamwidth. For the full \(A\)--\(B\)--\(C\)--\(D\) switching sequence, the information beam is further confined to approximately \(-14^\circ\leq\theta\leq16^\circ\), giving a \(30^\circ\) beamwidth. Thus, adding additional calibrated switching states preserves the intended broadside recovery direction while increasing the switching-induced state diversity that confines the BER-defined information beam. The corresponding low-BER regions near \(\theta=\pm180^\circ\) are the rear-side continuation of the same \(E\)-plane angular response of the planar radiator.

Outside these recoverable sectors, the measured BER increases by several orders of magnitude and, in some angular regions, the magnitude and phase errors become so large that the vector signal analyzer could not reliably demodulate the received waveform. These samples, for which demodulation failed, are plotted at BER \(=0.5\) in Fig.~\ref{fig:measured_ber_switching} to indicate outage-like information loss; the associated magnitude and phase errors are omitted because the constellation statistics are no longer reliable once demodulation fails. Importantly, the degradation does not coincide with a simple received-power null. In the valid off-broadside samples, the received SNR remains high enough for demodulation, while the magnitude and phase errors grow strongly with angle. This confirms that the BER degradation is produced primarily by switching-induced constellation distortion in the radiated field rather than by insufficient received power.

\begin{figure*}[!t]
	\centering
	\subfloat[]{\includegraphics[width=3.32in]{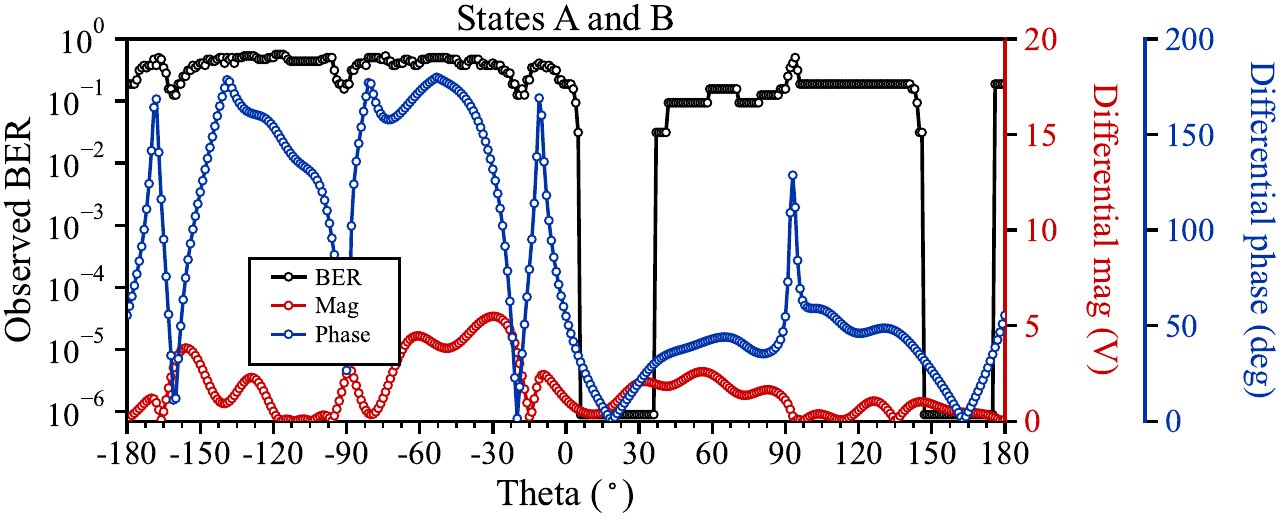}}\hfil
	\subfloat[]{\includegraphics[width=3.32in]{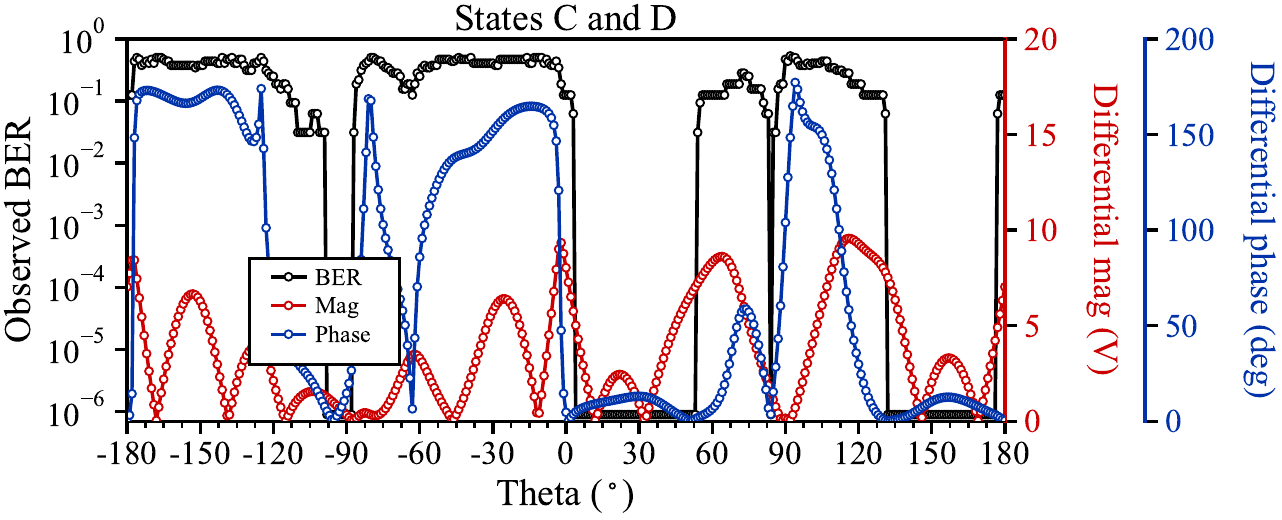}}\\
	\subfloat[]{\includegraphics[width=3.32in]{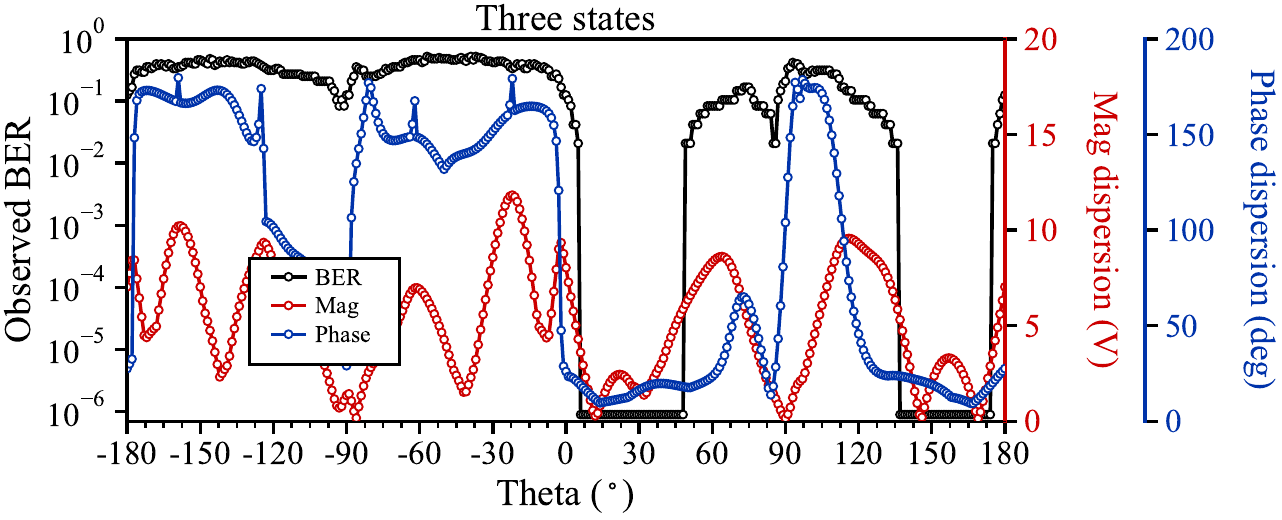}}\hfil
	\subfloat[]{\includegraphics[width=3.32in]{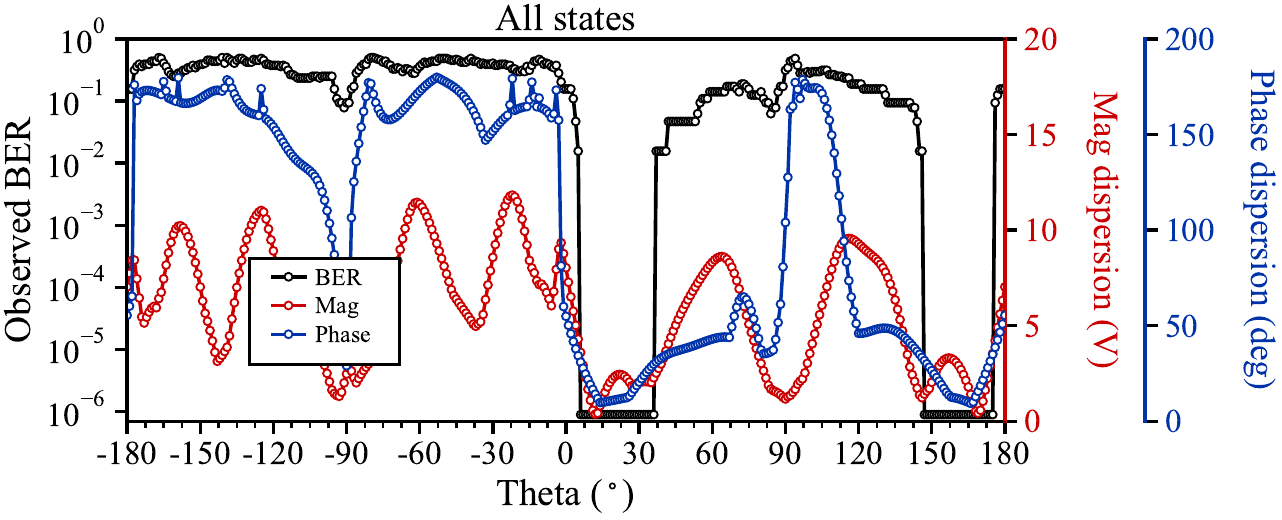}}
	\caption{Simulated 16-QAM BER and switching-induced field perturbation of the proposed array for a \(25^\circ\) steered information beam. (a) \(A\)--\(B\) switching, (b) \(C\)--\(D\) switching, (c) \(A\)--\(C\)--\(D\) switching, and (d) full \(A\)--\(B\)--\(C\)--\(D\) switching. For the two-state \(A\)--\(B\) and \(C\)--\(D\) cases, the red and blue curves denote the differential magnitude and differential phase between the two switching states, respectively. For the three-state and four-state cases, the red and blue curves denote the magnitude dispersion and phase dispersion among the active switching states, respectively.}
	\label{fig:beamsteering_30deg_switching}
\end{figure*}
\begin{figure*}[!t]
	\centering
	\subfloat[]{\includegraphics[width=3.32in]{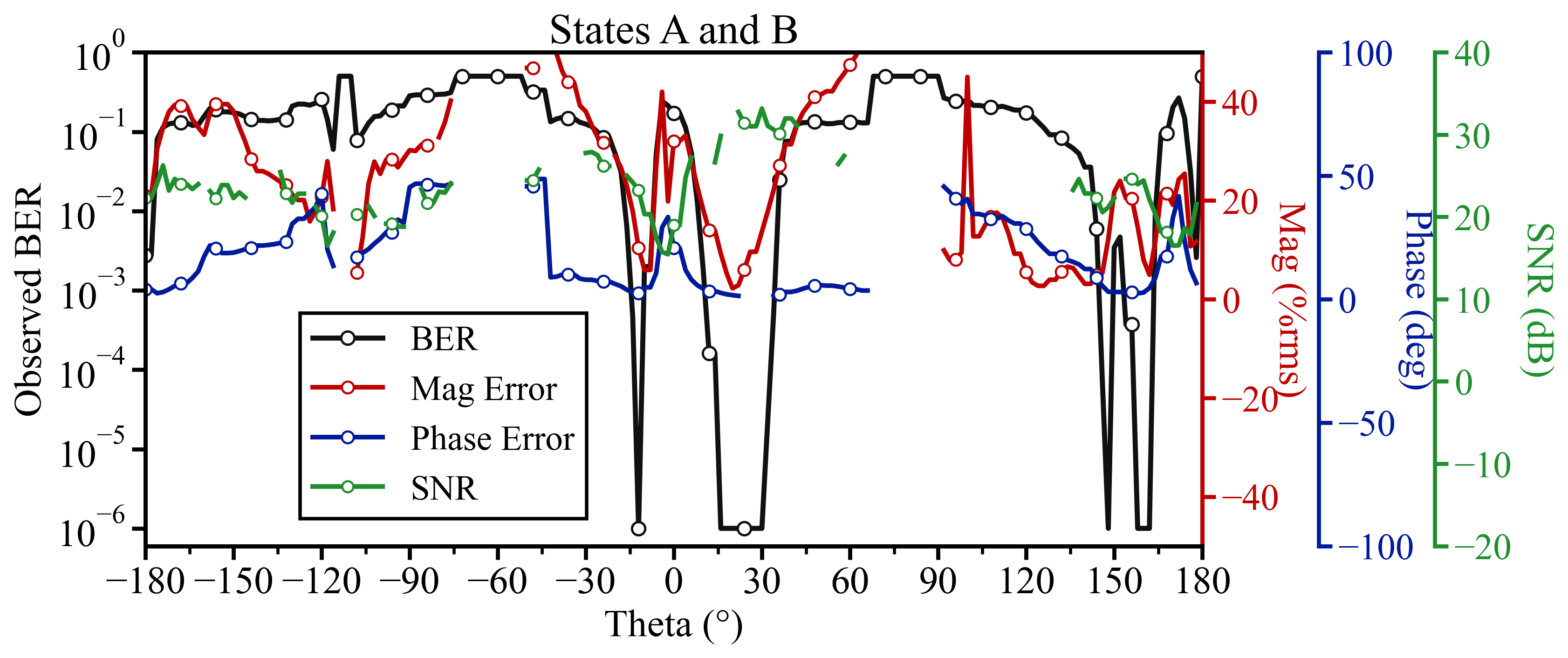}}\hfil
	\subfloat[]{\includegraphics[width=3.32in]{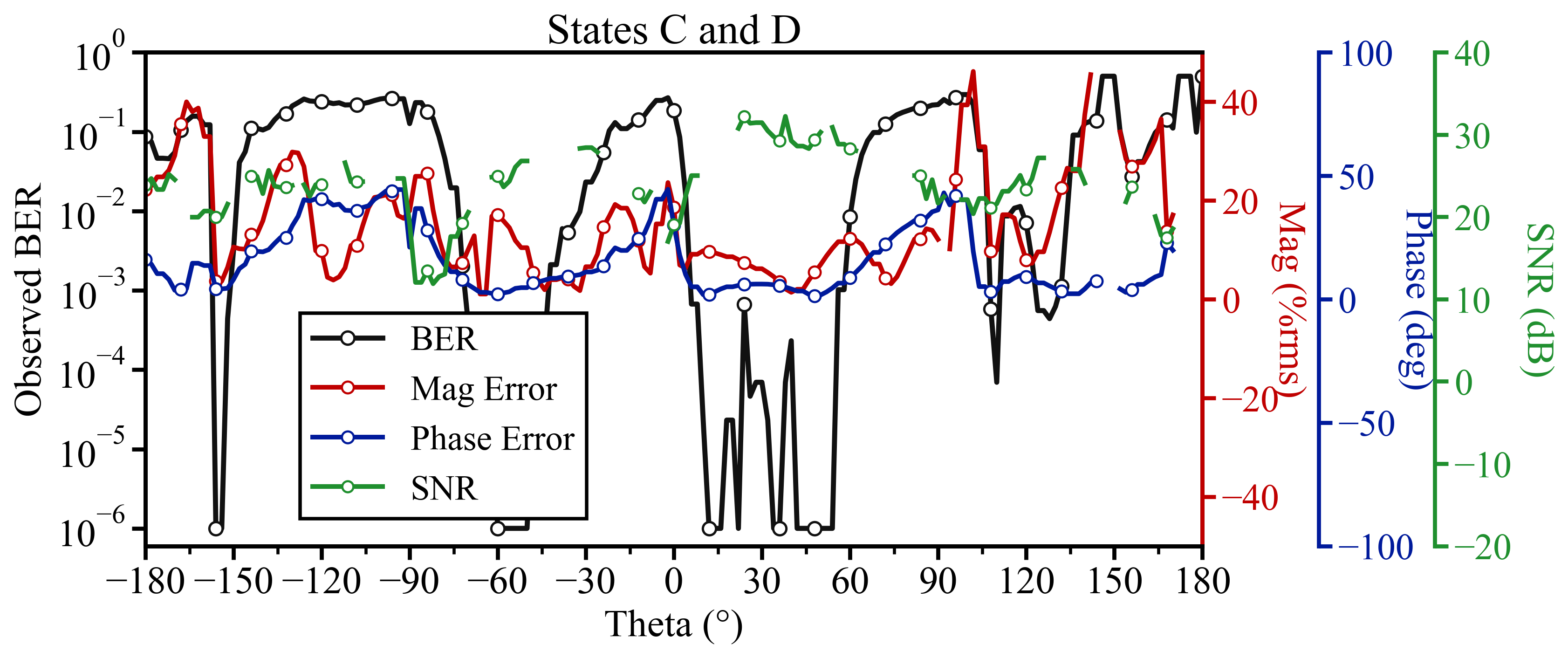}}\\
	\subfloat[]{\includegraphics[width=3.32in]{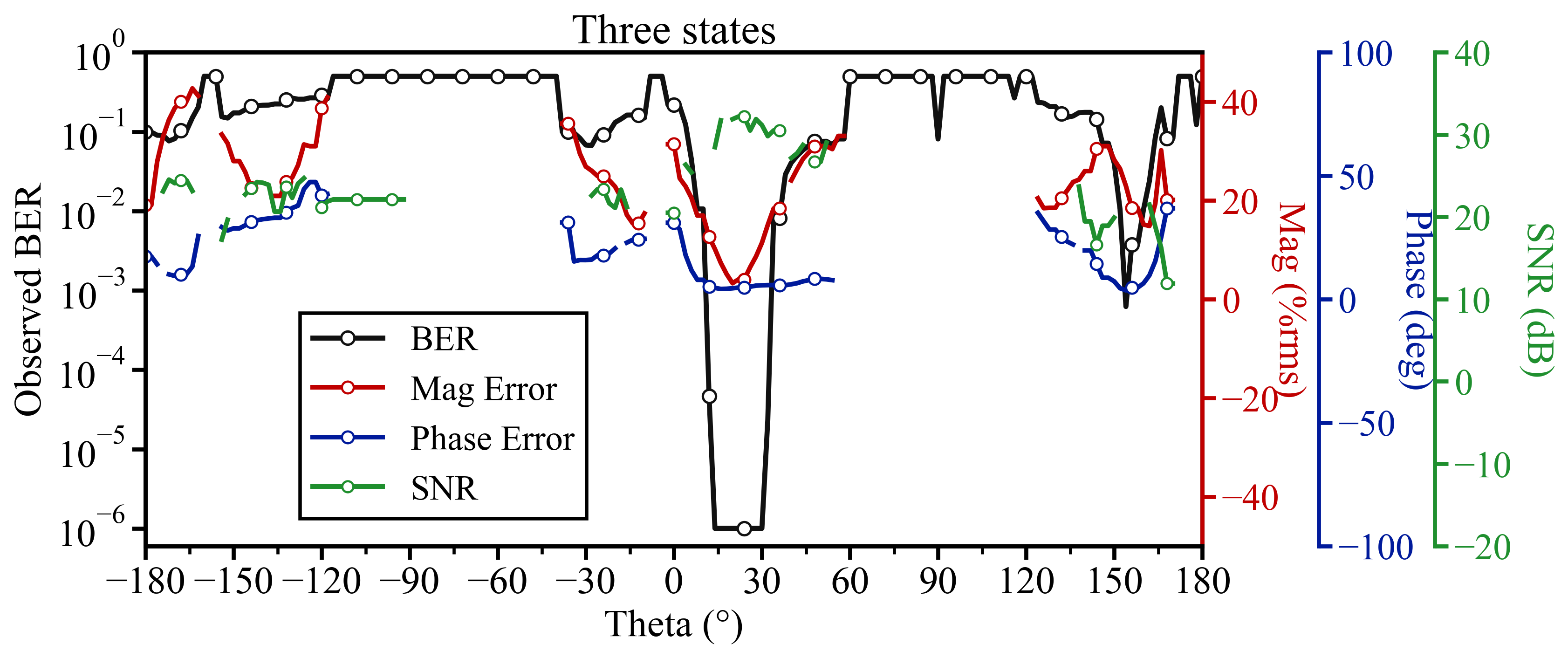}}\hfil
	\subfloat[]{\includegraphics[width=3.32in]{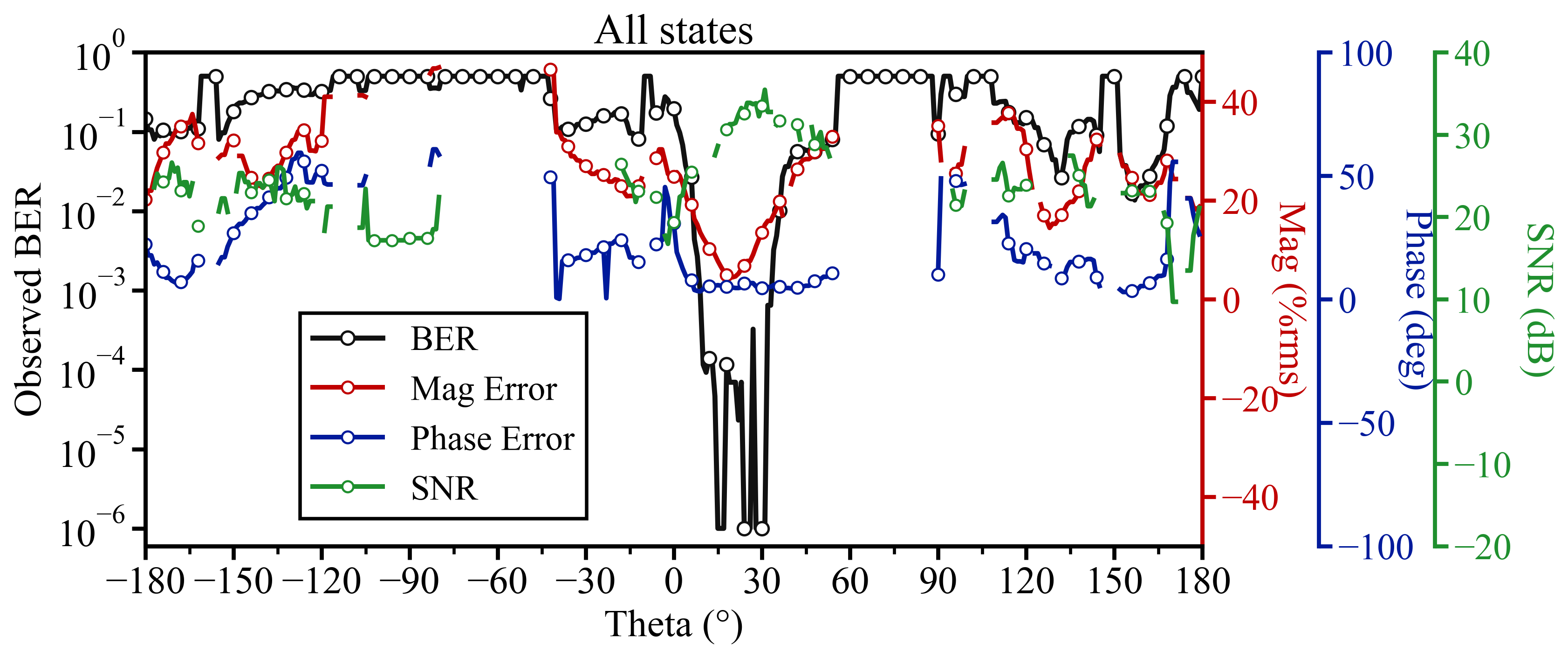}}
	\caption{Measured 16-QAM BER, magnitude error, phase error, and SNR of the proposed dynamic array under \(25^\circ\) beam-steered operation at 5.05~GHz. (a) \(A\)--\(B\) switching, (b) \(C\)--\(D\) switching, (c) \(A\)--\(C\)--\(D\) switching, and (d) full \(A\)--\(B\)--\(C\)--\(D\) switching.}
	\label{fig:measured_beamsteering_switching}
\end{figure*}
\begin{table*}[t]
	\caption{Comparison of representative directional modulation and dynamic-antenna approaches.}
	\label{tab:dm_comparison}
	\centering
	\renewcommand{\arraystretch}{1.15}
	\setlength{\tabcolsep}{4.0pt}
	\begin{tabular}{p{0.6cm} p{2.7cm} p{1.4cm} p{3.0cm} p{3.3cm} p{4.4cm}}
		\hline\hline
		\textbf{Ref.} & \textbf{Platform} & \textbf{Freq (GHz)} & \textbf{Antenna Form} & \textbf{Dynamic Quantity / Mechanism} & \textbf{Reported Result} \\
		\hline
		\cite{5422702}
		& Four-element array
		& 7
		& Microstrip patch
		& Element phase-state control
		& BER-selective directional modulation experiment \\
		
		\cite{5439878}
		& Four-element array
		& 6.9
		& Spiral microstrip
		& Element pattern-state switching
		& Beam-steered directional modulation experiment \\
		
		\cite{9665259}
		& Two-element array
		& 1.5
		& Dipole and log-periodic
		& Dynamic node separation
		& BER evaluated under distributed-array motion \\
		
		\cite{parron2021multiportdm}
		& Four-port antenna
		& 2.435
		& Circular patch
		& Multiport modal excitation
		& 360$^\circ$ steerable directional modulation \\
		
		\cite{10161710}
		& Single-element antenna
		& 2.3
		& Dipole
		& Switched current distribution
		& Measured single-element information-beam behavior \\
		
		\cite{9674846}
		& Five-element array
		& 2.4
		& Monopoles
		& Antenna-index switching
		& Energy-efficient directional modulation using one active element \\
		
		\cite{10286341}
		& Two-element array
		& 2.5
		& Microstrip patch
		& Amplitude-ratio phase-center dynamics
		& Measured information beam versus amplitude ratio \\
		
		\textbf{This work}
		& \textbf{Planar array}
		& \textbf{5.05}
		& \textbf{Four-element meander-line monopole array}
		& \textbf{Single-RF-input switching-controlled phase and magnitude perturbation}
		& \textbf{Measured E-plane information beam with omnidirectional H-plane recovery} \\
		\hline\hline
	\end{tabular}
\end{table*}
\subsection{Beamsteering Performance}
The information-recoverable sector can also be steered by adding a progressive phase distribution to the four feed paths. Because the meander-line radiators are not uniformly spaced, the applied phase offsets are selected from the effective adjacent element spacings \(d_{12}=d_{34}=20.4~\mathrm{mm}\) and \(d_{23}=19.6~\mathrm{mm}\), rather than from a single uniform array spacing. These spacings are based on phase-center locations calculated from the exported HFSS far-field phase responses, rather than only on the physical trace centers. For a target steering angle \(\theta_s\), the adjacent phase increments are chosen approximately as \(\Delta\beta_{ij}\approx-kd_{ij}\sin\theta_s\cos\phi\). This shifts the angles where the residual pairwise phase terms are small, and therefore shifts the direction where the switching states appear most similar to the receiver.

The practical steering range is also constrained by the narrowest half-power beamwidth among the static radiation states included in a switching sequence. In this array, the inside state provides the widest power beamwidth, whereas the outside state provides the narrowest power beamwidth. Therefore, switching styles that rely strongly on the inside--outside pair, or the full four-state sequence that includes this pair, are less attractive for large steering offsets because the recoverable-information sector can become limited by the narrowest constituent power beam. For this reason, the \(A\)--\(B\) and \(C\)--\(D\) two-state cases are the more practical steering modes, while the \(A\)--\(C\)--\(D\) and full \(A\)--\(B\)--\(C\)--\(D\) cases are retained mainly to illustrate the effect of adding state diversity.

The simulated steered response is shown in Fig.~\ref{fig:beamsteering_30deg_switching} for an intended information beam near \(\theta=25^\circ\). In all four switching cases, the low-BER region moves from broadside to the positive-angle steering direction, while off-beam angles retain large switching-induced field discrepancies. The two-state cases show the expected difference between the dominant perturbation channels: the \(A\)--\(B\) response is more phase-error limited, whereas the \(C\)--\(D\) response is more magnitude-error limited and therefore has a wider low-BER sector. The three-state and four-state cases combine both perturbation channels, producing stronger off-beam distortion and a more confined information-recoverable region.

The measured beam-steering results in Fig.~\ref{fig:measured_beamsteering_switching} follow the same trend. The primary measured low-BER sector spans approximately \(12^\circ\leq\theta\leq34^\circ\) for \(A\)--\(B\) switching, \(6^\circ\leq\theta\leq54^\circ\) for \(C\)--\(D\) switching, \(12^\circ\leq\theta\leq32^\circ\) for the \(A\)--\(C\)--\(D\) sequence, and \(9^\circ\leq\theta\leq33^\circ\) for the full \(A\)--\(B\)--\(C\)--\(D\) sequence. Outside these sectors, the measured BER increases by several orders of magnitude or is plotted at BER \(=0.5\) when the demodulation output is invalid. In the valid off-beam samples, the magnitude- and phase-error traces increase while the SNR remains finite, showing that the BER degradation is again produced primarily by constellation distortion rather than by a simple received-power null.

Overall, the simulation and measurement results demonstrate that the proposed antenna does not merely form a fixed broadside information beam. By applying the appropriate feed-phase distribution, the recoverable-information region can be shifted while preserving the same switching-dependent distortion mechanism that protects the off-beam directions.

Together with the comparison in Table~II, these measurements clarify the practical position of the proposed antenna among existing directional modulation and dynamic-antenna approaches. Conventional phased-array implementations provide strong spatial controllability but typically rely on multiple coherent RF channels, distributed feed networks, and precise phase control. Single-element dynamic antennas offer higher compactness, but their spatial selectivity is generally limited by the reduced number of controllable radiating degrees of freedom. The proposed antenna occupies an intermediate regime: it uses a compact planar four-element meander-line structure and a single RF input, while the calibrated switching network supplies enough state-dependent radiating degrees of freedom to form a measured information beam in the \(E\)-plane and maintain omnidirectional recovery in the \(H\)-plane. Therefore, the experimental results support the intended design tradeoff between implementation simplicity, planar integration, and antenna-level secure radiation capability.

\section{Conclusion}

A compact dynamic omnidirectional array for antenna-level physical-layer security has been presented and experimentally validated. The proposed antenna employs a four-element printed meander-line monopole array together with a single-RF-input, switching-controlled feeding network to realize directional modulation without conventional phased-array beamforming or multiple simultaneous RF chains. By using a duplicated two-cell structure, the array supports phase-dominant, magnitude-dominant, and combined phase--magnitude perturbation modes, which provide a clear physical interpretation of the resulting information beam and BER selectivity.

A theoretical framework was developed to relate the switching states, inter-cell spacing, and power ratio to the effective phase and magnitude perturbation channels. The analysis showed that the \(A\)--\(B\) switching pair is predominantly phase dominant, the \(C\)--\(D\) pair is predominantly magnitude dominant, and the \(A\)--\(C\)--\(D\) and full \(A\)--\(B\)--\(C\)--\(D\) switching sequences excite combined phase--magnitude perturbations. Full-wave simulations confirmed these trends, and measurements at 5.05~GHz using 16-QAM verified BER-defined \(E\)-plane information beamwidths of \(30^\circ\) to \(36^\circ\) for the calibrated switching modes, while quasi-static omnidirectional behavior is preserved in the \(H\)-plane.

These results demonstrate that compact planar directional modulation can be achieved through switching-induced perturbation of the radiated field rather than through conventional multi-channel beam synthesis. The proposed architecture therefore provides a hardware-efficient solution for secure wireless transmission in applications requiring low profile, planar integration, omnidirectional coverage in one principal plane, and angularly selective information recovery in another.

\bibliographystyle{ieeetr}
\bibliography{areferences}

\end{document}